\documentclass[printer]{aa}
\usepackage{graphicx}
\usepackage{txfonts}
\usepackage[colorlinks=true,linkcolor=blue,citecolor=blue,urlcolor=blue]{hyperref}
\usepackage{mathabx}
\usepackage[textsize=tiny]{todonotes}
\usepackage{verbatim}

\usepackage[]{xcolor}

\usepackage[normalem]{ulem}

\usepackage{siunitx}
\DeclareSIUnit\msun{M_\odot}
\DeclareSIUnit\year{yr}
\DeclareSIUnit\au{AU}
\DeclareSIUnit\eV{eV}
\DeclareSIUnit\erg{erg}

\begin{document}

   \title{Population study on MHD wind-driven disc evolution}

   \subtitle{Confronting theory and observation}

   \author{Jesse Weder \inst{\ref{unibe}}
          \and
          Christoph Mordasini \inst{\ref{unibe}}
          \and
          Alexandre Emsenhuber \inst{\ref{lmu}}
          }

   \institute{
   Physikalisches Institut, Universität Bern, Gesellschaftsstrasse 6, 3012 Bern, Switzerland\\ \email{jesse.weder@unibe.ch} \label{unibe}
   \and
   Universit\"ats-Sternwarte, Ludwig-Maximilians-Universit\"at M\"unchen, Scheinerstra\ss{}e 1, 81679 M\"unchen, Germany \\ \label{lmu}
    }

   \date{Received March 03, 2022 / Accepted March 31, 2023}

 
  \abstract
   {Current research has established magnetised disc winds as a promising way of driving accretion in protoplanetary discs.}
   {We investigate the evolution of large protoplanetary disc populations under the influence of magnetically driven disc winds as well as internal and external photoevaporation. We aim to constrain magnetic disc wind models through comparisons with observations.}
   {We ran 1D vertically integrated evolutionary simulations for low-viscosity discs, including magnetic braking and various outflows. The initial conditions were varied and chosen to produce populations that are representative of actual disc populations inferred from observations. We then compared the  observables from the simulations (e.g. stellar accretion rate, disc mass evolution, disc lifetime, etc.) with observational data.}
   {Our simulations show that  to reach stellar accretion rates comparable to those found by observations (\SI{\sim e-8}{\msun\per\year}), it is necessary to have access not only to strong magnetic torques, but weak magnetic winds as well. The presence of a strong magnetic disc wind, in combination with internal photoevaporation, leads to the rapid opening of an inner cavity early on, allowing the stellar accretion rate to drop while the disc is still massive. Furthermore, our model  supports the notion  that external photoevaporation via the ambient far-ultraviolet radiation of surrounding stars is a driving force in disc evolution and could  potentially exert a strong influence on planetary formation.}
   {Our disc population syntheses show that for a subset of magnetohydrodynamic wind models (weak disc wind, strong torque), it is possible to reproduce important statistical observational constraints. The magnetic disc wind paradigm thus represents a novel and appealing alternative to the classical $\alpha$-viscosity scenario.}


   \maketitle

%
%
%

\section{Introduction}
In recent years, the number of observed extrasolar planets has continued to rise exponentially. The growing amount of observational data offers the opportunity to put theoretical planet formation models to the test \citep[e.g.][]{Alessi2018,Emsenhuber2021a,Emsenhuber2021b}. In working to understand the processes involved with planet formation, we recognise the key role played by the evolution of the protoplanetary disc, as it delivers the material from which the planets grow. Currently, most global planet formation models rely on the viscous $\alpha$-disc model \citep{Shakura1973,LyndenBell1974}, where $\alpha$ is the scaling parameter for the effectiveness of angular momentum transport, whose physical origins are still under debate. Recently, the focus shifted towards magnetic fields as a promising new alternative to inducing angular momentum transport.

Protoplanetary discs are thought to be threaded by a large-scale poloidal magnetic field as a remnant of the parental molecular cloud core magnetic field. The presence of such a magnetic field is known to have strong influence on accretion processes inside the disc. Magnetorotational instability \citep[hereafter, MRI;][]{Balbus1991,Balbus1997} was regarded as a promising mechanism for inducing turbulence and, hence, angular momentum transport in protoplanetary discs, whereas the presence of a magnetic field can also drive a magneto centrifugal wind, removing both mass and angular momentum \citep{Blandford1982,Konigl2010}. Although low ionization rates are sufficient for the magnetic field to couple with the gas, non-ideal magnetohydrodynamics (hereafter, MHD) effects, such as ohmic dissipation and ambipolar diffusion, can render large swathes of the disc dead (MRI inactive). \cite{Gammie1996a} realised that ohmic dissipation would create a region near the midplane where MRI is suppressed (the so-called dead zone) and suggested that accretion is happening via MRI active layers above the midplane. When including ambipolar diffusion, the dead zone is extended to regions with low density, allowing only for a very thin layer to drive accretion \citep{Perez-Becker2011a,Perez-Becker2011b,Bai2013}. This leaves magneto-centrifugal winds as a promising mechanism to explain  the angular momentum removal. Far-ultraviolet (FUV) radiation can lead to the efficient ionization of the upper layer of the disc, while strongly coupling gas to the magnetic field. Gas is being loaded onto the field lines and centrifugally accelerated, removing both mass and angular momentum.

Observing magnetic fields and related outflows in protoplanetary discs is a challenging task that has only been accomplished very recently. There is evidence for hourglass-shaped magnetic fields in NGC 2024 and NGC 1333 IRAS4A \citep{Crutcher2012}. Furthermore, \cite{Harrison2021} recently measured upper limits on the magnetic field strength of AS 209, while \cite{Whelan2021} found evidence of an MHD disc wind emerging from the two accreting T Tauri stars: RU Lupi and AS 205 N.

Magnetically driven disc winds may not only have strong influence on angular momentum transport inside the protoplanetary disc, but they are also able to inhibit or even revert the type I migration of protoplanets \citep{Ogihara2015a,Ogihara2015,Suzuki2016}. The last of the mentioned works introduced a prescription for the removal of angular momentum and mass by MHD disc winds in a 1D disc model. However, the authors could not constrain all the parameters of their model because of the lack of observational detections of the process as well as the uncertainties in the underlying hydrodynamical simulations. Here, we aim to better constrain their prescription by looking at population-level results. We investigate the evolution of MRI-inactive (i.e. low-$\alpha$) protoplanetary discs under the combined effects of magnetically driven disc winds as well as internal and external photoevaporation. We use different model settings that were suggested in \citet{Suzuki2016} and we determine which ones match protoplanetary disc observations in terms of lifetimes, stellar accretion rates, and masses. This will allow us to investigate the imprint of magnetically driven disc winds on planet populations in future works.

This paper is structured as follows: In §\ref{sec:methodology} we give a full description of the evolution model and initial conditions. The results of disc population syntheses are shown in comparison with observational data in §\ref{sec:results}, followed by a discussion in §\ref{sec:discussion}. Our conclusions are listed in §\ref{sec:conclusion}.

\section{Methodology}
\label{sec:methodology}
We conducted 1D simulations of protoplanetary disc evolution, including the effects of magnetically driven disc winds and both internal and external photoevaporation, across a wide range of initial conditions (e.g. initial disc mass, inner disc truncation radius, ambient FUV field strength, etc.). The distributions of initial conditions are chosen to reflect the conditions found in young star forming regions. The disc evolution model is presented in §\ref{subsec:disc_evo_model} and initial conditions are given in §\ref{subsec:initial_conditions}.

\subsection{Disc evolution model}
\label{subsec:disc_evo_model}
The model at hand is an enhanced version of the gas disc model used in the Generation III \textit{Bern} global model of planetary formation and evolution \citep{Emsenhuber2021a}. The new model includes magnetically driven disc winds according to the model of \cite{Suzuki2016} to account for the removal of angular momentum and mass  originating from magnetic fields threading the disc. Furthermore, here we use a more physically motivated model for the external photoevaporation based on the FRIED grid from \cite{Haworth2018}. In the following, we describe the model, with an emphasis on the newly added parts.

The protoplanetary disc is treated as a 1D radial axissymmetric structure in cylindrical coordinates ($r,\phi,z$). The time evolution of the surface density $\Sigma(r) = \int\rho(r,z)dz$ is performed by numerically solving the evolution equation:
\begin{equation}\label{eq:disc_evolution}
   \begin{split}
      \frac{\partial \Sigma}{\partial t} =& \frac{1}{r}\frac{\partial}{\partial r}\left[ \frac{3}{r\Omega}\frac{\partial}{\partial r}(r^2\Sigma \overline{\alpha_{r\phi}} c_\mathrm{s}^2) \right] + \frac{1}{r}\frac{\partial}{\partial r}\left[ \frac{2}{\Omega}r \overline{\alpha_{\phi z}}(\rho c_\mathrm{s}^2)_\mathrm{mid} \right] \\
      &- \dot{\Sigma}_\mathrm{MDW} - \dot{\Sigma}_\mathrm{PEW,int} - \dot{\Sigma}_\mathrm{PEW,ext}.
  \end{split}
\end{equation}
Here, we use standard notation where $r$ is the distance from the host star, $t$ is the time,   and $\Omega$ corresponds to the angular velocity which is assumed to be Keplerian $\Omega = \sqrt{G \cdot M_\star/r^3}$, with  $G$ being the gravitational constant, $c_\mathrm{s}=\sqrt{k_\mathrm{B}T_\mathrm{mid}/(\mu m_\mathrm{H})}$ is the sound speed, and the molecular weight is $\mu=2.24$, hydrogen atom mass, $m_\mathrm{H}$, with $k_\mathrm{B}$ being the Boltzmann constant, and $\rho$ is the gas density. The subscript $()_\mathrm{mid}$ denotes values at the midplane. $\overline{\alpha_{r\phi}}$ and $\overline{\alpha_{\phi z}}$ are parametrisations for the redistribution and removal of angular momentum.

The equation incorporates classical viscous diffusion, advection through angular momentum removal and sink terms for removal of mass. It is based on the equation derived in Appendix A of \cite{Suzuki2016}. Here, $\overline{\alpha_{r\phi}}$ corresponds to the effective viscosity used in the classical $\alpha$-disc model \citep{Shakura1973}; it is mathematically connected to a turbulent viscosity via $\nu = \overline{\alpha_{r\phi}} c_\mathrm{s}H$.\footnote{We note that our $\overline{\alpha_{r\phi}}$ differs by $3/2$ from the one used in \cite{Suzuki2016} since we used different definitions of $\overline{\alpha_{r\phi}}$ and $H=c_\mathrm{s}/\Omega$.} However, in the picture of magnetic fields, we ought to think of it as a parameter for angular momentum transport through MRI. The advection is parameterised by $\overline{\alpha_{\phi z}}$, which corresponds to the non-dimensional stress acting as a torque on the disc, removing angular momentum and driving accretion inside the disc. Furthermore, we have various sink terms for mass removal via magnetically driven disc winds ($\dot{\Sigma}_\mathrm{MDW}$), internal photoevaporation ($\dot{\Sigma}_\mathrm{PEW,int}$), and external photoevaporation ($\dot{\Sigma}_\mathrm{PEW,ext}$). This is similar to \citet{Kunitomo2020}, although we concurrently include internal and external photoevaporation. A cold MHD wind and photoevaporation are both of different physical nature and, strictly speaking, we would expect a single wind of an intermediate nature \citep{Bai2016a}. Here, we represent the view where both winds play important roles at different stages of the disc evolution \citep[e.g.][]{Pascucci2022}.

The equation is solved on a logarithmically-spaced grid with \num{3400} grid cells between the inner disc edge, given by an initial condition (§ \ref{subsubsec:inner_disc_edge}), and \SI{1000}{\au}. We use an advection-diffusion algorithm derived in Appendix A of \cite{Birnstiel2010a}. We set a minimum value surface density of $\Sigma_\mathrm{min}=\SI{e-4}{\gram\per\square\centi\meter}$ throughout the grid and impose Dirichlet boundary conditions $\Sigma_\mathrm{min}$ at the disc edges.

\subsubsection{Thermal structure}
The thermal structure of the disc is evaluated at each time evolution step of the disc, following the approach in \cite{Nakamoto1994}:
\begin{equation}\label{eq:therm_structure}
   \sigma_\mathrm{SB}T_\mathrm{mid}^4 = \frac{1}{2} \left( \frac{3}{8}\tau_\mathrm{R} + \frac{1}{2\tau_\mathrm{P}} \right)F_\mathrm{rad} + \sigma_\mathrm{SB}T_\mathrm{ext}^4.
\end{equation}
Here, $\sigma_\mathrm{SB}$ is the Stefan-Boltzmann constant. $T_\mathrm{mid}$ corresponds to the temperature at the midplane. $F_\mathrm{rad}$ is the transferred energy from viscous dissipation and liberated gravitational energy; $T_\mathrm{ext}$ accounts for external sources of heating. The Planck mean optical depth is given by $\tau_\mathrm{P}=2.4\tau_\mathrm{R}$, while the Rosseland mean optical depth is expressed as $\tau_\mathrm{R}=\kappa(\rho_\mathrm{mid},T_\mathrm{mid})\Sigma$, where the opacity $\kappa$ is a function of midplane density and temperature. We use the maximum of the values given by the expressions from \cite{Bell1994} and \cite{Freedman2014}; the latter for a solar composition.

External heating is considered by disc surface irradiation- and direct irradiation through the midplane by the host star and heating from the surrounding molecular cloud;
\begin{equation}
   T_\mathrm{ext}^4 = T_\mathrm{irr,surf}^4 + T_\mathrm{irr,direct}^4 + T_\mathrm{cloud}^4.
\end{equation}
For the irradiation of the disc surface by the host star, we follow the calculations for flared discs \citep{Ruden1991,Hueso2005}:
\begin{equation}
   T_\mathrm{irr,surf}^4 = T_\star^4 \left[ \frac{2}{3\pi} \left( \frac{R_\star}{r} \right)^3 + \frac{1}{2} \left( \frac{R_\star}{r}\right)^2 \left(\frac{H}{r}\right) \left( \frac{\partial \ln(H)}{\partial(r)}-1 \right) \right],
\end{equation}
where we adopt $\partial \ln(H)/\partial \ln(r) = 9/7$ from \cite{Chiang1997}. Since the inner edge of the disc is directly exposed to irradiation from the host star, the disc midplane gets additionally heated, as follows:
\begin{equation}
   T_\mathrm{irr,direct}^4 = \frac{L_\star}{16\pi r^2 \sigma_\mathrm{SB}}e^{-\tau_\mathrm{mid}},
\end{equation}
while taking into account the optical depth through the midplane, given as $\tau_\mathrm{mid} = \int_0^r \rho_\mathrm{mid}\kappa_\mathrm{mid}(\rho_\mathrm{mid},T_\mathrm{mid})dr'$. To account for heating of the disc surface by the surrounding molecular cloud, we add a constant value of $T_\mathrm{cloud} = 10\mathrm{K}$. This is a strong simplification, since we neglected variations in the heating from the stellar cluster environment \citep[e.g.][]{Ndugu2018}.

\subsubsection{Magnetic disc winds}
\label{subsubsec:magnetic_disc_winds}
Magnetically driven disc winds are incorporated using the model from \cite{Suzuki2016}. The model considers both removal of angular momentum (magnetic braking) and mass.

The magnetic braking results in the advective term in the evolution equation (Eq. \ref{eq:disc_evolution}) and is parametrised by $\overline{\alpha_{\phi z}}$. Strength and time evolution of the magnetic field is still not well understood. \cite{Bai2013a} reported a positive dependence on the strength of the magnetic field $\overline{\alpha_{\phi z}} \propto (B_z^2/8\pi(\rho c_\mathrm{s}^2)_\mathrm{mid})^{0.66}$, with torques between $\overline{\alpha_{\phi z}}\sim10^{-5}-10^{-3}$ from local MHD simulations. We adopt the two parametrisations for $\overline{\alpha_{\phi z}}$ presented in \cite{Suzuki2016}.

   Firstly, in the case where the magnetic field diffuses outwards with decreasing $\Sigma$, the torque will stay approximately constant due to the dependency of $\rho$ on $\Sigma$ and we therefore chose $\overline{\alpha_{\phi z}} = \mathrm{const}$.
   Secondly, if the magnetic field stays constant, only $\rho$ is dependent on $\Sigma$. This can be parametrised by \begin{equation}\overline{\alpha_{\phi z}} = \overline{\alpha_{\phi z, 0}} \cdot(\Sigma/\Sigma_\mathrm{init})^{-0.66},\end{equation} where $\overline{\alpha_{\phi z, 0}}$ is the initial value of $\overline{\alpha_{\phi z}}$ and $\Sigma_\mathrm{init}$ is the initial surface density at the given location.

The sink term for mass removal by magnetic disc winds is given by
\begin{equation}
   \dot{\Sigma}_\mathrm{wind} = C_\mathrm{w}(\rho c_\mathrm{s})_\mathrm{mid},
\end{equation}
with $C_\mathrm{w}$ being a non-dimensional mass loss rate. This non-dimensional mass loss rate is constrained by the disc wind energetics $C_\mathrm{w,e}$ (Eqs. \ref{eq:strong_cwe} \& \ref{eq:weak_cwe}) and confined by an upper limit of $C_\mathrm{w,0}=10^{-5}$ to avoid very high mass fluxes,
\begin{equation}
   C_\mathrm{w} = \min(C_\mathrm{w,0},C_\mathrm{w,e}).
\end{equation}

From the conservation of total MHD energy, we can derive an energy constrain on the disc winds; speaking in terms of variables this means that the mass flux $C_\mathrm{w}$ can be expressed in terms of $\overline{\alpha_{r\phi}}$, $\overline{\alpha_{\phi z}}$, and other local quantities. We adopted two approaches, namely: 'strong disc wind' and 'weak disc wind', as discussed in \cite{Suzuki2016}.

\paragraph{Strong disc wind:}
The strong disc wind scenario assumes that all liberated gravitational energy contributes to launching a wind, while viscous heating is transferred to radiation (i.e. disc heating). This leads to:
\begin{align}
   C_{w,e} =& \max\left[ \frac{3}{r^3\Omega(\rho c_\mathrm{s})_\mathrm{mid}}\frac{\partial}{\partial r}(r^2\Sigma \overline{\alpha_{r\phi}}c_\mathrm{s}^2) + \frac{2c_\mathrm{s}}{r\Omega}\overline{\alpha_{\phi z}}, 0 \right] \label{eq:strong_cwe}, \\
   F_\mathrm{rad} =& \max\left[ -\frac{3}{2r}\frac{\partial}{\partial r}(r^2\Sigma \Omega \overline{\alpha_{r\phi}}c_\mathrm{s}^2), 0 \right].
\end{align}

\paragraph{Weak disc wind:}
In the weak disc wind scenario, the liberated energy contributes to both launching a wind and heating of the disc. The percentage that contributes to disc heating is controlled by the parameter $\epsilon_\mathrm{rad} \in [0,1]$. Here, we have
\begin{align}
   C_{w,e} =& (1-\epsilon_\mathrm{rad})\left[ \frac{9\sqrt{2\pi}c_\mathrm{s}^2}{2r^2\Omega^2}\overline{\alpha_{r\phi}} + \frac{2c_\mathrm{s}}{r\Omega}\overline{\alpha_{\phi z}} \right] \label{eq:weak_cwe}, \\
   F_\mathrm{rad} =& \epsilon_\mathrm{rad} \left[ \frac{9}{4}\Omega \Sigma \overline{\alpha_{r\phi}}c_\mathrm{s}^2 + r\Omega\overline{\alpha_{\phi z}}(\rho c_\mathrm{s}^2)_\mathrm{mid} \right].
\end{align}
It can be seen that in the limit of no disc winds ($\epsilon_\mathrm{rad}=1$ and $\overline{\alpha_{\phi,z}}=0$), we get back to the equation for the viscous dissipation rate of $F_\mathrm{rad}=\frac{9}{4}\Omega \Sigma \overline{\alpha_{r\phi}} c_\mathrm{s}^2$ \citep{Nakamoto1994,Emsenhuber2021a}. For the scenario of a weak wind, we followed \cite{Suzuki2016} and adopted $\epsilon_\mathrm{rad}=0.9$, meaning that only $\num{10}\%$ of the liberated energy goes into launching the wind.

\subsubsection{Photoevaporation}
Protoplanetary discs are exposed to high-energy radiation from the central star and the surrounding cluster. The MHD wind model at hand describes a cold, magnetocentrifugal wind which is launched by the liberated accretion energy. The presence of high energy radiation would lead to a single wind of possibly intermediate nature, namely, magneto-thermal wind \citep[e.g.][]{Bai2016a,Wang2019,Rodenkirch2020}. However, the physical nature of such a wind has yet to be fully characterised. Therefore, we investigate the scenario, where internal photoevaporation is suppressed by the emerging magnetic wind in an early phase and dominates only the final stage of the disc evolution \citep{Lesur2022,Pascucci2022}. While such a separation in time would not work for external photoevaporation, it seems likely that the presence of an external UV radiation field would lead to a considerably enhanced mass loss rate at the outer disc compared to the cold wind scenario; this is a hypothesis that will have to be investigated with magneto-thermal models in the future.

Given these assumptions, we consider both internal photoevaporation by the host star through extreme-ultraviolet radiation (EUV; $13.6\mathrm{eV} < h\nu$) and external photoevaporation by far-ultraviolet radiation (FUV; $\SI{6}{\eV} < h\nu < \SI{13.6}{\eV}$) from the surrounding cluster. The model for internal photoevaporation is the same as used in \cite{Emsenhuber2021a}, except for the EUV flux scaling and shielding by disc winds. For external photoevaporation we now use a new model based on precalculated values from \cite{Haworth2018}.
Recent work on internal photoevaporation showed that X-ray photoevaporation is expected to excel EUV photoevaporation by order of magnitudes \cite[e.g.][]{Jennings2018}. \cite{Emsenhuber2023} compared the evolution of viscous disc populations including both internal X-ray and external FUV photoevaporation. They found that the high mass-loss rates of the two leads towards too short lived discs, which would be inconsistent with observations. This coupled with the fact that X-ray can penetrate much higher column densities (making the desired separation in time by radiation shielding ineffective)\ only motivate the inclusion of a simple EUV photoevaporation prescription.

\paragraph{Internal photoevaporation:}
We used a simple parametrisation for the internal photoevaporation following the \cite{Clarke2001} model, that is, one based on the weak stellar wind model from \cite{Hollenbach1994}. EUV radiation from the host star heats the surface of the gas disc and creates a layer of ionized hydrogen with a temperature of $T_\mathrm{II}\approx 10^4\,\mathrm{K}$ and a mean molecular weight of $\mu_{\mathrm{II}}=0.68$.\footnote{Note: subscript II accounts for ionized hydrogen.} In the model from \cite{Hollenbach1994}, this heated layer is expected to launch a thermal wind at sound speed beyond some distance from the host star, where the sound speed of the ionized gas, $c_\mathrm{s,II}=\sqrt{k_\mathrm{B}T_\mathrm{II}/\mu_\mathrm{II}m_\mathrm{H}}$, is greater than the escape velocity, $v_\mathrm{e}=\sqrt{2GM_\star/r}$, and is not gravitationally bound any more (i.e. the gravitational radius; $r_\mathrm{g,II}=GM_\star/c^2_\mathrm{s,II}$). Similarly to \cite{Alexander2012}, we reduced the gravitational radius to a critical radius, $r_\mathrm{crit}$, by a factor of $\beta_\mathrm{II}=0.14$ \citep[see also][]{Liffman2003}.

Following \cite{Clarke2001}, we defined the base density of the wind at the critical radius as $n_0(r_\mathrm{crit,14})=5.7\cdot 10^{4} \Phi_{41}^{1/2}r_\mathrm{crit,14}^{-3/2}$, where $r_\mathrm{crit,14}=\beta_\mathrm{II}r_\mathrm{g,II}/10^{14}\mathrm{cm}$ is the scaling radius and $\Phi_{41}$ corresponds to the ionizing photon luminosity in units of $10^{41} \,\mathrm{s}^{-1}$. The ionizing photon luminosity is scaled with stellar mass according to $\Phi_{\mathrm{EUV}}=10^{40.7}(M_\star/\mathrm{M}_\odot)^{1.5}$ \cite[see Table 2,][]{Gorti2009}. This scaling law holds true for stellar masses $\lesssim 3 \mathrm{M}_\odot$, which is the case in our populations. The base density outside the critical radius is assumed to follow the power law $n_0(r)=n_0(r_\mathrm{crit,14})\cdot (r/r_\mathrm{crit})^{-5/2}$. Putting everything together leaves us with the sink term for the internal photoevaporation:
\begin{equation}
   \dot{\Sigma}_\mathrm{PEW,int} = 2c_\mathrm{s,II}n_0m_\mathrm{H} \qquad \mathrm{for} \quad r> \beta_\mathrm{II} r_\mathrm{g,II}.
\end{equation}
The total mass loss rate scales with stellar mass as $\dot{M}_\mathrm{PEW,int} \propto M_\star^{1.25}$, which is below recent calculations from \cite{Komaki2021}, who derived a stellar mass dependence of $\dot{M}_\mathrm{PEW,int} \propto M_\star^{2}$ considering not only EUV but also FUV and X-ray irradiation by the host star.

Photoevaporative winds can only be launched when high energy photons manage to heat the disc beyond the critical radius $r_\mathrm{crit}$. This becomes inherently difficult with a strong magnetic wind emerging from the inner disc $<1\mathrm{AU}$ and it is therefore expected that internal photoevaporation is not effective in the early stage of disc evolution \citep[][]{Pascucci2022}. In order to take this shielding effect into account, we calculate the column density of the emerging magnetic disc wind along the disc surface. This gives
\begin{equation}
    n_\mathrm{wind}(r) = \int_{r_\mathrm{in}}^{r} \frac{\dot{\Sigma}_\mathrm{wind}(r')}{2 \cdot \mu_\mathrm{wind} \cdot m_\mathrm{H} \cdot v_\mathrm{wind}} dr',    
\end{equation}
where $v_\mathrm{wind}=\SI{70}{\kilo\meter\per\second}$ is the typical velocity of the emerging wind \citep{Pascucci2022} and $\mu_\mathrm{wind}=2.23$ is the mean molecular weight of the wind in terms of the hydrogen atom mass, $m_\mathrm{H}$. The factor of $2$ takes into account that the wind emerges from both sides of the disc. We assume that the EUV radiation can penetrate only a column density of $<10^{19}\, \mathrm{cm^{-2}}$ (\SI{<e20}{\per\square\centi\meter} \citealp{Ercolano2009}; $\lesssim 10^{19-20}\, \mathrm{cm^{-2}}$ \citealp{Pascucci2022}). For regions where $n_\mathrm{wind}(r)>10^{-19}\, \mathrm{cm^{-2}}$, the internal photoevaporation is suppressed (i.e., $\Sigma_\mathrm{PEW, int}(r)=0$). This represents a simple, but physically motivated approach to describe the interplay between MHD and photoevaporative winds.

\paragraph{External photoevaporation:}
Star formation usually occurs in groups of hundreds to thousands of stars. Nearby young massive stars emit ultraviolet radiation that can heat up the gas disc, which results in a thermally driven wind. We use the  \textbf{F}UV \textbf{R}adiation \textbf{I}nduced \textbf{E}vaporation of \textbf{D}iscs (FRIED) grid from \cite{Haworth2018} to obtain mass loss rates from far-ultraviolet irradiation, which is considered to be the dominant driver of external photoevaporation \citep{Adams2004}. The grid consists of precalculated mass loss rates for various stellar masses ($0.05-\SI{1.9}{\msun}$), FUV field strengths ($10-10^4\,\mathrm{G}_0$)\footnote{The FUV flux is given by $\mathrm{G}_0$ \cite[Habing unit;][]{Habing1968}, with $1\mathrm{G}_0$ corresponding to \SI{1.6e3}{\erg\per\square\centi\meter\per\second} for the range of \num{6} to \SI{13.6}{\eV}.}, disc masses ($\num{3.2e-5}- 0.2 M_\star$ of the stellar mass), and sizes ($1-\SI{400}{\au}$). 
Linear interpolation was used to retrieve mass loss rates for given stellar masses and FUV field strengths, as well as disc sizes and masses in between the grid points.\footnote{The latter is connected to the outer surface density through the relation: $\Sigma_\mathrm{out}= M_\mathrm{disc}/(2\pi R^2_\mathrm{disc})$ \citep[Eqs. 3 \& 4 in][]{Haworth2018}. We thus used $\Sigma(r)$ to define the disc mass for a given disc size, $r$.} For the FUV field strength, we carried out a linear interpolation in log-log. For the stellar mass, we carried out a linear interpolation in lin-log (linear in stellar mass).

While stellar mass and FUV field strength are given relatively straightforward, this is not the case for the disc radius and the corresponding mass of the disc. As discussed in \cite{Sellek2020a} the effective outer radius of the disc is located at the transition between the optically thick and optically thin regime. Analogously to \cite{Sellek2020a}, we evaluate the mass loss rate for all disc grid cells and set the disc radius to where the obtained mass loss rate $\dot{M}_\mathrm{PEW,ext}$ is maximal.

The FRIED grid includes a floor value of \SI{e-10}{\msun\per\year} for the evaporation rate. This is a significant rate and prevents the proper investigation of weak FUV field environments. Nonetheless, in striving to  study them, we subtracted the floor value of \SI{e-10}{\msun\per\year} from the mass-loss rate returned by the interpolation, while ensuring a negligible minimum mass-loss rate of \SI{e-15}{\msun\per\year}. Further, we extended the range of possible FUV field strengths down to $1\mathrm{G}_0$ by linearly interpolating the value returned from the grid at $10\mathrm{G}_0$ and a new floor value of \SI{e-15}{\msun\per\year} at $1\mathrm{G}_0$.

The 2D calculations from \cite{Haworth2019} suggest that the mass loss rates from the FRIED grid have to be regarded as lower limits. Their simulations furthermore illustrated that the mass loss rate is set entirely by the outer half of the disc and originates mostly from within the outer 10\% of the disc outer edge, $R_\mathrm{edge}$, which we define as where the outer surface density reaches $\Sigma_\mathrm{min}$, our minimum value set throughout the grid. We therefore assume that the mass is removed uniformly from the outer 10\% of the disc ($\beta_\mathrm{ext}=0.9$):
\begin{equation}
   \dot{\Sigma}_\mathrm{PEW,ext} =
   \begin{cases}
       0&\qquad \mathrm{for} \quad r< \beta_\mathrm{ext} R_\mathrm{edge}\\
       \frac{\dot{M}_\mathrm{PEW,ext}}{\pi(R_\mathrm{edge}^2 - \beta_\mathrm{ext}^2R_\mathrm{edge}^2)}& \qquad \mathrm{for} \quad r \geq \beta_\mathrm{ext} R_\mathrm{edge}.
   \end{cases}
\end{equation}
This new model for external photoevaporation replaces our old method described in \cite{Emsenhuber2021a,Emsenhuber2021b}, where the mass loss rate $\dot{M}_\mathrm{PEW,ext}$ was chosen such that the synthetic disc lifetimes fitted the observations. With the loss of this tuning parameter, disc lifetimes are entirely given by the initial conditions, inferred from observations.

\begin{figure*}[h]
   \centering
   \includegraphics[width=0.32\hsize]{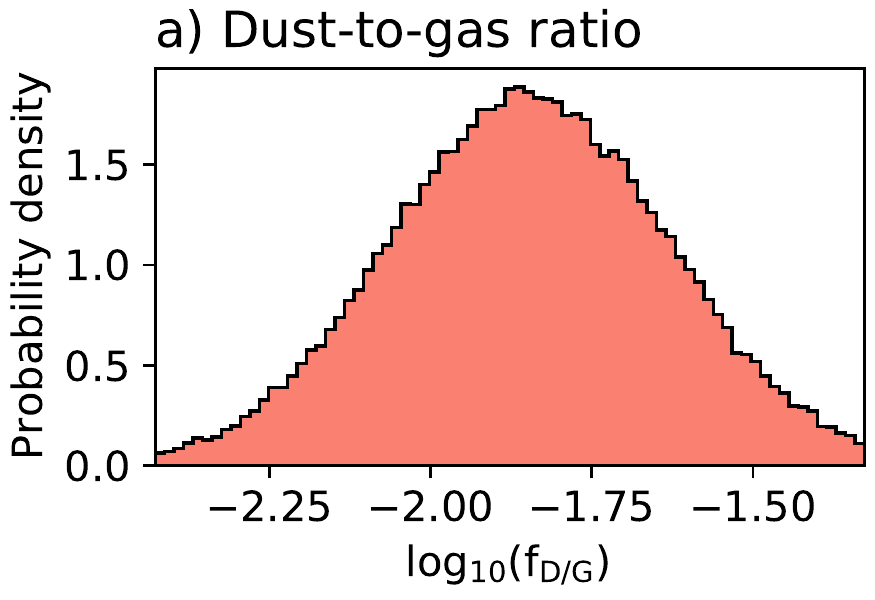}
   \includegraphics[width=0.32\hsize]{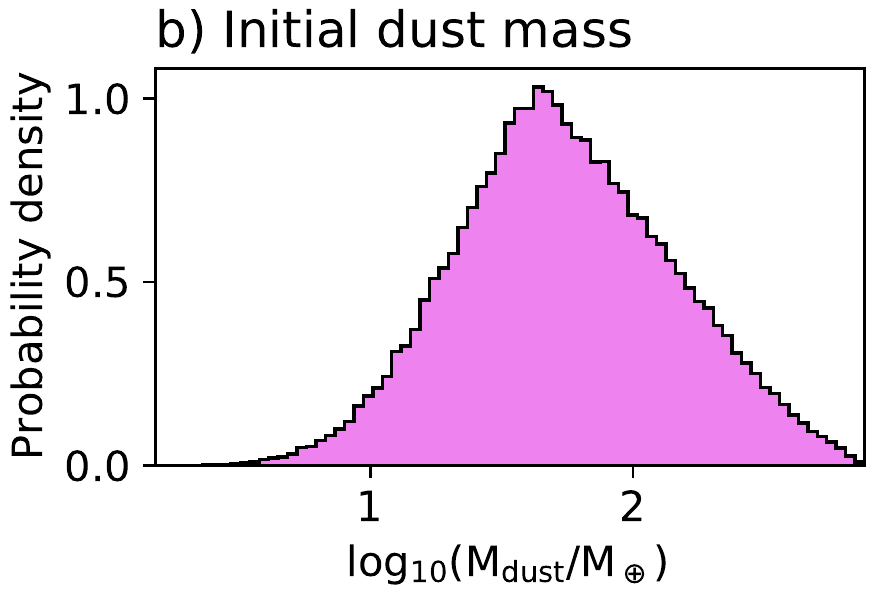}
   \includegraphics[width=0.32\hsize]{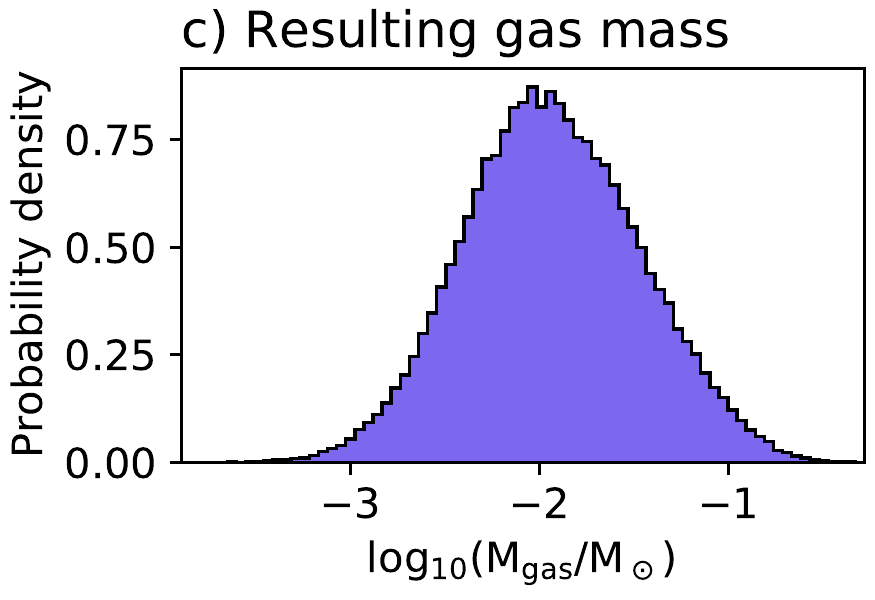}
   \\
   \includegraphics[width=0.32\hsize]{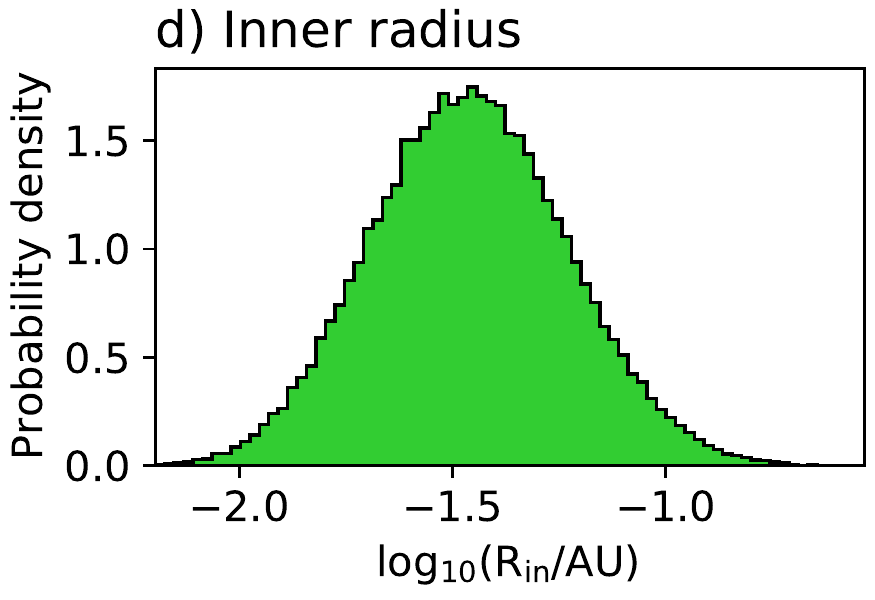}
   \includegraphics[width=0.32\hsize]{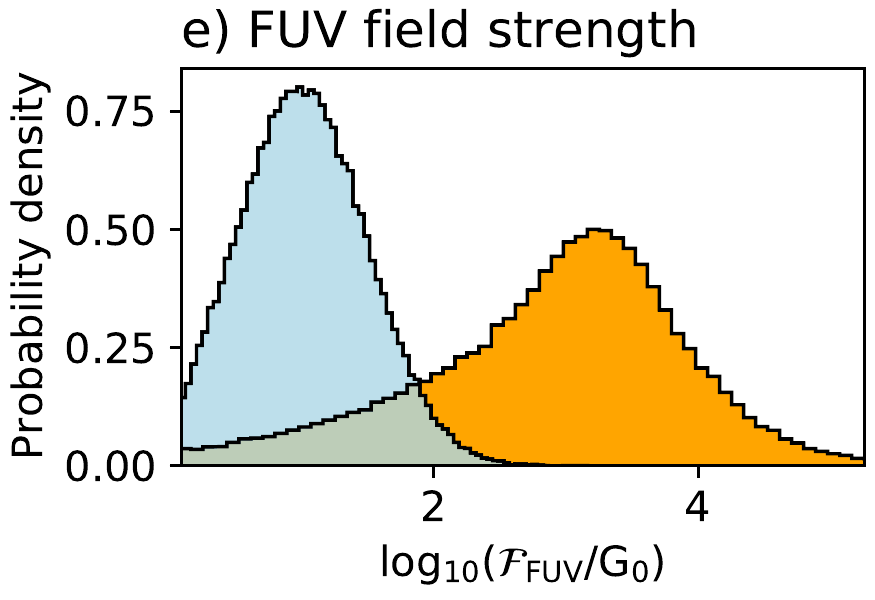}
   \includegraphics[width=0.32\hsize]{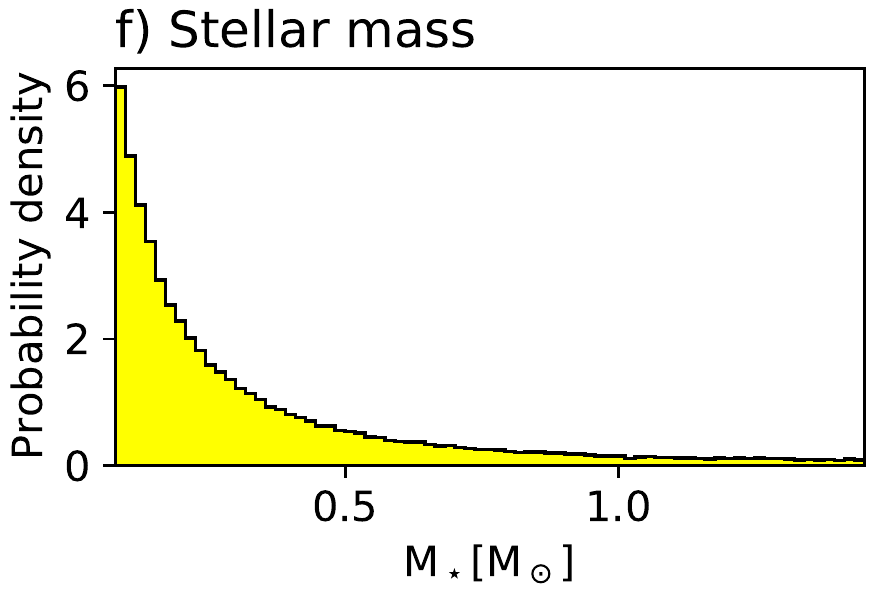}
   \caption{Adopted distributions of the Monte Carlo variables for our disc populations: dust-to-gas ratio (panel a), initial dust masses (panel b), resulting gas mass (panel c), inner disc radius (panel d), FUV field strength for weak-FUV and strong-FUV environments (panel e), and stellar mass (panel f).}
   \label{fig:init_dist}
\end{figure*}
\begin{figure}
   \centering
   \includegraphics[width=\hsize]{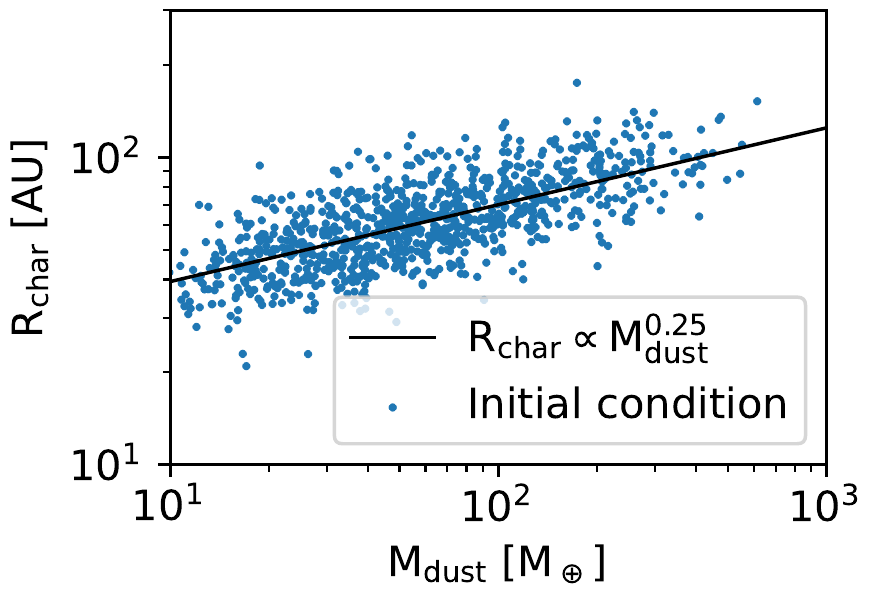}
   \caption{Initial relation of disc dust mass and characteristic radius for a synthetic population of 1000 discs, following Fig. 12 of \cite{Tobin2020}.}
   \label{fig:characteristic_radius}
\end{figure}

\subsection{Initial conditions}
\label{subsec:initial_conditions}
We used the approach of evolving large populations of protoplanetary discs, exploring a wide parameter space of initial conditions \cite[e.g.][]{Lodato2017,Mulders2017,Schib2020,Emsenhuber2021b,Tabone2022b}. We conducted disc population syntheses for different combinations of disc wind scenarios and ambient FUV field strengths, summarised in Table \ref{tab:init_cond_list}. These scenarios cover the four possible combination of strong or weak disc winds in combination with either a constant torque (decreasing magnetic field) or increasing torque (constant magnetic field), as described in Section \ref{subsubsec:magnetic_disc_winds}. Following the discussion above, all discs are assumed to be MRI inactive with $\overline{\alpha_{r\phi}}=5.3\cdot10^{-5}$ \citep{Suzuki2016}.
\begin{table*}[h]
   \begin{center}
      \caption{List of initial parameter combinations for different disc wind scenarios considered in the main text. Values are adopted from \cite{Suzuki2016} with slightly stronger initial torque strengths for the $\Sigma$-dependent cases.}
      \label{tab:init_cond_list}
      \begin{tabular}{l l l c c}
         \hline\hline
         DW-type & $\overline{\alpha_{\phi z}}$ & $\mathcal{F}_\mathrm{FUV}$ & Results \\
         \hline
         Strong DW + $\Sigma$-dep. torque & $3 \cdot$ $ 10^{-5}\{\Sigma/\Sigma_\mathrm{init}\}^{-0.66}$ & weak & Table \ref{tab:numbers_lowFUV}, Fig. \ref{fig:mass_acc_lowFUV} \& Fig. \ref{fig:discfraction} \\
         Weak DW + $\Sigma$-dep. torque & $3 \cdot$ $ 10^{-5}\{\Sigma/\Sigma_\mathrm{init}\}^{-0.66}$ & weak & Table \ref{tab:numbers_lowFUV}, Fig. \ref{fig:mass_acc_lowFUV} \& Fig. \ref{fig:discfraction} \\
         Strong DW + const. torque & $10^{-4}$ & weak & Table \ref{tab:numbers_lowFUV}, Fig. \ref{fig:mass_acc_lowFUV} \& Fig. \ref{fig:discfraction} \\
         Weak DW + const. torque & $10^{-4}$ & weak & Table \ref{tab:numbers_lowFUV}, Fig. \ref{fig:mass_acc_lowFUV} \& Fig. \ref{fig:discfraction} \vspace{0.1cm} \\
         Strong DW + $\Sigma$-dep. torque & $3 \cdot$ $ 10^{-5}\{\Sigma/\Sigma_\mathrm{init}\}^{-0.66}$ & strong & Table \ref{tab:numbers_adams}, Fig. \ref{fig:mass_acc_adams} \& Fig. \ref{fig:discfraction} \\
         Weak DW + $\Sigma$-dep. torque & $3 \cdot$ $ 10^{-5}\{\Sigma/\Sigma_\mathrm{init}\}^{-0.66}$ & strong & Table \ref{tab:numbers_adams}, Fig. \ref{fig:mass_acc_adams} \& Fig. \ref{fig:discfraction} \\
         Strong DW + const. torque & $10^{-4}$ & strong & Table \ref{tab:numbers_adams}, Fig. \ref{fig:mass_acc_adams} \& Fig. \ref{fig:discfraction} \\
         Weak DW + const. torque & $10^{-4}$ & strong & Table \ref{tab:numbers_adams}, Fig. \ref{fig:mass_acc_adams} \& Fig. \ref{fig:discfraction} \\
         \hline
      \end{tabular}
   \end{center}
\end{table*}
Besides these specified initial parameters, we also have a set of Monte Carlo variables that are distributed in order to reproduce the various conditions found in star-forming clusters, namely: 1) the dust-to-gas ratio ($f_\mathrm{D/G}$, Sect.~\ref{subsubsec:dust_to_gas}), 2) the initial disc mass ($M_\mathrm{disc,init}$, Sect.~\ref{subsubsec:init_disc_mass}),
    3) the disc's characteristic radius ($R_\mathrm{char}$, Sect.~\ref{subsubsec:characteristic_disc_radius}),
    4) the disc's inner edge ($R_\mathrm{in}$, Sect.~\ref{subsubsec:inner_disc_edge}),
    5) the stellar mass ($M_\star$, Sect.~\ref{subsubsec:stellar_mass}), and
    6) the FUV field strength ($\mathcal{F}_\mathrm{FUV}$, Sect.~\ref{subsubsec:FUV_field_strength}).

The resulting distributions of the initial parameters are shown in Fig. \ref{fig:init_dist}. With the exception of the rescaling of the disc mass with stellar mass, we assume the Monte Carlo variables to be independent of each other. However, it may well be that there is some relation between the stellar mass and the FUV field strength, for example.

\subsubsection{Initial surface density}
\label{subsubsec:init_surface_density}
We assume an initial gas surface density similar to \cite{Veras2004}, with a power-law index $\beta$ and an exponential cut-off with characteristic radius, $R_\mathrm{char}$. The inner disc is assumed to be truncated by the magnetic field of the host star through magnetospheric accretion; therefore, $R_\mathrm{in}$  is taken to be the corrotational radius. The initial surface density profile is set with
\begin{equation} \label{eq:init_surface_density}
   \Sigma_\mathrm{init}(r) = \Sigma_\mathrm{0,5.2AU} \left( \frac{r}{5.2\mathrm{AU}} \right)^{-\beta} e^{-\left( \frac{r}{R_\mathrm{char}} \right)^{2-\beta}} \left( 1-\sqrt{\frac{R_\mathrm{in}}{r}} \right).
\end{equation}
Here, $\Sigma_\mathrm{0,5.2AU}$ is the initial surface density at 5.2AU, and $\beta=0.9$ was chosen according to \cite{Andrews2010}. The value of of $\Sigma_\mathrm{0,5.2AU}$ is fully determined by the integral of Eq.~(\ref{eq:init_surface_density}), ignoring the decrease at the inner edge, which is negligible (Eq. 14 in \citealp{Emsenhuber2021a}).

\subsubsection{Dust-to-gas ratio}
\label{subsubsec:dust_to_gas}
Disc masses are usually obtained from dust continuum emission measurements. The dust-to-gas ratio, $f_\mathrm{D/G}$, is a necessary quantity to convert these initial dust masses to initial gas masses. Following \cite{Emsenhuber2021b}, we assume that stellar and disc metallicities $[\mathrm{Fe}/\mathrm{H}]$ are identical and, thus, we used the relation \citep{Murray2001}:
\begin{equation}
   \frac{f_\mathrm{D/G}}{f_\mathrm{D/G,\odot}} = 10^{[\mathrm{Fe/H}]},
\end{equation}
with the Sun's dust-to-gas ratio being $f_\mathrm{D/G,\odot}=0.0149$ \citep{Lodders2003}. For the distribution of the metallicity $[\mathrm{Fe/H}],$ we used data from the Coralie RV search sample \citep{Santos2005}, where the metallicity is normal distributed with $\mu=-0.02$ and $\sigma=0.22$. We note that we limited the metallicity to $-0.6<[\mathrm{Fe/H}]<0.5$ to avoid unrealistically high or low values in comparison to the solar neighbourhood.

\subsubsection{Initial disc mass}
\label{subsubsec:init_disc_mass}
We used a log-normal fit to dust disc masses of Class 1 protoplanetary discs in the Perseus star-forming region from \cite{Tychoniec2018} with $\log_{10}(\mu/\mathrm{M}_\Earth)=2.03$ and $\sigma=0.35\,\mathrm{dex}$. As the host star masses of the sample are not known, we assume that it is representative for stellar masses of $\mathrm{M}_\star \sim 0.3 \, \mathrm{M}_\odot$ \citep[see discussion in][]{Tobin2016}. We further assumed a simple linear correlation between the disc and stellar masses ($\mathrm{M}_\mathrm{disc} \propto \, \mathrm{M}_\star$) \citep{Raymond2007,Andrews2013}. \cite{Somigliana2022} inferred a slightly steeper initial correlation from comparing observations and numerical models of viscous disc evolution ($\mathrm{M}_\mathrm{disc} \propto \mathrm{M}_\star^{1.2 - 2.1}$). Furthermore, the relation has found to be steepening with age \citep{Pascucci2016,Somigliana2022}. We use the previously defined dust-to-gas ratio to convert the dust masses to gas masses. Again, we limited the distribution to disc masses of $\num{4e-4}M_\star < M_\mathrm{disc} < 0.16M_\star$, where the upper limit ensures self-gravitational stability.

\subsubsection{Characteristic disc radius}
\label{subsubsec:characteristic_disc_radius}
The characteristic disc radius, $R_\mathrm{char}$, is inferred from the disc dust mass using the relation $R_\mathrm{char} = 70 \cdot \left[ M_\mathrm{dust}/\SI{100}{\mathrm{M}_\Earth} \right]^{0.25}$, with an applied spread of $0.1\,\mathrm{dex}$ to achieve a similar distribution as in Fig. 12 of \cite{Tobin2020}. The resulting correlation between dust masses and characteristic radii is shown in Fig.~\ref{fig:characteristic_radius}.

\subsubsection{Inner disc edge}
\label{subsubsec:inner_disc_edge}
Following \cite{Emsenhuber2021b} once again, we assume that the inner disc is truncated by the stellar magnetic field (i.e. magnetospheric accretion). The inner radius can thus be inferred from rotation rates of young stellar objects by calculating the corotational radius $r_\mathrm{co}=(G \cdot M_\star/\Omega)^{1/3}$, with $\Omega=2\pi/P$ being the angular velocity and $P$ the rotation period of the central star. We used a log-normal distribution fit to observed rotation rates of young stars in the NGC 2264 open cluster from \cite{Venuti2017}, $\log_{10}(\mu/d)=0.676$ and $\sigma=0.306\,\mathrm{dex}$. To avoid having inner radii smaller than the initial stellar radius, we set a lower bound to $R_\mathrm{in}= \SI{1.65e-2}{\au}$ \citep[as in][]{Emsenhuber2021b}.

\subsubsection{Stellar mass}
\label{subsubsec:stellar_mass}
Both internal and external photoevaporation processes are sensitive to the mass of the host star. Observed protoplanetary discs usually surround stars with stellar masses below $1\,\mathrm{M}_\odot$. We therefore employed a commonly used parametrization for the initial mass function by \cite{Chabrier2003}, with a log-normal distribution for $M_\star \leq 1\,\mathrm{M}_\odot$ and a power-law decay above $M_\star > 1\,\mathrm{M}_\odot$. Our stellar mass distribution spans a range from $0.08\,\mathrm{M}_\odot$ to $1.5\,\mathrm{M}_\odot$, with a median mass of $0.21\,\mathrm{M}_\odot$, which is in agreement with observations of stars hosting a protoplanetary disc \cite[e.g.][]{Rigliaco2011,Tobin2020}.

\subsubsection{FUV field strength}
\label{subsubsec:FUV_field_strength}
The local FUV flux, $\mathcal{F}_\mathrm{FUV}$, which drives external photoevaporation, is determined by the location inside the stellar cluster. We used two approaches to characterise either strong or weak FUV field environments. For the strong FUV field approach, we used an ensemble distribution for FUV fluxes experienced by stars in a cluster, calculated by \citet[their Fig. 9]{Adams2006}. The distribution shows approximately log-normal behaviour with a tail towards lower values of $\mathrm{G}_0$.
We approached the weak FUV field by assuming a simple log-normal distribution with $\log_{10} \left(\mu / \mathrm{G}_0 \right) = 1$ and $\sigma=0.5\,\mathrm{dex}$. Values beyond $10^4\,\mathrm{G}_0$ are treated as the boundary value. This step is  not expected to have any consequence in this case, since discs in a FUV field of $10^4\,\mathrm{G}_0$ or more have already evaporated after a few $10^5$ years.

\section{Results}
\label{sec:results}

For each of the eight combinations listed in Table \ref{tab:init_cond_list}, we calculated the temporal evolution of a population, comprising \num{1000} discs each. The resulting observables (e.g. stellar accretion rate, disc mass, disc lifetime, etc.) can thus be compared with observational data.

\subsection{Secular evolution: Exemplary cases}
\label{subsec:secular_evolution}

\begin{figure*}
   \centering
   \includegraphics[width=\hsize]{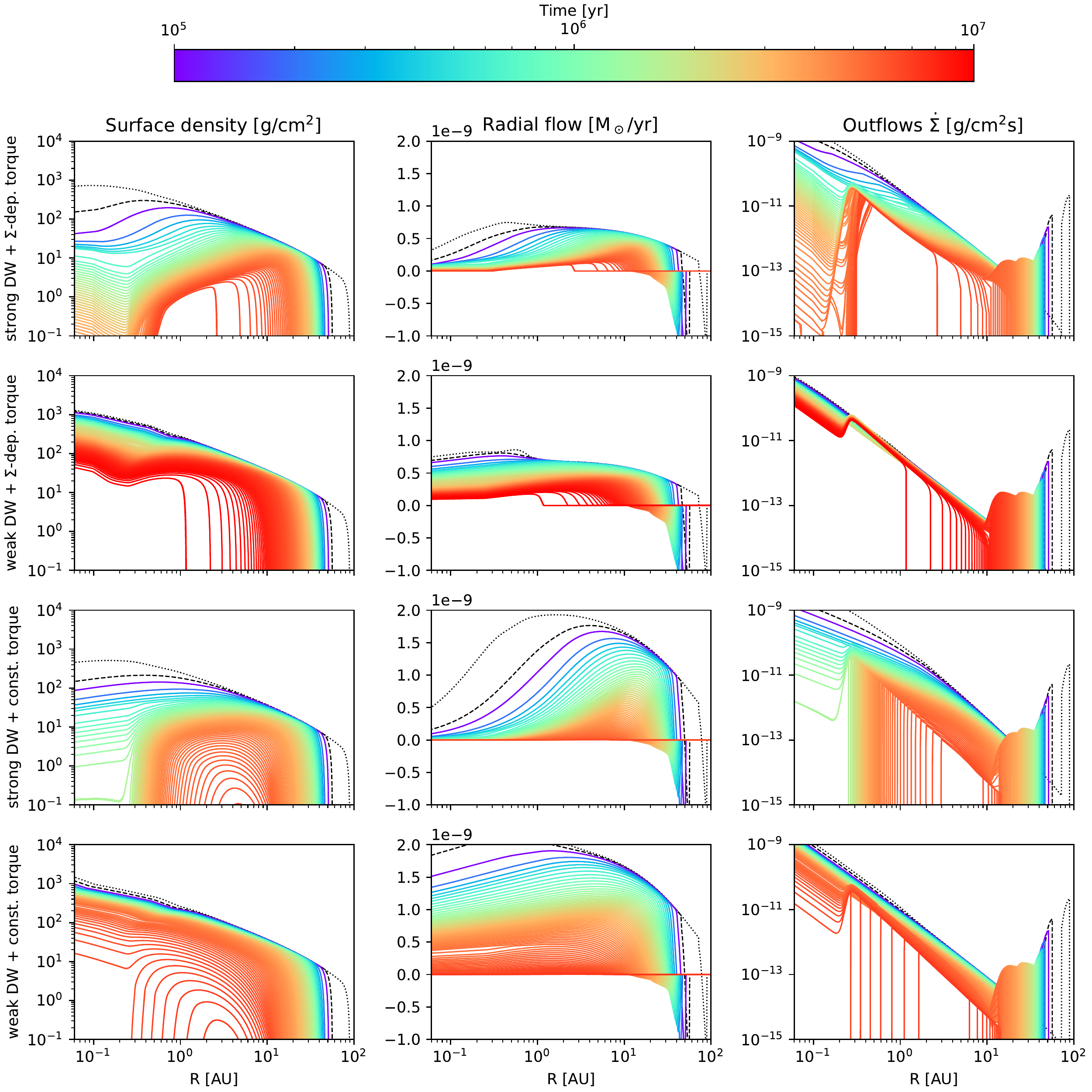}
   \caption{Exemplary disc evolution shown for initial conditions: $M_\mathrm{disc}=0.1\,\mathrm{M}_\star$ with $\mathrm{M}_\star=0.25\,\mathrm{M}_\odot$, $\mathrm{R}_\mathrm{in}=0.04\,\mathrm{AU}$ and $\mathcal{F}_\mathrm{FUV}=10\,\mathrm{G}_0$. We show temporal evolution of the surface density, radial mass flow rate and rate of change in surface density due to outflows (i.e. magnetic disc winds, internal- and external photoevaporation) for all disc wind scenarios. The individual contributions can be distinguished roughly when considering the outflows $\dot{\Sigma}$. External photoevaporation corresponds to the outermost peak while internal photoevaporation acts at $\sim\,0.3\mathrm{AU}$ to a few $\mathrm{AU}$. MHD winds remove mass from the whole disc and increase in strength towards the inner disc. Dotted lines show the evolution at $10^{4}\,\mathrm{yrs}$, dashed lines at $5\cdot10^{4}\,\mathrm{yrs}$. Coloured lines are spaced by $10^{5}\,\mathrm{yrs}$.}
   \label{fig:exemplary_case_10g0}
\end{figure*}

\begin{figure*}
   \centering
   \includegraphics[width=\hsize]{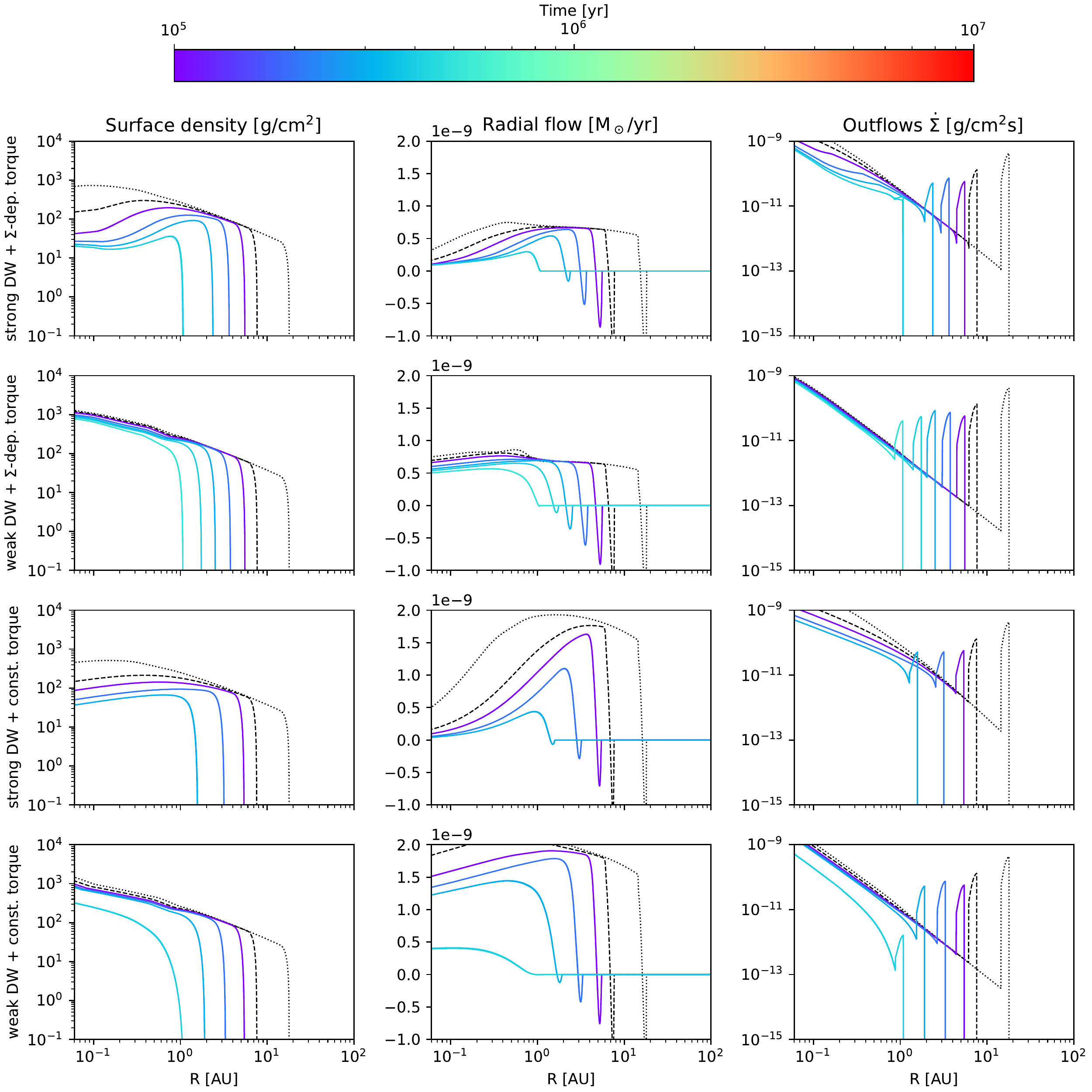}
   \caption{Exemplary disc evolution, analogous to Fig. ~\ref{fig:exemplary_case_10g0} but for a strong ambient FUV field strength of $5000\,\mathrm{G}_0$.}
   \label{fig:exemplary_case_5e3g0}
\end{figure*}

Before going into the results of the population syntheses, we offer a brief discussion of some general characteristic features of our model. We selected a $0.1\, \mathrm{M}_\star$ disc around a $0.25\, \mathrm{M}_\odot$ star, which gives an initial disc mass of $0.025\,\mathrm{M}_\odot$, and we evolved it for different disc wind scenarios. Figure \ref{fig:exemplary_case_10g0} shows the evolution of the disc in a $10\,\mathrm{G}_0$ FUV field, while Fig.~\ref{fig:exemplary_case_5e3g0} shows the evolution of the same disc in a $5000\,\mathrm{G}_0$ environment. For the eight resulting discs, we show the detailed surface density, radial mass flow, and outflow evolution. We compute the radial mass flow as:
\begin{equation} \label{eq:acc_rate}
   \dot{M}_\mathrm{acc}(r) = \frac{6\pi}{r\Omega}\frac{\partial}{\partial r}(r^2\Sigma \overline{\alpha_{r\phi}}c_s^2) + \frac{4\pi}{\Omega}r\overline{\alpha_{\phi z}}(\rho c_s^2)_\mathrm{mid},
\end{equation}
where the first term accounts for contributions of the diffusive term and the second term represents accretion driven by magnetic braking \citep{Suzuki2016}. The outflows contain the sum of the three different components (magnetic winds and internal and external photoevaporation).

In the weak $\mathcal{F}_\mathrm{FUV}$ case (Fig.~\ref{fig:exemplary_case_10g0}), the evolution of the surface density is similar to the model of \citet{Kunitomo2020}, with the exception of the outer disc truncation due to external photoevaporation. Over a large part of the disc (except for the outer disc), magnetic winds initially dominate the mass loss. With strong winds, the inner disc is exposed to a strong wind, resulting in a rapid decrease in the surface density in the inner region. This also means that only a fraction of the material reaches the inner disc, thus the stellar accretion rate always remains  low ($\sim\SI{e-10}{\msun\per\year}$). With weak disc winds, the contrast of the radial mass flow in the inner and outer region is much lower, leading to larger stellar accretion rates. Furthermore, the torque strength has a strong influence on the radial mass flow. With the $\Sigma$-dependent torque prescription we use here (which has an initial torque strength of $\overline{\alpha_{\phi z, 0}}=3\cdot \num{e-5}$), the radial mass flow in the disc is $\lesssim\SI{1e-9}{\msun\per\year}$; whereas for a strong constant torque ($\overline{\alpha_{\phi z}}=\num{e-4}$), it is larger initially. The accretion rate in the $\Sigma$-dependent case decreases less rapidly with time. For this set of initial conditions, only the combination of weak disc winds and a strong constant torque is able to produce a stellar accretion rate larger than \SI{e-9}{\msun\per\year}. The $\Sigma$-dependent torques with larger initial values of $\overline{\alpha_{\phi z, 0}}$ could also produce larger stellar accretion rates (see §\ref{subsec:varAlpha} for some insights on higher initial $\overline{\alpha_{\phi z,0}}$ values). In all but one case, an inner cavity opens up towards the end of the evolution, leading to a complete stop of accretion onto the star.

For a field of $5000\,\mathrm{G}_0$ (Fig.~\ref{fig:exemplary_case_5e3g0}), the discs are dispersed very rapidly $\lesssim 1\,\mathrm{Myr}$ outside-in;  thus, there is not enough time for an inner cavity to open up. Here, internal photoevaporation does not play a role in disc dispersal, as the discs are quickly dispersed from the outside. As external photoevaporation has little effect in the inner region of the disc, we still find that only the combination of weak disc winds and strong constant torques produces a stellar accretion rate of \SI{e-9}{\msun\per\year} or more.

We note that the radial flow diagrams show a rapid outflow at the outer edge of the disc. This is due to the rapid fall off in surface density and the low-$\alpha$ viscosity imposed. However, we do not witness any viscous spreading of the disc, as external photoevaporation (even in the case of  weak $\mathcal{F}_\mathrm{FUV}$) compensates for this effect.


\subsection{Stellar accretion rate and disc dispersal processes}
\label{subsec:macc_disc}

\begin{table*}[h]
   \begin{center}
      \caption{Characteristics for different weak FUV field populations.}
      \label{tab:numbers_lowFUV}
      \begin{tabular}{l c c c r r r r}
         \hline\hline
            DW scenario & $t_{1/2}$\textsuperscript{a)} & $f_\mathrm{disc,2Myr}$\textsuperscript{b)} & $f_\mathrm{cavity}$\textsuperscript{c)} & $f_{M_\mathrm{acc}}$\textsuperscript{d)} & $f_{M_\mathrm{PEW,int}}$\textsuperscript{d)} & $f_{M_\mathrm{PEW,ext}}$\textsuperscript{d)} & $f_{M_\mathrm{MDW}}$\textsuperscript{d)} \\
                        & [\SI{}{\mega\year}] & [\SI{}{\percent}] & [\SI{}{\percent}] & [\SI{}{\percent}] & [\SI{}{\percent}] & [\SI{}{\percent}] & [\SI{}{\percent}] \\
         \hline
            strong DW + $\Sigma$-dep. torque  & 1.19   & 32 &  57 & 0.7   & 11.4 & 52.5 & 35.5 \\
            weak DW + $\Sigma$-dep. torque    & 6.45   & 88 &  35 &  10.8   & 19.0 & 53.6 & 16.6 \\
            strong DW + const. torque         & 0.80  & 10 &  63 &  0.1  & 8.7 & 49.3 & 41.9 \\
            weak DW + const. torque           & 6.48  & 81 &  43 & 16.6   & 13.6 & 49.9 & 19.9 \\
         \hline
      \end{tabular}
   \end{center}
   Notes: a) time at which half of the discs are dispersed, b) fraction of discs remaining at \SI{2}{\mega\year}, c) fraction of discs opening up an inner cavity, d) contribution of the different processes to disc dispersal in terms of mass loss percentage (mean value over all simulations)
\end{table*}
\begin{figure*}[h]
   \centering
   \includegraphics[width=0.9\hsize]{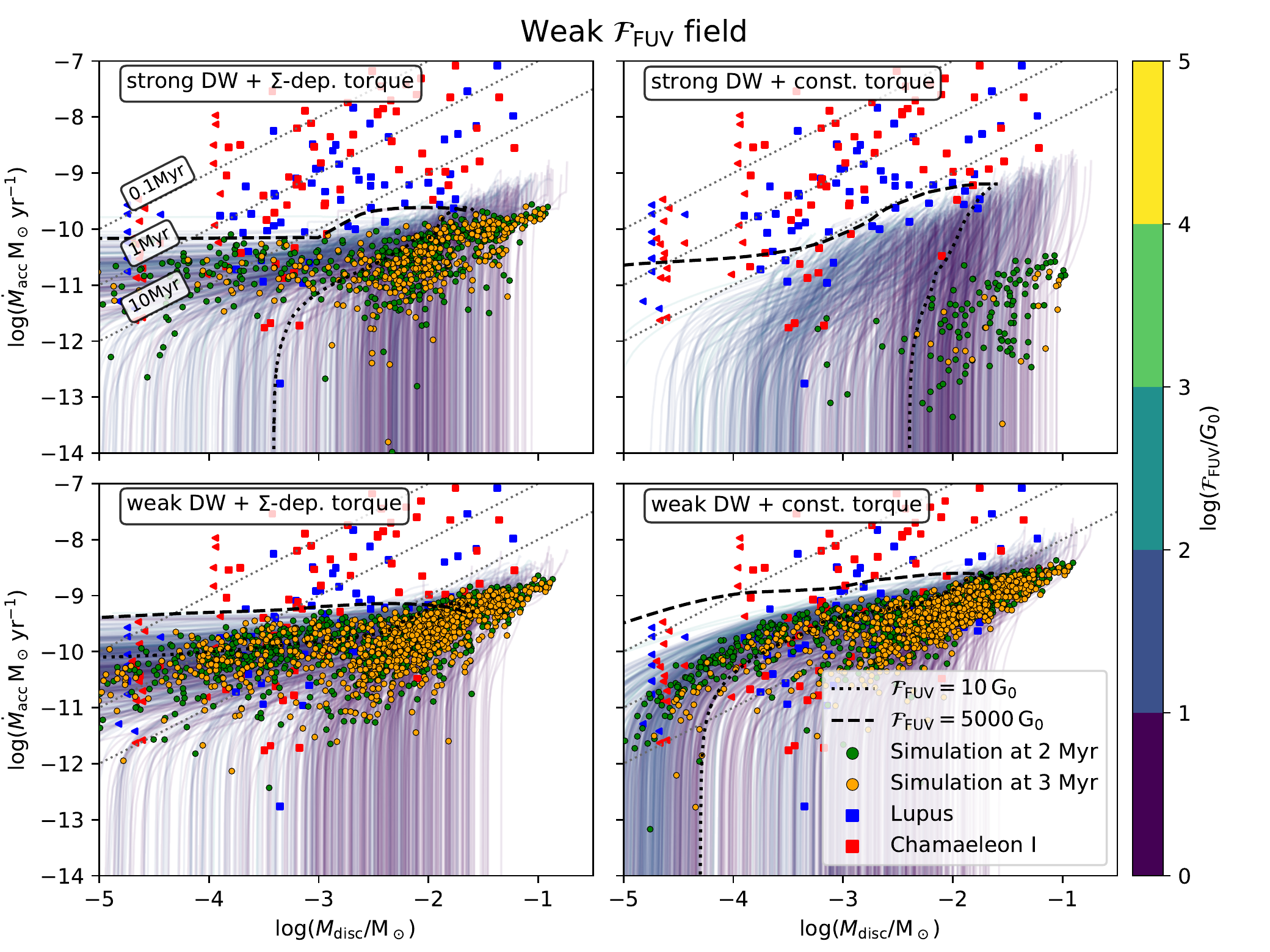}
   \caption{Stellar accretion rate $\dot{M}_\mathrm{acc}$ vs. gas disc mass $M_\mathrm{disc}$ shown for all disc wind scenarios (see Table \ref{tab:init_cond_list}) with a weak FUV field strength distribution. Evolution tracks of individual systems, coloured by $\mathcal{F}_\mathrm{FUV}$, and snapshots at $2\,\mathrm{Myr}$ (green circles) and $3\,\mathrm{Myr}$ (orange circles) are displayed. Tracks of exemplary cases are highlighted with black dashed and dotted lines. We show the detailed evolution of those exemplary discs in Figs.~\ref{fig:exemplary_case_10g0} and \ref{fig:exemplary_case_5e3g0}. We compare our simulations with observed populations in Lupus and Chamaeleon I. Observational dust disc masses and stellar accretion rates are taken from \cite{Manara2019} and dust masses are converted to gas masses by assuming the standard dust-to-gas ratio of $0.01$. Triangles denote upper limits on disc mass. Lines of constant $M_\mathrm{disc}/\dot{M}_\mathrm{acc}$ are shown for 0.1, 1 and 10 Myr (thin dotted lines).}
   \label{fig:mass_acc_lowFUV}
\end{figure*}

\begin{table*}[h]
   \begin{center}
      \caption{Characteristics for different strong FUV field populations.}
      \label{tab:numbers_adams}
      \begin{tabular}{l c c c r r r r}
         \hline\hline
            DW scenario & $t_{1/2}$\textsuperscript{a)} & $f_\mathrm{disc,2Myr}$\textsuperscript{b)} & $f_\mathrm{cavity}$\textsuperscript{c)} & $f_{M_\mathrm{acc}}$\textsuperscript{d)} & $f_{M_\mathrm{PEW,int}}$\textsuperscript{d)} & $f_{M_\mathrm{PEW,ext}}$\textsuperscript{d)} & $f_{M_\mathrm{MDW}}$\textsuperscript{d)} \\
                        & [\SI{}{\mega\year}] & [\SI{}{\percent}] & [\SI{}{\percent}] & [\SI{}{\percent}] & [\SI{}{\percent}] & [\SI{}{\percent}] & [\SI{}{\percent}] \\
         \hline
            strong DW + $\Sigma$-dep. torque  & 0.60   &  11 &  30 & 0.3   & 2.5 & 86.5 &  10.7 \\
            weak DW + $\Sigma$-dep. torque    & 1.27   & 37 &  11 &  4.2   & 4.0 & 87.4 &  4.4 \\
            strong DW + const. torque         & 0.44   &  4 &  34 &  0.1   & 2.0 & 84.6 & 13.3 \\
            weak DW + const. torque           & 0.90   & 30 &  21 &  6.4   & 2.7 & 85.3 &  5.7 \\
         \hline
      \end{tabular}
   \end{center}
   Notes: a) time at which half of the discs are dispersed, b) fraction of discs remaining at \SI{2}{\mega\year}, c) fraction of discs opening up an inner cavity, d) contribution of the different processes to disc dispersal in terms of mass loss percentage (mean value over all simulations)
\end{table*}
\begin{figure*}[h]
   \centering
   \includegraphics[width=0.9\hsize]{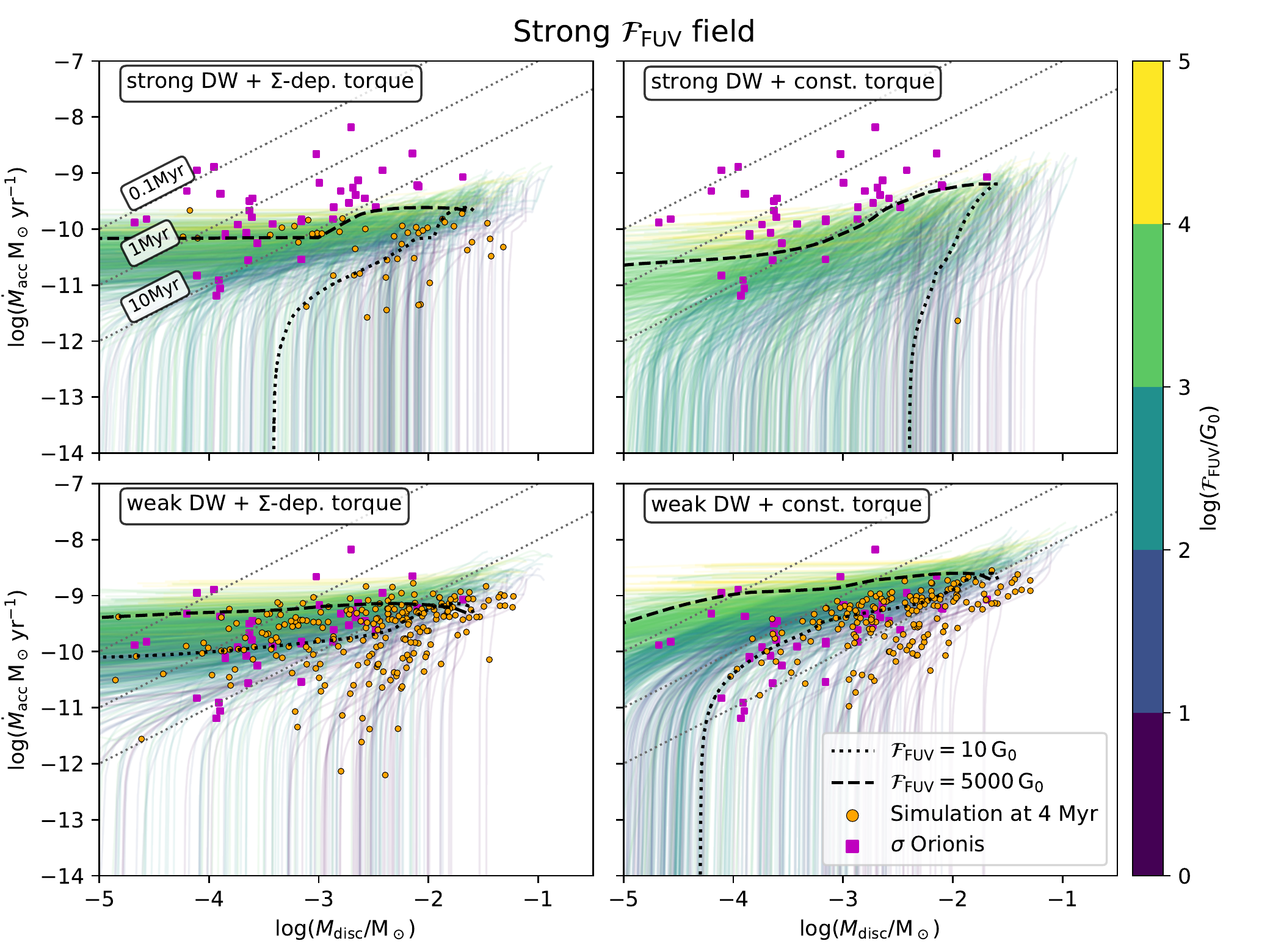}
   \caption{Stellar accretion rate $\dot{M}_\mathrm{acc}$ vs. gas disc mass $M_\mathrm{disc}$, analogous to Fig. \ref{fig:mass_acc_lowFUV}, but for populations with a strong FUV field distribution (see Table \ref{tab:init_cond_list}). A snapshot for the individual systems is shown at $4\,\mathrm{Myr}$ (orange circles) in comparison with observational data from $\sigma$ Orionis. Observational dust disc masses are taken from \cite{Ansdell2017a} and converted to gas masses by assuming the standard dust-to-gas ratio of $0.01$. Stellar accretion rates are taken from \cite{Rigliaco2011}.}
   \label{fig:mass_acc_adams}
\end{figure*}

The stellar accretion rate is an important observable that is correlated with the disc mass and age of the system \citep{Manara2016,Hartmann2016,Testi2022a}. Determining the stellar accretion rates, disc masses, and stellar ages is very challenging \cite[see][for a review]{Hartmann2016}. As the determination of the age is affected by large uncertainty and the choice of a time of 'zero' in our simulations is not necessarily in agreement, we only look at the stellar accretion rates and disc masses. The stellar accretion rate is usually derived from the accretion luminosity, assuming magnetospheric accretion \cite[e.g.][]{Gullbring1998,Alcala2017a}, while the mass of the disc is usually derived from dust continuum measurements and with the assumption of $f_\mathrm{D/G}=0.01$ \cite[e.g.][]{Tychoniec2018,Manara2019}.

In our simulations the stellar accretion rate corresponds to the value of the radial mass flow (Eq. \ref{eq:acc_rate}) at the inner boundary of the grid (magnetospheric accretion, Sect.~\ref{subsubsec:inner_disc_edge}). In Figs. \ref{fig:mass_acc_lowFUV} and \ref{fig:mass_acc_adams}, we show the stellar accretion rate and the gas disc mass of the systems as a side-by-side comparison of the different disc wind scenarios and FUV field strength distributions. Observational data for comparison is taken from \cite{Rigliaco2011} and \cite{Ansdell2017a} for $\sigma$ Orionis and \cite{Manara2019} for Lupus and Chamaeleon I. In Tables \ref{tab:numbers_lowFUV} and \ref{tab:numbers_adams}, we give some characteristic numbers on time scales and mass-loss from different processes for each population.

Evolution tracks in the $M_\mathrm{disc}$-$\dot{M}_\mathrm{acc}$ plane show essentially two distinct behaviours of the evolutionary path.

Firstly, there are systems where the stellar accretion rate stays above $\gtrsim \SI{e-12}{\msun\per\year}$, while the disc mass declines to masses below $\lesssim \SI{e-5}{\msun}$, caused by large parts of the mass being removed from the outer disc by external photoevaporation, similar to the case shown in Fig.~\ref{fig:exemplary_case_5e3g0}. This corresponds to quasi horizontal tracks. We refer to this process as 'outside-in dispersion' \cite[e.g.][]{Scally2001,Koepferl2013}.
   
   Secondly, the other systems show a rapid decrease in stellar accretion rate while still possessing a disc of a mass of \SI{>e-5}{\msun}. This corresponds to tracks which ultimately move nearly vertically downwards. This is caused by an inner cavity opening up, preventing stellar accretion, in a similar way to the case of the discs shown in Fig.~\ref{fig:exemplary_case_10g0}. These discs are referred to as transitional discs, which show inside-out dispersal \cite[e.g.][]{Alexander2014}.

As discussed in Sect.~\ref{subsec:secular_evolution}, there is a strong dependence of the evolutionary path on the FUV field strength. This is reflected in the results shown in Fig.~\ref{fig:mass_acc_adams}, where discs exposed to strong FUV fields beyond approximately $3000\,\mathrm{G}_0$ sustain strong mass loss from external photoevaporation. These discs are rapidly dispersed before an inner hole can open. In the presence of weaker FUV fields, the external photoevaporation leads to a lower mass-loss rate, giving magnetically driven disc winds and internal photoevaporation time to open up an inner cavity. The inner cavity in transitional discs is created by preventing the inner disc from being resupplied by inward flowing gas from the mass reservoir in the outer disc. The inner cavities in our simulations usually initially open up at less than $\lesssim 1\, \mathrm{AU}$.

In Tables \ref{tab:numbers_lowFUV} and \ref{tab:numbers_adams}, we list the fraction of discs opening up an inner cavity. We define discs going through such a transitional phase as ones that are optically thin ($1>\tau$) at radii $\gtrsim 1\,\mathrm{AU,}$ with part of the outer disc being optically thick ($1<\tau$). We get fractions of \SI{35}{\percent} to \SI{63}{\percent} of all discs going through a transitional phase in weak ambient FUV fields, whereas discs in a strong FUV environment are less likely to open up a cavity (\SI{11}{\percent} to \SI{34}{\percent}). Discs exposed to a strong magnetic disc wind are more likely to go through a transitional phase.

To compare our synthetic populations with observations, we offer snapshots of the populations at fixed ages.
For a weak ambient FUV field, we display snapshots at \SI{2}{\mega\year} (orange circles) and \SI{3}{\mega\year} (green circles). We note that in the presence of strong disc winds, discs are dispersed much faster. Thus, at \SI{2}{\mega\year,} there are merely \SI{10}{\percent} to \SI{32}{\percent} left, compared to substantial \SI{81}{\percent} to \SI{88}{\percent} for weak disc winds. Hence, the number of synthetic data points depends on the magnetic disc wind scenario. Furthermore, we note that the location of the bulk in the $M_\mathrm{disc}-\dot{M}_\mathrm{acc}$ plane does not show strong dependence on time.
For the strong FUV field distribution we show a snapshot at \SI{4}{\mega\year} (orange circles). At that time, discs exposed to very strong FUV fields have already dispersed. Discs still present at that time are exposed to FUV fields of $\lesssim \num{1000}\,\mathrm{G}_0$ (see Fig. \ref{fig:lifetime_UVFS}). We note again the low number of data points for strong disc winds.

Since the magnetic disc winds are strongest within $< 1\,\mathrm{AU}$ from the inner edge \citep{Suzuki2016,Kunitomo2020}, the evolution of the inner disc is strongly affected by the magnetically driven disc winds. This can be seen in the ensemble evolution of the different disc wind scenarios (see Figs. \ref{fig:exemplary_case_10g0} and \ref{fig:exemplary_case_5e3g0}). The combination of a strong magnetic disc wind in the early phase and the onset of internal photoevaporation at a later stage, when the column density of the magnetic wind drops, can lead to rapid depletion of the inner disc region, causing the stellar accretion rate to decrease early on during the evolution. This behaviour is also reflected by the larger fraction of discs opening up a cavity in the strong disc wind scenarios ($\sim 57-63\%$ and $\sim 30-34\%$ for a weak and a strong FUV field, respectively; see Tables \ref{tab:numbers_lowFUV} and \ref{tab:numbers_adams}). 

In Tables \ref{tab:numbers_lowFUV} and \ref{tab:numbers_adams} we show the mean fraction of initial mass lost through different processes. 
We see that external photoevaporation is by far the most important process in dispersing the disc, both for weak (\SI{\sim50}{\percent}) and strong (\SI{80}{\percent} to \SI{90}{\percent}) FUV field environments. Values are larger in the cases with $\Sigma$-dependent torques. This is the effect of a lesser amount of radial mass flow in this scenario (Figs.~\ref{fig:exemplary_case_10g0} and \ref{fig:exemplary_case_5e3g0}), leaving more gas in the outer disc, which is then more susceptible to external photoevaporation.
The mass fraction removed by magnetic winds is dependent on the disc wind scenario. Strong disc winds lead to larger losses through this process, which can reach up to \SI{\sim42}{\percent} in weak and \SI{\sim13}{\percent} in strong FUV fields. Constant torque scenarios also lead to greater losses through disc winds, although the difference is lower than that among strong and weak disc winds.
Internal photoevaporation is less effective except for the case of weak disc winds and $\Sigma$-dependent torques. The combination of low disc winds (from the weak disc winds scenario) and lower initial radial material redistribution (from $\Sigma$-dependent torques) lead to a lower column density of the magnetic wind so that internal photoevaporation is active earlier on.

Stellar accretion plays only a subordinate role for disc dispersal in our simulations. With strong disc winds, the losses are so large before the gas can reach the inner edge that the amount of the disc being accreted onto the star does not exceed on average \SI{1}{\percent}. Weak disc winds allow for more mass to be accreted by the star (up to 17\%).

Observational data from star forming regions Lupus (age \SI{\sim2}{\mega\year}, \citealp{Testi2022a}), Chamaeleon I (age \SI{\sim3}{\mega\year}, \citealp{Testi2022a}) and $\sigma$ Orionis (age $\num{\sim 3}-\SI{5}{\mega\year}$, \citealp{Ansdell2017a}) show a wide spread in accretion timescales $\dot{M}_\mathrm{acc}/M_\mathrm{disc}$ between approximately \num{0.1} and \SI{10}{\mega\year}. Lupus and Chamaeleon I are known to have weak FUV field strengths\footnote{\cite{Cleeves2016} estimated the FUV field in Lupus to be as low as $\mathcal{F}_\mathrm{FUV} \geq 4\mathrm{G}_0$.} and we thus compare them to our weak FUV field populations. We note that $\sigma$ Orionis does contain an OB system ($\sigma$ Ori) and its FUV fluxes are expected to be high ($\sim 8000\;\mathrm{G}_0$ within $1\;\mathrm{pc}$ of $\sigma$ Ori, according to \citealp{Ansdell2017a}). Therefore, the strong FUV field populations were compared to data from $\sigma$ Orionis. We note that the constituents of the star forming regions show a large spread in ages (according to Table 1 of \citealp{Testi2022a}).

A good agreement with the observational data was obtained only for a weak magnetic disc wind. However, accretion rates beyond \SI{e-8}{\msun\per\year}, as observed in the young star-forming regions Lupus and Chamaeleon I, were not reproduced by our model.

\subsection{Disc lifetime}
\label{subsec:disc_lifetime}
\begin{figure}
   \centering
   \includegraphics[width=\hsize]{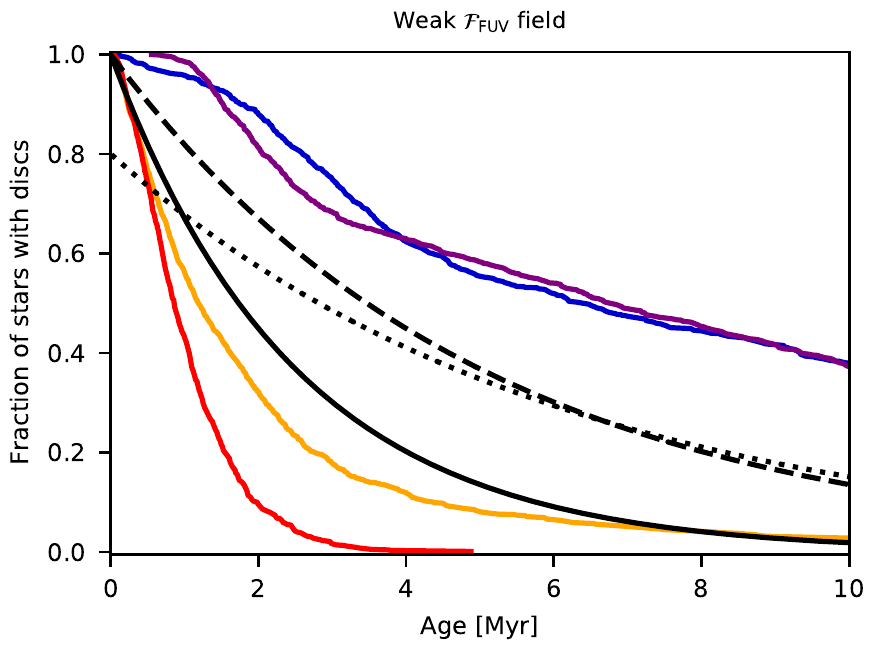} \\
   \includegraphics[width=\hsize]{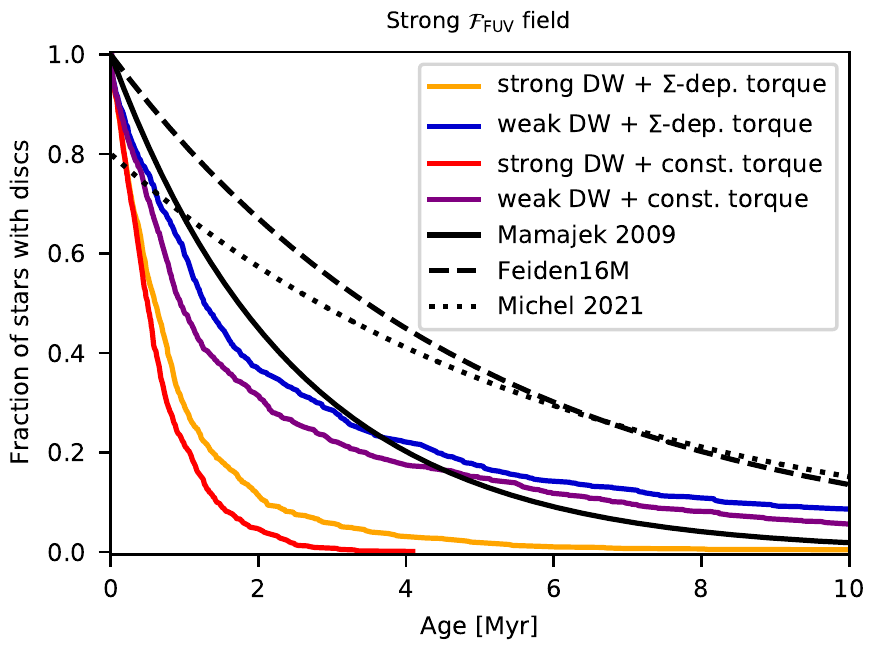}
   \caption{Fraction of stars possessing a circumstellar disc as a function of time for both weak and strong FUV field distributions. Evolution tracks of the simulations are shown in comparison with exponential decay fits to observational data from \cite{Mamajek2009,Richert2018,Michel2021}. Note: the fit from \cite{Richert2018} is denoted Feiden16M, since it relies on magnetic PMS models from \cite{Feiden2016} to determine the cluster ages.}
   \label{fig:discfraction}
\end{figure}
As we have seen, the combination of stellar accretion, photoevaporation, and magnetically driven disc winds can lead to a rapid dispersal of the disc. We consider a disc to be dispersed when it becomes unobservable in the near-infrared (NIR). Following the dispersal condition of \cite{Kimura2016} the disc is considered unobservable in NIR when the optical depth $\tau$ drops below unity in the regions where the midplane temperature is $T_\mathrm{mid}>300\mathrm{K}$.

\begin{figure}
   \centering
   \includegraphics[width=\hsize]{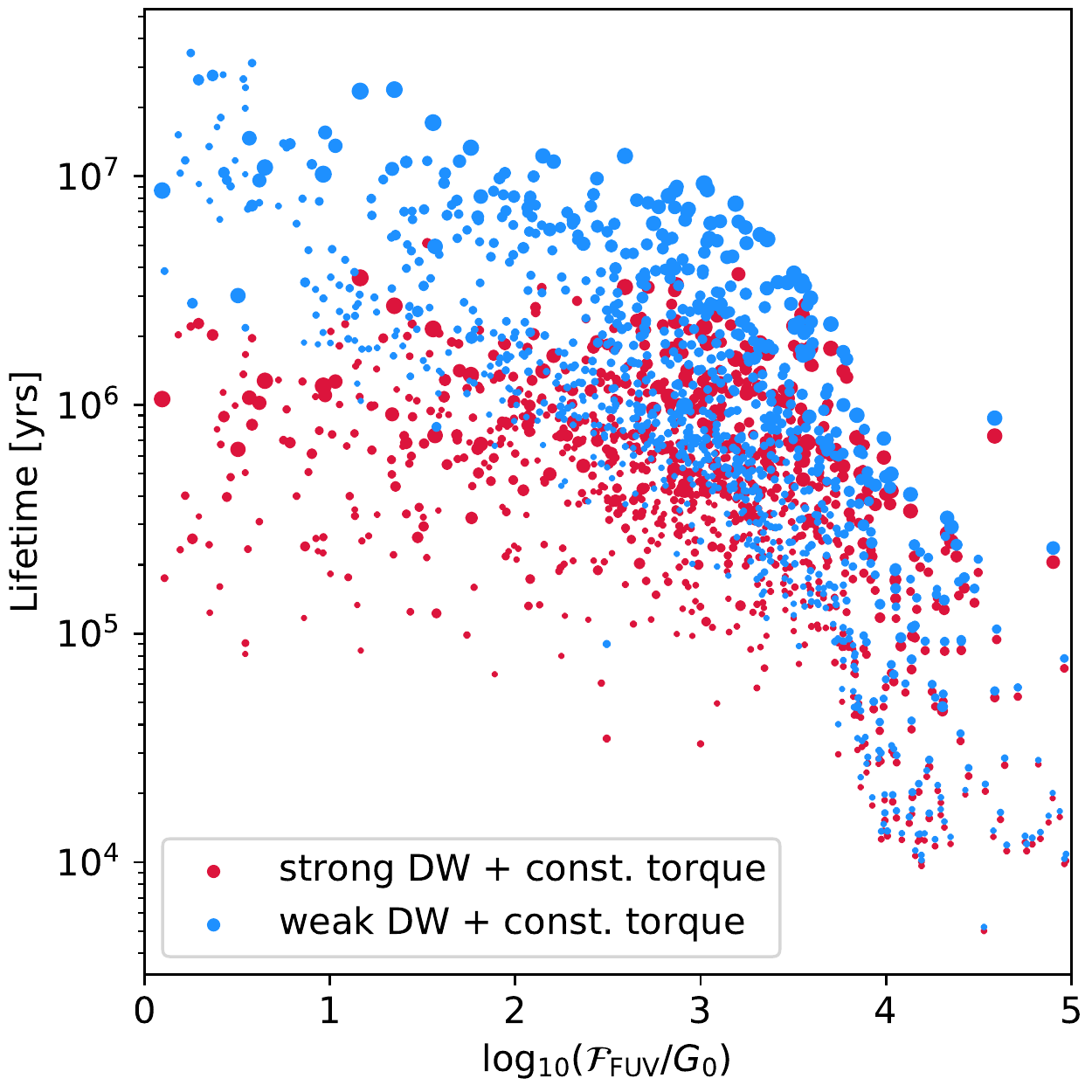}
   \caption{Disc lifetime with regard to the ambient FUV field strength $\mathcal{F}_\mathrm{FUV}$. Point sizes indicate the mass of the host star. We note that our grid only extends to $10^4\mathrm{G}_0$ and values beyond are treated as the respective boundary value, which is also seen in the results,  given that beyond $10^4\mathrm{G}_0$, the lifetimes do not further decrease.}
   \label{fig:lifetime_UVFS}
\end{figure}

Figure \ref{fig:discfraction} shows the time evolution of the fraction of stars possessing a disc. An exponential decay of the following form is typically used,
\begin{equation}
   f_\mathrm{disc} = f_0 \cdot \exp(-t/\tau_\mathrm{disc}),
\end{equation}
to describe the evolution, with $f_0$ being the initial fraction of stars possessing a disc and $\tau_\mathrm{disc}$ being the characteristic time scale\footnote{Note: the characteristic timescale is connected to the half life time by $t_{1/2}=\tau_\mathrm{disc} \cdot \ln(2)$.}. We show fits from \citet{Mamajek2009}; $f_0=1$ and $\tau_\mathrm{disc}=\SI{2.5}{\mega\year}$, \citet{Richert2018}; $f_0=1$ and $\tau_\mathrm{disc}=\SI{5}{\mega\year}$ and \citet{Michel2021}; $f_0=0.8$ and $\tau_\mathrm{disc}=\SI{6.4}{\mega\year}$. Furthermore, \citet{Richert2018} made use of magnetic pre-main sequence (PMS) models from \citet{Feiden2016} for determining the cluster ages and we thus refer to it as Feiden16M. Both \citet{Mamajek2009} and \citet{Richert2018} have assumed that initially all stars possess discs, whereas \citet{Michel2021} assumed an initial fraction of $f_0=0.8$ to account for close binary systems \citep{Kraus2012}. The differences in characteristic time scales, $\tau_\mathrm{disc}$, emerge partially from different modelling approaches for the ages of the stellar clusters and partially from the choice of clusters considered. While the sample of \cite{Mamajek2009} also includes clusters containing O/B type stars (e.g. $\sigma\mathrm{Ori}$, $\lambda\mathrm{Ori}$, $25\mathrm{Ori}$), \cite{Michel2021} intentionally excluded them, since the presence of such stars implies a strong FUV field, which can have significant impact on the disc lifetime (see below and Fig. \ref{fig:lifetime_UVFS}).

Our populations show very rapid dispersal for strong disc winds, regardless of the ambient FUV field distribution. Here, the lifetimes are below the values from observations. For weak disc winds, disc lifetimes are longer and show a dependence on the ambient FUV field distribution. For a strong FUV field, the fraction of stars with a protoplanetary discs  roughly follows the data from \citet{Mamajek2009}, whereas for a weak FUV field, the fraction of stars declines much more slowly, namely, in a way that is comparable to the data from \citet{Richert2018} and \citet{Michel2021}. The dependency on the strength of the magnetic disc wind can be explained by the dispersal condition we used, where the evolution of the inner disc is the most important factor. When assuming strong disc winds, the decrease in surface density is much faster. When the wind gets weaker, internal photoevaporation sets in, preventing the inner disc from being replenished with gas, leading to rapid dispersal (Figs. \ref{fig:exemplary_case_10g0} and \ref{fig:exemplary_case_5e3g0}). Again, we stress that in our model, the resulting disc lifetimes are a direct prediction of the model. This is in contrast to past approaches \citep[e.g.][]{Mordasini2009,Emsenhuber2021b}, where the external photoevaporation rates were adjusted such that disc lifetimes were in agreement with observations.

As mentioned previously, the presence of nearby massive stars can have a large impact on the lifetime of circumstellar discs \cite[e.g.][]{Adams2004,Adams2006,Clarke2007,Fang2012}. Figure \ref{fig:lifetime_UVFS} shows the dependence of the lifetime on the ambient FUV field strength and stellar mass for two example disc wind scenarios of populations with strong FUV field strength distributions: strong disc wind with constant torque and weak disc wind with constant torque. Populations with weak FUV field distributions can be regarded as a subset ($1\,\mathrm{G}_0 < \mathcal{F}_\mathrm{FUV} < 10^2\,\mathrm{G}_0$) of these simulations.

There is a strong dependence of the disc lifetime on the ambient FUV field strength for weak disc winds. Also the mass of the host star shows impact on the disc lifetime, with lifetime tending to increase with higher stellar mass.\footnote{We note here that the initial disc mass is correlated with stellar mass as an initial condition.} This is in contrast with the results of observations, which show that disc fractions decrease with increasing stellar mass \citep[e.g.][]{Bayo2012}. However, \cite{Komaki2021} and \cite{Picogna2021} both suggested that this is a result of strong internal photoevaporation. For strong disc winds, the lifetime is only correlated for very high FUV field strengths ($>10^{3}\,\mathrm{G}_0$). This can explain why the disc fractions for strong disc winds in Fig. \ref{fig:discfraction}   depend only weakly on the FUV field distribution.

\subsection{Additional cases: Variable $\overline{\alpha_{\phi z,0}}$}
\label{subsec:varAlpha}
Our simulations still fall short of the high accretors, especially in the case of a weak FUV field (Fig. \ref{fig:mass_acc_lowFUV}). In order to investigate the dependency on the initial torque strength, we ran additional populations with $\overline{\alpha_{\phi z, 0}}$ distributed uniformly in log between $10^{-5.5}$ and $10^{-2.5}$ for two fixed FUV field strengths, $10 \mathrm{G}_0$ and $10^3\mathrm{G}_0$. The results are shown in Appendix \ref{app:varAlpha}.

The tracks in Fig. \ref{fig:mdisc_macc_varAlpha_lowFUV} show that high accretors can be reached by high initial torques. However, when comparing the simulations after a few million years with the observations, regions where $\dot{M}_\mathrm{acc}>10^{-8}\mathrm{M}_\odot \mathrm{yr}^{-1} $ and $M_\mathrm{disc}/\dot{M}_\mathrm{acc} \lesssim 0.1 \,\mathrm{Myr}$ are almost never achieved. This has to do with the fact that discs beyond $M_\mathrm{disc}/\dot{M}_\mathrm{acc} \sim 0.1 \,\mathrm{Myr}$ are dispersed very quickly. The simulation tracks show a strong correlation with the initial torque strength $\overline{\alpha_{\phi z,0}}$ and the large spread in initial torque strength results in a large spread in accretion rates, similar to observations. The majority of simulations with a $\Sigma$-dependent torque stay at almost constant accretion rates and move horizontal in the $\log(M_\mathrm{disc})$-$\log(\dot{M}_\mathrm{acc})$ plane.

Regarding the disc lifetimes, the choice of the torque (i.e. constant or $\Sigma$-dependent) has less impact than the choice of the disc wind scenario (strong or weak). Discs exposed to a strong disc wind are short lived, which is in line with our previous findings.

\subsection{Additional cases: Varying internal photoevaporation}
\label{subsec:add_cases_varying_int_PEW}
Our model incorporates both MHD winds and photoevaporation. The interplay between these two types of outflows is currently still under debate (see §\ref{subsec:mhd_vs_pew}). In order to take this into account, we ran the two limiting cases with 'No EUV radiation shielding' and 'No internal photoevaporation'. We show the main results from §\ref{subsec:macc_disc} and §\ref{subsec:disc_lifetime} in Appendices \ref{app:noshielding} and \ref{app:nointPEW}.

We find that internal photoevaporation only plays an important role at a late stage of the disc evolution by reducing the accretion rate and opening the inner cavity. In the absence of internal photoevaporation (Appendix \ref{app:nointPEW}), discs do not open an inner cavity. Thus, the star remains accreting gas and, as a result, the NIR lifetimes are much longer. This can be seen also in the evolutionary paths in the $M_\mathrm{disc}-\dot{M}_\mathrm{acc}$ plane. Instead of having a gap forming, simulations follow lines of constant accretion or are dispersed outside-in. Also note that in the case of a constant magnetic field ($\Sigma$-dependent torque), simulations tend to have constant accretion towards the end, as a result of the increasing torque. However, it does not change the location where we find the simulations in the $M_\mathrm{disc}$-$\dot{M}_\mathrm{acc}$ plane and, in addition, the best correspondence is still found for weak disc winds with a strong torque. Thus, the presence of internal photoevaporation has mainly influence on the NIR lifetimes. This necessity of the interplay between MHD wind and internal photoevaporation to explain the inner disc lifetime has already been reported by \cite{Kunitomo2020}.

\section{Discussion}
\label{sec:discussion}
Our results give some first insights on the interplay of magnetically driven disc winds, magnetic braking, and internal and external photoevaporation in MRI inactive discs, as well as their influences on and their agreement with observables. However, there are several simplifications and caveats that have to be discussed.

\subsection{Stellar accretion}
Our model considers magnetospheric accretion driven by turbulence (MRI) and magnetic braking (Eq. \ref{eq:acc_rate}). We find that stellar accretion makes up only a minor contribution to disc dispersal, which is in contrast to the common point of view \cite[e.g.][]{Ercolano2017}. We obtained upper limits on the stellar accretion rates of \SI{e-10}{\msun\per\year} to \SI{e-8}{\msun\per\year}, dependent on the disc wind scenario, which is on the lower side of what we expect from observations. Especially the young star forming regions Lupus and Chamaeleon I show stellar accretion rates beyond \SI{e-8}{\msun\per\year}. Higher magnetic torques could explain these high accretors, but it is still difficult to find such systems after a few million years. More massive and compact discs would also lead to higher accretion rates.

There may be other processes at work that could contribute to the accretion rates observed. \cite{Takasao2018} conducted 3D MHD simulations of magnetised accretion discs and found that a failed magnetic disc wind can drive fast accretion onto high latitudes, namely, so-called funnel-wall accretion. This new accretion process can coexist with the usual accretion processes, but is expected to drive much lower accretion rates. Thus, the discrepancy between our simulations and observed accretion rates above $\gtrsim\SI{e-8}{\msun\per\year}$ persists.

We also note that some observations of transitional discs show relatively high accretion rates \cite[$\sim 10^{-8}\,\mathrm{M}_\odot/\mathrm{yr}$,][]{Owen2016}, despite the presence of an inner cavity. \cite{Wang2017} showed that such high accretion rates could, in principle, be obtained by a strong magnetised wind for moderately good coupling of the magnetic field. In our model, such high accretion rates can no longer be supported after opening the cavity.

\subsection{Magnetic versus photoevaporative winds}
\label{subsec:mhd_vs_pew}
In our model, we assume the presence of both magnetic and photoevaporative winds. By calculating the column density, we take into account the shielding of the disc from EUV radiation by the magnetic wind. As soon as the column density falls below a critical value, internal photoevaporation is added to the outflows. In reality, the interplay between MHD-driven and photoevaporative winds is expected to be much more complicated. \cite{Bai2016a} found a magneto-thermal wind (also known as magneto-photoevaporation) which is a thermally assisted magnetic outflow. \cite{Wang2019} found that for magnetic winds, the mass loss rates are relatively robust with regard to high-energy photon luminosities and EUV photons can even reduce the mass-loss rates. At this point, the transition from MHD wind to photoevaporation driven outflow remains unsolved. The present opinion tend to give both processes important roles at different evolutionary stages \cite[see e.g. recent reviews by][]{Pascucci2022,Lesur2022}.

\begin{figure}
   \centering
   \includegraphics[width=\hsize]{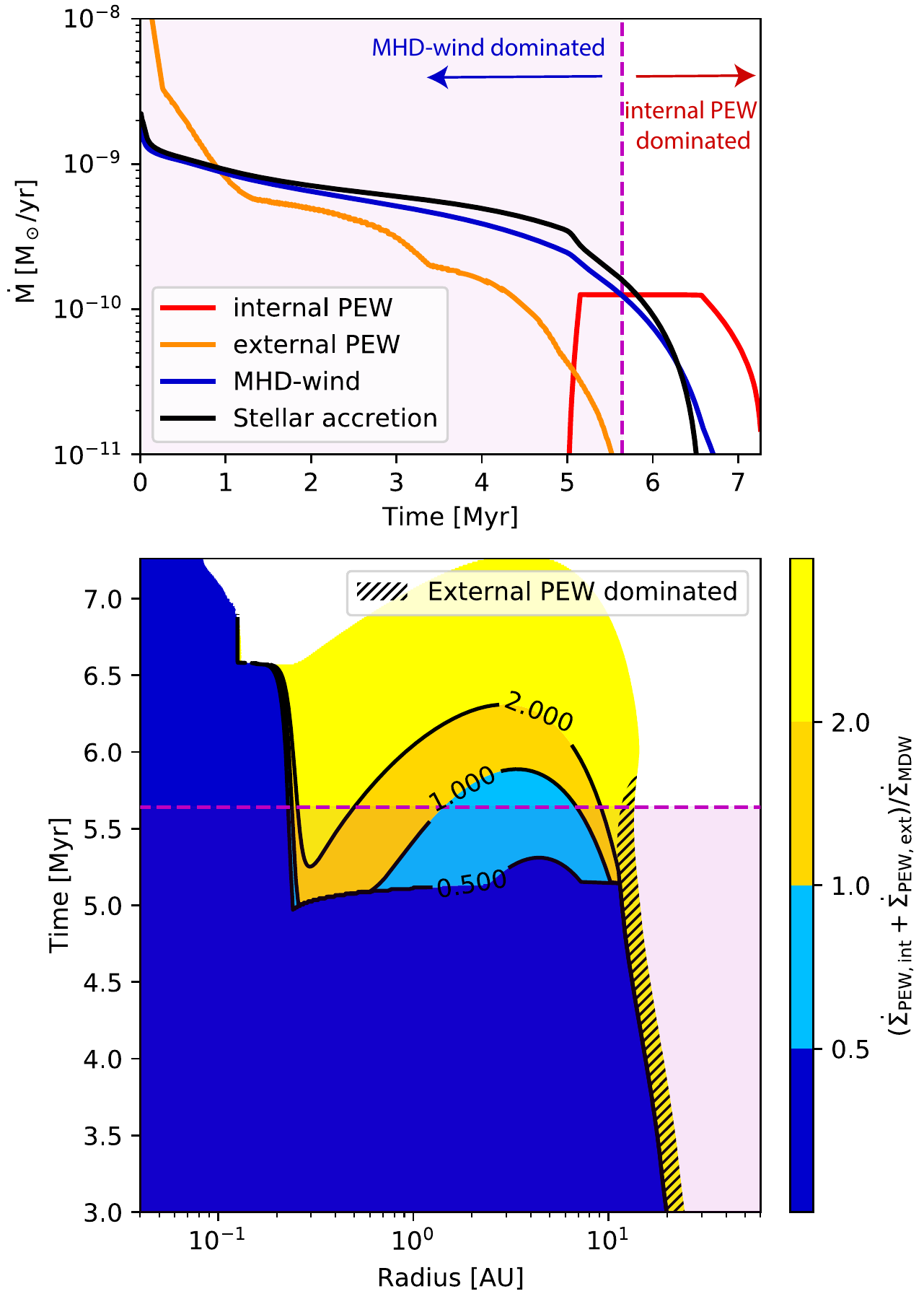}
   \caption{Relative importance of different mass-loss processes for the exemplary case with weak DW, strong constant torque and $\mathcal{F}_\mathrm{FUV}=10\,\mathrm{G}_0$ (bottom row of Fig. \ref{fig:exemplary_case_10g0}). Top: Time evolution of mass-loss rates for different processes. Purple dashed line distinguishes MHD-wind dominated and internal PEW dominated period (analogous to Fig. 2 c in \citealp{Kunitomo2020}). Bottom: Ratios of $\dot{\Sigma}$ for different mass loss processes shown for radial extend and time. Blue areas are dominated by MHD-wind while yellow areas indicate PEW dominated. The hatched area highlight external PEW-dominated regions.}
   \label{fig:rel_importance}
\end{figure}

Our simulations show that at a given time, the evolution of large parts of the disc are either dominated by magnetic winds or internal photoevaporation. This is shown by Fig. \ref{fig:rel_importance}, which compares the mass-loss rates and strengths of different winds by location and time. As depicted in Fig. \ref{fig:rel_importance} in the early phase, magnetic winds dominate the mass-loss rates, whereas at a later stage, the internal photoevaporation gains importance. This aspect was already reported on and discussed by \cite{Kunitomo2020}. Although external photoevaporation is strong initially, its mass-loss rates are very high and concentrated in the outer disc only, where they dominate magnetic winds by several orders of magnitude (as shown in Figs.~\ref{fig:exemplary_case_10g0} and \ref{fig:exemplary_case_5e3g0}). In Appendices \ref{app:noshielding} and \ref{app:nointPEW}, we show the two extreme cases where we either add up MHD wind and internal photoevaporation or completely disable internal photoevaporation throughout the entire disc evolution. We see that the internal photoevaporation plays an important role in setting the disc life time, but it does not have big influence on other observables. We therefore do not expect our results to be very strongly affected by the interplay of magnetic and photoevaporative winds.

\subsection{Dust evolution}
Our model does not include dust evolution and we directly compare obtained gas disc masses with observed dust masses converted to gas mass, with $f_\mathrm{D/G}=1/100$. However, \cite{Sellek2020b} found that growth and radial drift of dust can lower the dust-to-gas ratio significantly over time. This implies that gas disc masses inferred from sub-mm observations stand as the lower limits, rather than actual disc masses. They further note that including dust evolution can also increase the scatter in accretion rates.

\subsection{Disc dispersal}
We essentially observed  two different disc dispersal modes (i.e. outside-in and inside-out). Our simulations suggest that the dispersal mode is strongly correlated with the external FUV field strength. Furthermore, it depends on the disc wind scenario chosen, as a strong wind removes mass at a higher rate from the inner disc than a weak wind. \cite{Koepferl2013} classified over 1500 sources in nearby star-forming regions and found the inside-out clearing of discs to be the preferred way of disc dispersal. In our simulations, systems in a weak $\mathcal{F}_\mathrm{FUV}$ environment survive long enough, so that an inner cavity will eventually open up. As can be seen in Appendix \ref{app:nointPEW}, the opening of an inner cavity is strongly dependent in the internal photoevaporation.

We would also like to point out that outside-in dispersal via external photoevaporation happens to be a very efficient process and disc lifetimes in strong FUV environments are often below \SI{1}{\mega\year}, which would suggest that only very few disc are caught by observations during outside-in dispersal.

We also neglected X-ray as source of internal photoevaporation. While the EUV prescription we use has a integrated mass loss rate of about \SI{e-10}{\msun\per\year} (Fig.~\ref{fig:rel_importance}), X-ray internal photoevaporation can produce mass loss rates of the order of \SI{e-8}{\msun\per\year} \citep{Owen2010,Owen2011,Owen2012}. Including X-ray photoevaporation in our model would considerably increase mass loss rate by internal photoevaporation, which would cause a reduction of loss by MHD winds and stellar accretion. This would, in turn, make it even more difficult to match the observed stellar accretion rate and would lead to further reduced disc lifetimes.

\subsection{Comparison with previous works}
There are several models describing MHD wind-driven disc evolution in 1D \cite[e.g.][]{Armitage2013,Bai2016b,Suzuki2016,Tabone2022a}.
Our model closely follows  \cite{Kunitomo2020}, who combined MHD winds based on \cite{Suzuki2016} with internal EUV and X-ray photoevaporation, but with internal EUV and external FUV instead of internal X-ray photoevaporation, along with a consistent treatment of EUV shielding. We also find that disc dispersal is a result of both magnetic winds and photoevaporation, since without internal photoevaporation and considering weak external FUV fields, disc lifetimes are much longer than inferred from observations (Appendix \ref{app:nointPEW}). This is in contrast to \cite{Armitage2013}, \cite{Bai2016b}, and \cite{Tabone2022b} who suggested that disc dispersal could be driven solely by magnetic disc winds.

The approach of running disc populations syntheses has gained popularity in recent years \citep[e.g.][]{Lodato2017,Mulders2017,Tabone2022b,Somigliana2022}. However, models and approaches differ. 
\cite{Lodato2017} investigated the self-similar solution of the classical $\alpha$-disc model \citep{LyndenBell1974} for varying initial disc masses and viscous times. They were able to reproduce the correlation between accretion rate and disc mass for the Lupus star-forming region under the assumption that the efficiency of angular momentum transport is an increasing function of radius.
\cite{Tabone2022b} managed to reproduce the $\dot{M}_\mathrm{acc}-M_\mathrm{disc}$ correlation and the spread around the mean trend in Lupus, as a result of MHD-wind driven disc evolution. They used a distribution of accretion timescales that was obtained by fitting observed disc fractions. This allowed them to obtain disc lifetimes that comply with observations without considering photoevaporation. We note that variations in accretion timescales is equivalent to variations in their $\alpha_\mathrm{DW}$ (similar to our $\overline{\alpha_{\phi z}}$) and naturally lead to a larger spread in accretion rates.

In our fiducial simulations, the results (i.e. spread in disc lifetimes and $\dot{M}_\mathrm{acc}$-$M_\mathrm{disc}$ evolution) are a direct consequence of the variations in initial conditions (i.e. stellar mass, disc properties and ambient FUV-field strength). In contrast to the models mentioned above, we do not assume variations in accretion or viscous timescales (i.e. variations in $\alpha$). In Appendix \ref{app:varAlpha}, we show additional cases, where we varied the initial torque strength $\overline{\alpha_{\phi z}}$, which introduced a large spread in observed accretion rates (see also § \ref{subsec:varAlpha} for a discussion).

\section{Summary and conclusions}
\label{sec:conclusion}
We constructed a global model for protoplanetary disc evolution including internal redistribution of angular momentum via a turbulent viscosity ($\alpha$-disc) and angular momentum removal through magnetic braking, magnetically driven disc winds,  and internal and external photoevaporation. We further developed a new, physically consistent (albeit simple) model for the shielding of EUV-driven photoevaporation by MHD winds. Taking into account the recent discussion of non-ideal MHD effects in the literature and suppressing MRI in large parts of the disc, we studied the impact of four different magnetic disc wind scenarios on MRI inactive discs for a wide range of initial conditions, resembling conditions found in young star-forming regions. We then compared observables such as stellar accretion rates, disc mass, disc dispersal mode, and disc lifetime with observational data.

We present the following conclusions based on our model:
\begin{enumerate}
   \item We find that discs are primarily dispersed by the combination of outflows (magnetically driven disc winds, as well as internal and external photoevaporation), while only a minor fraction of the mass is accreted onto the star.
   \item Weak wind-driven mass-loss rates (weak disc winds) are favoured over high mass-loss rates (strong winds) to reproduce the results of observations. Strong magnetic disc winds combined with internal photoevaporation lead to fast depletion of the inner disc region, the opening of an inner cavity and a swift decrease in stellar accretion rate.
   \item A strong torque can support stellar accretion in MRI inactive discs up to $\dot{M}_\mathrm{acc} \sim \SI{e-8}{\msun\per\year}$ or higher, as observed. Our comparison with observations therefore indicates that we need weak MHD-driven mass loss, but strong torques for agreement. This allows us to identify the subtype of MHD-wind models (weak disc wind + strong torque) that best reproduces the  observations.
   \item The inclusion of EUV internal photoevaporation only affects the late stage of the disc evolution, where it can lead to rapid disc dispersal. This has an important influence on disc lifetimes in our simulations.
   \item We found that for weak disc winds, the disc lifetimes are strongly influenced by the ambient FUV field strength, driving external photoevaporation. Initial disc mass and stellar mass seem to play merely secondary roles. Discs under the influence of weak disc winds have lifetimes which are in line with observations for both weak- and strong ambient FUV field strengths. For strong disc winds, disc lifetimes are too short when compared with observations of NIR excess.
\end{enumerate}
These findings support the recent shift towards magnetically driven accretion. However, there is further research underway towards improving our understanding the dynamics of the magnetic field evolution and more detailed modelling of the strength and evolution of the disc wind torque needed. Further investigation of the interplay between magnetic and photoevaporative winds is highly encouraged as both processes are expected to play important roles in disc evolution.

Future works will address planet formation in these wind-driven accretion discs and possible imprints on planet populations. Regarding external photoevaporation, we now have a model that is dependent on the FUV field strength, which is considered one of the most important quantities that affects disc evolution in star-forming clusters. This will enable us to also investigate the influence of the cluster environment on planetary formation, which is a hot topic in recent discussions \cite[e.g.][]{Winter2020,Adibekyan2021a}.

\begin{acknowledgements}
   We want to thank Oliver Schib, Remo Burn, Thomas Haworth, and Ilaria Pascucci for constructive discussions. This work has been carried out within the framework of the National Centre of Competence in Research PlanetS supported by the Swiss National Science Foundation (SNSF) under grants 51NF40\_182901 and 51NF40\_205606. The authors acknowledge the financial support of the SNSF. J.W. and C.M. acknowledge the support from the SNSF under grant 200021\_204847 ``PlanetsInTime''. We thank the anonymous referee for a thorough read and a constructive report that was very helpful for improving the manuscript.
\end{acknowledgements}

\bibliographystyle{aa.bst}
\bibliography{references.bib}

\begin{thebibliography}{103}
\expandafter\ifx\csname natexlab\endcsname\relax\def\natexlab#1{#1}\fi

\bibitem[{Adams {et~al.}(2006)Adams, Proszkow, Fatuzzo, \& Myers}]{Adams2006}
Adams, F., Proszkow, E., Fatuzzo, M., \& Myers, P. 2006, Astrophysical Journal,
  641, 504

\bibitem[{Adams {et~al.}(2004)Adams, Hollenbach, Laughlin, \&
  Gorti}]{Adams2004}
Adams, F.~C., Hollenbach, D., Laughlin, G., \& Gorti, U. 2004, The
  Astrophysical Journal, 611, 360

\bibitem[{Adibekyan {et~al.}(2021)Adibekyan, Santos, Demangeon, Faria, Barros,
  Oshagh, Figueira, Delgado~Mena, Sousa, Israelian, Campante, \&
  Hakobyan}]{Adibekyan2021a}
Adibekyan, V., Santos, N.~C., Demangeon, O. D.~S., {et~al.} 2021, Astronomy
  {\&} Astrophysics, 649, A111

\bibitem[{Alcal{\'{a}} {et~al.}(2017)Alcal{\'{a}}, Manara, Natta, Frasca,
  Testi, Nisini, Stelzer, Williams, Antoniucci, Biazzo, Covino, Esposito,
  Getman, \& Rigliaco}]{Alcala2017a}
Alcal{\'{a}}, J.~M., Manara, C.~F., Natta, A., {et~al.} 2017, Astronomy {\&}
  Astrophysics, 600, A20

\bibitem[{Alessi \& Pudritz(2018)}]{Alessi2018}
Alessi, M. \& Pudritz, R.~E. 2018, Monthly Notices of the Royal Astronomical
  Society, 478, 2599

\bibitem[{Alexander {et~al.}(2014)Alexander, Pascucci, Andrews, Armitage, \&
  Cieza}]{Alexander2014}
Alexander, R., Pascucci, I., Andrews, S., Armitage, P., \& Cieza, L. 2014, in
  Protostars and Planets VI (University of Arizona Press)

\bibitem[{Alexander \& Pascucci(2012)}]{Alexander2012}
Alexander, R.~D. \& Pascucci, I. 2012, Monthly Notices of the Royal
  Astronomical Society: Letters, 422, L82

\bibitem[{Andrews {et~al.}(2013)Andrews, Rosenfeld, Kraus, \&
  Wilner}]{Andrews2013}
Andrews, S.~M., Rosenfeld, K.~A., Kraus, A.~L., \& Wilner, D.~J. 2013, The
  Astrophysical Journal, 771, 129

\bibitem[{Andrews {et~al.}(2010)Andrews, Wilner, Hughes, Qi, \&
  Dullemond}]{Andrews2010}
Andrews, S.~M., Wilner, D.~J., Hughes, A.~M., Qi, C., \& Dullemond, C.~P. 2010,
  The Astrophysical Journal, 723, 1241

\bibitem[{Ansdell {et~al.}(2017)Ansdell, Williams, Manara, Miotello, Facchini,
  Marel, Testi, \& Dishoeck}]{Ansdell2017a}
Ansdell, M., Williams, J.~P., Manara, C.~F., {et~al.} 2017, The Astronomical
  Journal, 153, 240

\bibitem[{Armitage {et~al.}(2013)Armitage, Simon, \& Martin}]{Armitage2013}
Armitage, P.~J., Simon, J.~B., \& Martin, R.~G. 2013, The Astrophysical
  Journal, 778, L14

\bibitem[{Bai(2013)}]{Bai2013a}
Bai, X.-N. 2013, The Astrophysical Journal, 772, 96

\bibitem[{Bai(2016)}]{Bai2016b}
Bai, X.-N. 2016, The Astrophysical Journal, 821, 80

\bibitem[{Bai \& Stone(2013)}]{Bai2013}
Bai, X.-N. \& Stone, J.~M. 2013, The Astrophysical Journal, 769, 76

\bibitem[{Bai {et~al.}(2016)Bai, Ye, Goodman, \& Yuan}]{Bai2016a}
Bai, X.-N., Ye, J., Goodman, J., \& Yuan, F. 2016, The Astrophysical Journal,
  818, 152

\bibitem[{Balbus \& Hawley(1991)}]{Balbus1991}
Balbus, S.~A. \& Hawley, J.~F. 1991, The Astrophysical Journal, 376, 214

\bibitem[{Balbus \& Hawley(1997)}]{Balbus1997}
Balbus, S.~A. \& Hawley, J.~F. 1997, International Astronomical Union
  Colloquium, 163, 90

\bibitem[{Bayo {et~al.}(2012)Bayo, Barrado, Hu{\'{e}}lamo,
  Morales-Calder{\'{o}}n, Melo, Stauffer, \& Stelzer}]{Bayo2012}
Bayo, A., Barrado, D., Hu{\'{e}}lamo, N., {et~al.} 2012, Astronomy {\&}
  Astrophysics, 547, A80

\bibitem[{Bell \& Lin(1994)}]{Bell1994}
Bell, K.~R. \& Lin, D. N.~C. 1994, The Astrophysical Journal, 427, 987

\bibitem[{Birnstiel {et~al.}(2010)Birnstiel, Dullemond, \&
  Brauer}]{Birnstiel2010a}
Birnstiel, T., Dullemond, C.~P., \& Brauer, F. 2010, Astronomy and
  Astrophysics, 513, A79

\bibitem[{Blandford \& Payne(1982)}]{Blandford1982}
Blandford, R.~D. \& Payne, D.~G. 1982, Monthly Notices of the Royal
  Astronomical Society, 199, 883

\bibitem[{Chabrier(2003)}]{Chabrier2003}
Chabrier, G. 2003, Publications of the Astronomical Society of the Pacific,
  115, 763

\bibitem[{Chiang \& Goldreich(1997)}]{Chiang1997}
Chiang, E.~I. \& Goldreich, P. 1997, The Astrophysical Journal, 490, 368

\bibitem[{Clarke(2007)}]{Clarke2007}
Clarke, C.~J. 2007, Monthly Notices of the Royal Astronomical Society, 376,
  1350

\bibitem[{Clarke {et~al.}(2001)Clarke, Gendrin, \& Sotomayor}]{Clarke2001}
Clarke, C.~J., Gendrin, A., \& Sotomayor, M. 2001, Monthly Notices of the Royal
  Astronomical Society, 328, 485

\bibitem[{Cleeves {et~al.}(2016)Cleeves, {\"{O}}berg, Wilner, Huang, Loomis,
  Andrews, \& Czekala}]{Cleeves2016}
Cleeves, L.~I., {\"{O}}berg, K.~I., Wilner, D.~J., {et~al.} 2016, The
  Astrophysical Journal, 832, 110

\bibitem[{Crutcher(2012)}]{Crutcher2012}
Crutcher, R.~M. 2012, Annual Review of Astronomy and Astrophysics, 50, 29

\bibitem[{Emsenhuber {et~al.}(2023)Emsenhuber, Burn, Weder, Monsch, Picogna,
  Ercolano, \& Preibisch}]{Emsenhuber2023}
Emsenhuber, A., Burn, R., Weder, J., {et~al.} 2023, Astronomy {\&}
  Astrophysics, in press

\bibitem[{Emsenhuber {et~al.}(2021{\natexlab{a}})Emsenhuber, Mordasini, Burn,
  Alibert, Benz, \& Asphaug}]{Emsenhuber2021a}
Emsenhuber, A., Mordasini, C., Burn, R., {et~al.} 2021{\natexlab{a}}, Astronomy
  {\&} Astrophysics, 656, A69

\bibitem[{Emsenhuber {et~al.}(2021{\natexlab{b}})Emsenhuber, Mordasini, Burn,
  Alibert, Benz, \& Asphaug}]{Emsenhuber2021b}
Emsenhuber, A., Mordasini, C., Burn, R., {et~al.} 2021{\natexlab{b}}, Astronomy
  {\&} Astrophysics, 656, A70

\bibitem[{Ercolano {et~al.}(2009)Ercolano, Clarke, \& Drake}]{Ercolano2009}
Ercolano, B., Clarke, C.~J., \& Drake, J.~J. 2009, The Astrophysical Journal,
  699, 1639

\bibitem[{Ercolano \& Pascucci(2017)}]{Ercolano2017}
Ercolano, B. \& Pascucci, I. 2017, Royal Society Open Science, 4, 170114

\bibitem[{Fang {et~al.}(2012)Fang, Van~Boekel, King, Henning, Bouwman, Doi,
  Okamoto, Roccatagliata, \& Sicilia-Aguilar}]{Fang2012}
Fang, M., Van~Boekel, R., King, R.~R., {et~al.} 2012, Astronomy and
  Astrophysics, 539, 1

\bibitem[{Feiden(2016)}]{Feiden2016}
Feiden, G.~A. 2016, Astronomy {\&} Astrophysics, 593, A99

\bibitem[{Freedman {et~al.}(2014)Freedman, Lustig-Yaeger, Fortney, Lupu,
  Marley, \& Lodders}]{Freedman2014}
Freedman, R.~S., Lustig-Yaeger, J., Fortney, J.~J., {et~al.} 2014, The
  Astrophysical Journal Supplement Series, 214, 25

\bibitem[{Gammie(1996)}]{Gammie1996a}
Gammie, C.~F. 1996, The Astrophysical Journal, 457, 355

\bibitem[{Gorti \& Hollenbach(2009)}]{Gorti2009}
Gorti, U. \& Hollenbach, D. 2009, The Astrophysical Journal, 690, 1539

\bibitem[{Gullbring {et~al.}(1998)Gullbring, Hartmann, Briceno, \&
  Calvet}]{Gullbring1998}
Gullbring, E., Hartmann, L., Briceno, C., \& Calvet, N. 1998, The Astrophysical
  Journal, 492, 323

\bibitem[{Habing(1968)}]{Habing1968}
Habing, H.~J. 1968, Bull. Astron. Inst. Netherlands,, 19, 421

\bibitem[{Harrison {et~al.}(2021)Harrison, Looney, Stephens, Li, Teague,
  Crutcher, Yang, Cox, Fern{\'{a}}ndez-L{\'{o}}pez, \& Shinnaga}]{Harrison2021}
Harrison, R.~E., Looney, L.~W., Stephens, I.~W., {et~al.} 2021, The
  Astrophysical Journal, 908, 141

\bibitem[{Hartmann {et~al.}(2016)Hartmann, Herczeg, \& Calvet}]{Hartmann2016}
Hartmann, L., Herczeg, G., \& Calvet, N. 2016, Annual Review of Astronomy and
  Astrophysics, 54, 135

\bibitem[{Haworth \& Clarke(2019)}]{Haworth2019}
Haworth, T.~J. \& Clarke, C.~J. 2019, Monthly Notices of the Royal Astronomical
  Society, 485, 3895

\bibitem[{Haworth {et~al.}(2018)Haworth, Clarke, Rahman, Winter, \&
  Facchini}]{Haworth2018}
Haworth, T.~J., Clarke, C.~J., Rahman, W., Winter, A.~J., \& Facchini, S. 2018,
  Monthly Notices of the Royal Astronomical Society, 481, 452

\bibitem[{Hollenbach {et~al.}(1994)Hollenbach, Johnstone, Lizano, \&
  Shu}]{Hollenbach1994}
Hollenbach, D., Johnstone, D., Lizano, S., \& Shu, F. 1994, The Astrophysical
  Journal, 428, 654

\bibitem[{Hueso \& Guillot(2005)}]{Hueso2005}
Hueso, R. \& Guillot, T. 2005, Astronomy {\&} Astrophysics, 442, 703

\bibitem[{Jennings {et~al.}(2018)Jennings, Ercolano, \& Rosotti}]{Jennings2018}
Jennings, J., Ercolano, B., \& Rosotti, G.~P. 2018, Monthly Notices of the
  Royal Astronomical Society, 477, 4131

\bibitem[{Kimura {et~al.}(2016)Kimura, Kunitomo, \& Takahashi}]{Kimura2016}
Kimura, S.~S., Kunitomo, M., \& Takahashi, S.~Z. 2016, Monthly Notices of the
  Royal Astronomical Society, 461, 2257

\bibitem[{Koepferl {et~al.}(2013)Koepferl, Ercolano, Dale, Teixeira, Ratzka, \&
  Spezzi}]{Koepferl2013}
Koepferl, C.~M., Ercolano, B., Dale, J., {et~al.} 2013, Monthly Notices of the
  Royal Astronomical Society, 428, 3327

\bibitem[{Komaki {et~al.}(2021)Komaki, Nakatani, \& Yoshida}]{Komaki2021}
Komaki, A., Nakatani, R., \& Yoshida, N. 2021, The Astrophysical Journal, 910,
  51

\bibitem[{K{\"{o}}nigl \& Salmeron(2010)}]{Konigl2010}
K{\"{o}}nigl, A. \& Salmeron, R. 2010, in Physical Processes in Circumstellar
  Disks around Young Stars, ed. P.~J.~V. Garcia (Chicago: University of Chicago
  Press), 283--354

\bibitem[{Kraus {et~al.}(2012)Kraus, Ireland, Hillenbrand, \&
  Martinache}]{Kraus2012}
Kraus, A.~L., Ireland, M.~J., Hillenbrand, L.~A., \& Martinache, F. 2012, The
  Astrophysical Journal, 745, 19

\bibitem[{Kunitomo {et~al.}(2020)Kunitomo, Suzuki, \& Inutsuka}]{Kunitomo2020}
Kunitomo, M., Suzuki, T.~K., \& Inutsuka, S.-i. 2020, Monthly Notices of the
  Royal Astronomical Society, 11, 3849

\bibitem[{Lesur {et~al.}(2022)Lesur, Ercolano, Flock, Lin, Yang, Barranco,
  Benitez-Llambay, Goodman, Johansen, Klahr, Laibe, Lyra, Marcus, Nelson,
  Squire, Simon, Turner, Umurhan, \& Youdin}]{Lesur2022}
Lesur, G., Ercolano, B., Flock, M., {et~al.} 2022, arXiv:2203.09821

\bibitem[{Liffman(2003)}]{Liffman2003}
Liffman, K. 2003, Publications of the Astronomical Society of Australia, 20,
  337

\bibitem[{Lodato {et~al.}(2017)Lodato, Scardoni, Manara, \& Testi}]{Lodato2017}
Lodato, G., Scardoni, C.~E., Manara, C.~F., \& Testi, L. 2017, Monthly Notices
  of the Royal Astronomical Society, 472, 4700

\bibitem[{Lodders(2003)}]{Lodders2003}
Lodders, K. 2003, The Astrophysical Journal, 591, 1220

\bibitem[{Lynden-Bell \& Pringle(1974)}]{LyndenBell1974}
Lynden-Bell, D. \& Pringle, J.~E. 1974, Monthly Notices of the Royal
  Astronomical Society, 168, 603

\bibitem[{Mamajek {et~al.}(2009)Mamajek, Usuda, Tamura, \& Ishii}]{Mamajek2009}
Mamajek, E.~E., Usuda, T., Tamura, M., \& Ishii, M. 2009, in AIP Conference
  Proceedings (AIP), 3--10

\bibitem[{Manara {et~al.}(2019)Manara, Mordasini, Testi, Williams, Miotello,
  Lodato, \& Emsenhuber}]{Manara2019}
Manara, C.~F., Mordasini, C., Testi, L., {et~al.} 2019, Astronomy {\&}
  Astrophysics, 631, L2

\bibitem[{Manara {et~al.}(2016)Manara, Rosotti, Testi, Natta, Alcal{\'{a}},
  Williams, Ansdell, Miotello, Van Der~Marel, Tazzari, Carpenter, Guidi,
  Mathews, Oliveira, Prusti, \& Van~Dishoeck}]{Manara2016}
Manara, C.~F., Rosotti, G., Testi, L., {et~al.} 2016, Astronomy and
  Astrophysics, 591, 3

\bibitem[{Michel {et~al.}(2021)Michel, van~der Marel, \& Matthews}]{Michel2021}
Michel, A., van~der Marel, N., \& Matthews, B.~C. 2021, The Astrophysical
  Journal, 921, 72

\bibitem[{Mordasini {et~al.}(2009)Mordasini, Alibert, \& Benz}]{Mordasini2009}
Mordasini, C., Alibert, Y., \& Benz, W. 2009, Astronomy {\&} Astrophysics, 501,
  1139

\bibitem[{Mulders {et~al.}(2017)Mulders, Pascucci, Manara, Testi, Herczeg,
  Henning, Mohanty, \& Lodato}]{Mulders2017}
Mulders, G.~D., Pascucci, I., Manara, C.~F., {et~al.} 2017, The Astrophysical
  Journal, 847, 31

\bibitem[{Murray {et~al.}(2001)Murray, Chaboyer, Arras, Hansen, \&
  Noyes}]{Murray2001}
Murray, N., Chaboyer, B., Arras, P., Hansen, B., \& Noyes, R.~W. 2001, The
  Astrophysical Journal, 555, 801

\bibitem[{Nakamoto \& Nakagawa(1994)}]{Nakamoto1994}
Nakamoto, T. \& Nakagawa, Y. 1994, The Astrophysical Journal, 421, 640

\bibitem[{Ndugu {et~al.}(2018)Ndugu, Bitsch, \& Jurua}]{Ndugu2018}
Ndugu, N., Bitsch, B., \& Jurua, E. 2018, Monthly Notices of the Royal
  Astronomical Society, 474, 886

\bibitem[{Ogihara {et~al.}(2015{\natexlab{a}})Ogihara, Kobayashi, Inutsuka, \&
  Suzuki}]{Ogihara2015a}
Ogihara, M., Kobayashi, H., Inutsuka, S.-i., \& Suzuki, T.~K.
  2015{\natexlab{a}}, Astronomy {\&} Astrophysics, 579, A65

\bibitem[{Ogihara {et~al.}(2015{\natexlab{b}})Ogihara, Morbidelli, \&
  Guillot}]{Ogihara2015}
Ogihara, M., Morbidelli, A., \& Guillot, T. 2015{\natexlab{b}}, Astronomy {\&}
  Astrophysics, 584, L1

\bibitem[{Owen(2016)}]{Owen2016}
Owen, J.~E. 2016, Publications of the Astronomical Society of Australia, 33,
  e005

\bibitem[{Owen {et~al.}(2012)Owen, Clarke, \& Ercolano}]{Owen2012}
Owen, J.~E., Clarke, C.~J., \& Ercolano, B. 2012, Monthly Notices of the Royal
  Astronomical Society, 422, 1880

\bibitem[{Owen {et~al.}(2011)Owen, Ercolano, \& Clarke}]{Owen2011}
Owen, J.~E., Ercolano, B., \& Clarke, C.~J. 2011, Monthly Notices of the Royal
  Astronomical Society, 412, 13

\bibitem[{Owen {et~al.}(2010)Owen, Ercolano, Clarke, \& Alexander}]{Owen2010}
Owen, J.~E., Ercolano, B., Clarke, C.~J., \& Alexander, R.~D. 2010, Monthly
  Notices of the Royal Astronomical Society, 401, 1415

\bibitem[{Pascucci {et~al.}(2022)Pascucci, Cabrit, Edwards, Gorti, Gressel, \&
  Suzuki}]{Pascucci2022}
Pascucci, I., Cabrit, S., Edwards, S., {et~al.} 2022, [arXiv:2203.10068]

\bibitem[{Pascucci {et~al.}(2016)Pascucci, Testi, Herczeg, Long, Manara,
  Hendler, Mulders, Krijt, Ciesla, Henning, Mohanty, Drabek-Maunder, Apai,
  Sz{\H{u}}cs, Sacco, \& Olofsson}]{Pascucci2016}
Pascucci, I., Testi, L., Herczeg, G.~J., {et~al.} 2016, The Astrophysical
  Journal, 831, 125

\bibitem[{Perez-Becker \& Chiang(2011{\natexlab{a}})}]{Perez-Becker2011b}
Perez-Becker, D. \& Chiang, E. 2011{\natexlab{a}}, The Astrophysical Journal,
  735, 8

\bibitem[{Perez-Becker \& Chiang(2011{\natexlab{b}})}]{Perez-Becker2011a}
Perez-Becker, D. \& Chiang, E. 2011{\natexlab{b}}, The Astrophysical Journal,
  727, 2

\bibitem[{Picogna {et~al.}(2021)Picogna, Ercolano, \& Espaillat}]{Picogna2021}
Picogna, G., Ercolano, B., \& Espaillat, C.~C. 2021, Monthly Notices of the
  Royal Astronomical Society, 508, 3611

\bibitem[{Raymond {et~al.}(2007)Raymond, Scalo, \& Meadows}]{Raymond2007}
Raymond, S.~N., Scalo, J., \& Meadows, V.~S. 2007, The Astrophysical Journal,
  669, 606

\bibitem[{Richert {et~al.}(2018)Richert, Getman, Feigelson, Kuhn, Broos,
  Povich, Bate, \& Garmire}]{Richert2018}
Richert, A. J.~W., Getman, K.~V., Feigelson, E.~D., {et~al.} 2018, Monthly
  Notices of the Royal Astronomical Society, 477, 5191

\bibitem[{Rigliaco {et~al.}(2011)Rigliaco, Natta, Randich, Testi, \&
  Biazzo}]{Rigliaco2011}
Rigliaco, E., Natta, A., Randich, S., Testi, L., \& Biazzo, K. 2011, Astronomy
  {\&} Astrophysics, 525, A47

\bibitem[{Rodenkirch {et~al.}(2020)Rodenkirch, Klahr, Fendt, \&
  Dullemond}]{Rodenkirch2020}
Rodenkirch, P.~J., Klahr, H., Fendt, C., \& Dullemond, C.~P. 2020, Astronomy
  {\&} Astrophysics, 633, A21

\bibitem[{Ruden \& Pollack(1991)}]{Ruden1991}
Ruden, S.~P. \& Pollack, J.~B. 1991, The Astrophysical Journal, 375, 740

\bibitem[{Santos {et~al.}(2005)Santos, Israelian, Mayor, Bento, Almeida, Sousa,
  \& Ecuvillon}]{Santos2005}
Santos, N.~C., Israelian, G., Mayor, M., {et~al.} 2005, Astronomy {\&}
  Astrophysics, 437, 1127

\bibitem[{Scally \& Clarke(2001)}]{Scally2001}
Scally, A. \& Clarke, C. 2001, Monthly Notices of the Royal Astronomical
  Society, 325, 449

\bibitem[{Schib {et~al.}(2021)Schib, Mordasini, Wenger, Marleau, \&
  Helled}]{Schib2020}
Schib, O., Mordasini, C., Wenger, N., Marleau, G.-D., \& Helled, R. 2021,
  Astronomy {\&} Astrophysics, 645, A43

\bibitem[{Sellek {et~al.}(2020{\natexlab{a}})Sellek, Booth, \&
  Clarke}]{Sellek2020a}
Sellek, A.~D., Booth, R.~A., \& Clarke, C.~J. 2020{\natexlab{a}}, Monthly
  Notices of the Royal Astronomical Society, 498, 2845

\bibitem[{Sellek {et~al.}(2020{\natexlab{b}})Sellek, Booth, \&
  Clarke}]{Sellek2020b}
Sellek, A.~D., Booth, R.~A., \& Clarke, C.~J. 2020{\natexlab{b}}, Monthly
  Notices of the Royal Astronomical Society, 492, 1279

\bibitem[{Shakura \& Sunyaev(1973)}]{Shakura1973}
Shakura, N.~I. \& Sunyaev, R.~A. 1973, Symposium - International Astronomical
  Union, 55, 155

\bibitem[{Somigliana {et~al.}(2022)Somigliana, Toci, Rosotti, Lodato, Tazzari,
  Manara, Testi, \& Lepri}]{Somigliana2022}
Somigliana, A., Toci, C., Rosotti, G., {et~al.} 2022, Monthly Notices of the
  Royal Astronomical Society, 514, 5927

\bibitem[{Suzuki {et~al.}(2016)Suzuki, Ogihara, Morbidelli, Crida, \&
  Guillot}]{Suzuki2016}
Suzuki, T.~K., Ogihara, M., Morbidelli, A., Crida, A., \& Guillot, T. 2016,
  Astronomy {\&} Astrophysics, 596, A74

\bibitem[{Tabone {et~al.}(2022{\natexlab{a}})Tabone, Rosotti, Cridland,
  Armitage, \& Lodato}]{Tabone2022a}
Tabone, B., Rosotti, G.~P., Cridland, A.~J., Armitage, P.~J., \& Lodato, G.
  2022{\natexlab{a}}, Monthly Notices of the Royal Astronomical Society, 512,
  2290

\bibitem[{Tabone {et~al.}(2022{\natexlab{b}})Tabone, Rosotti, Lodato, Armitage,
  Cridland, \& van Dishoeck}]{Tabone2022b}
Tabone, B., Rosotti, G.~P., Lodato, G., {et~al.} 2022{\natexlab{b}}, Monthly
  Notices of the Royal Astronomical Society: Letters, 512, L74

\bibitem[{Takasao {et~al.}(2018)Takasao, Tomida, Iwasaki, \&
  Suzuki}]{Takasao2018}
Takasao, S., Tomida, K., Iwasaki, K., \& Suzuki, T.~K. 2018, The Astrophysical
  Journal, 857, 4

\bibitem[{Testi {et~al.}(2022)Testi, Natta, Manara, de~Gregorio~Monsalvo,
  Lodato, Lopez, Muzic, Pascucci, Sanchis, Miranda, Scholz, De~Simone, \&
  Williams}]{Testi2022a}
Testi, L., Natta, A., Manara, C.~F., {et~al.} 2022, Astronomy {\&}
  Astrophysics, 663, A98

\bibitem[{Tobin {et~al.}(2016)Tobin, Looney, Li, Chandler, Dunham, Segura-Cox,
  Sadavoy, Melis, Harris, Kratter, \& Perez}]{Tobin2016}
Tobin, J.~J., Looney, L.~W., Li, Z.-Y., {et~al.} 2016, The Astrophysical
  Journal, 818, 73

\bibitem[{Tobin {et~al.}(2020)Tobin, Sheehan, Megeath,
  D{\'{i}}az-Rodr{\'{i}}guez, Offner, Murillo, van~’t Hoff, van Dishoeck,
  Osorio, Anglada, Furlan, Stutz, Reynolds, Karnath, Fischer, Persson, Looney,
  Li, Stephens, Chandler, Cox, Dunham, Tychoniec, Kama, Kratter, Kounkel,
  Mazur, Maud, Patel, Perez, Sadavoy, Segura-Cox, Sharma, Stephenson, Watson,
  \& Wyrowski}]{Tobin2020}
Tobin, J.~J., Sheehan, P.~D., Megeath, S.~T., {et~al.} 2020, The Astrophysical
  Journal, 890, 130

\bibitem[{Tychoniec {et~al.}(2018)Tychoniec, Tobin, Karska, Chandler, Dunham,
  Harris, Kratter, Li, Looney, Melis, P{\'{e}}rez, Sadavoy, Segura-Cox, \& van
  Dishoeck}]{Tychoniec2018}
Tychoniec, {\L}., Tobin, J.~J., Karska, A., {et~al.} 2018, The Astrophysical
  Journal Supplement Series, 238, 19

\bibitem[{Venuti {et~al.}(2017)Venuti, Bouvier, Cody, Stauffer, Micela, Rebull,
  Alencar, Sousa, Hillenbrand, \& Flaccomio}]{Venuti2017}
Venuti, L., Bouvier, J., Cody, A.~M., {et~al.} 2017, Astronomy {\&}
  Astrophysics, 599, A23

\bibitem[{Veras \& Armitage(2004)}]{Veras2004}
Veras, D. \& Armitage, P.~J. 2004, Monthly Notices of the Royal Astronomical
  Society, 347, 613

\bibitem[{Wang {et~al.}(2019)Wang, Bai, \& Goodman}]{Wang2019}
Wang, L., Bai, X.-N., \& Goodman, J. 2019, The Astrophysical Journal, 874, 90

\bibitem[{Wang \& Goodman(2017)}]{Wang2017}
Wang, L. \& Goodman, J.~J. 2017, The Astrophysical Journal, 835, 59

\bibitem[{Whelan {et~al.}(2021)Whelan, Pascucci, Gorti, Edwards, Alexander,
  Sterzik, \& Melo}]{Whelan2021}
Whelan, E.~T., Pascucci, I., Gorti, U., {et~al.} 2021, The Astrophysical
  Journal, 913, 43

\bibitem[{Winter {et~al.}(2020)Winter, Kruijssen, Longmore, \&
  Chevance}]{Winter2020}
Winter, A.~J., Kruijssen, J. M.~D., Longmore, S.~N., \& Chevance, M. 2020,
  Nature, 586, 528

\end{thebibliography}

\begin{appendix}

\section{Variable $\overline{\alpha_{\phi z,0}}$} \label{app:varAlpha}
\begin{figure}[h]
   \centering
   \includegraphics[width=0.9\hsize]{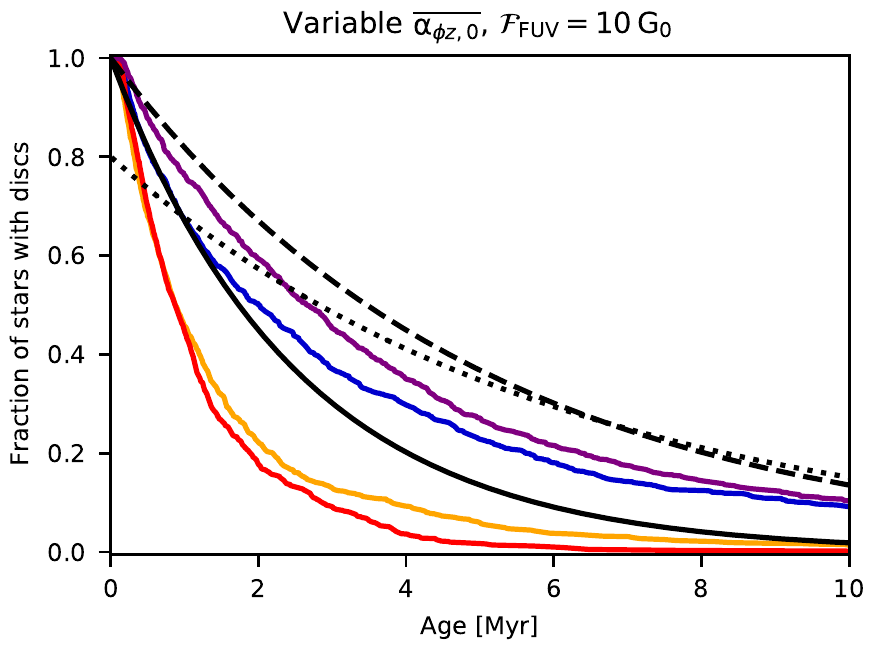} \\
   \includegraphics[width=0.9\hsize]{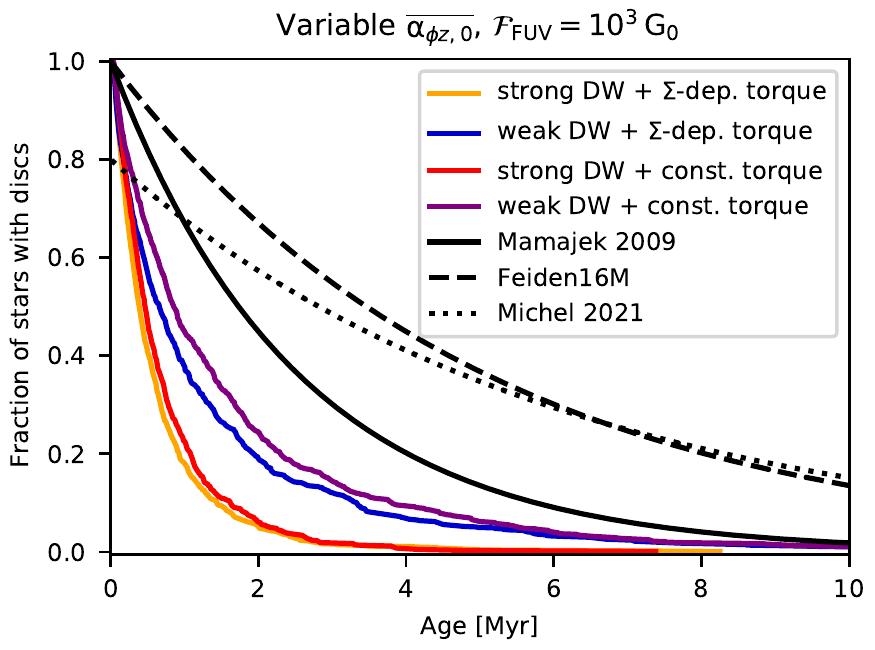}
   \caption{Fraction of stars possessing a circumstellar disc as a function of time, analogous to Fig. \ref{fig:discfraction}, but for populations with varying $\overline{\alpha_{\phi z, 0}}$. Top panel shows the disc fraction evolution for a weak $10\mathrm{G}_0$ field, while the bottom panel shows the same evolution for a $10^3 \mathrm{G}_0$ FUV field.}
   \label{fig:discfraction_varAlpha}
\end{figure}

\begin{figure*}[h]
    \centering
    \includegraphics[width=0.9\hsize]{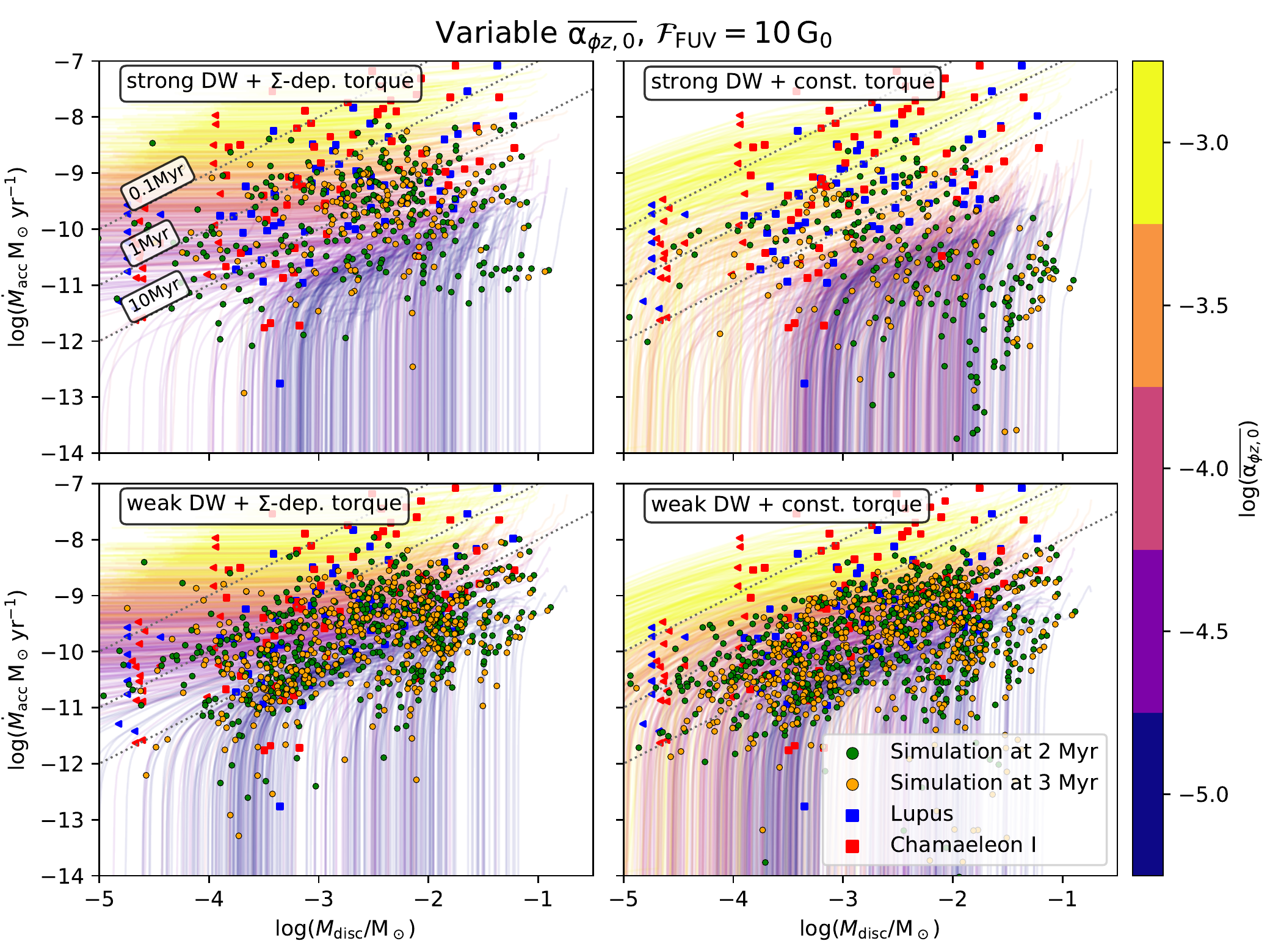}
    \caption{Stellar accretion rate $\dot{M}_\mathrm{acc}$ vs. gas disc mass $M_\mathrm{disc}$, analogous to Fig. \ref{fig:mass_acc_lowFUV}, but for an external FUV field of $10\mathrm{G}_0$ and varying $\overline{\alpha_{\phi z,0}}$. Evolution tracks of individual systems are coloured by $\overline{\alpha_{\phi z, 0}}$. Snapshots at $2\,\mathrm{Myr}$ (green circles) and $3\,\mathrm{Myr}$ (orange circles) are displayed. We compare our simulations with observed populations in Lupus and Chamaeleon I. Observational dust disc masses and stellar accretion rates are taken from \cite{Manara2019} and dust masses are converted to gas masses by assuming the standard dust-to-gas ratio of $0.01$. Triangles denote upper limits on disc mass. Lines of constant $M_\mathrm{disc}/\dot{M}_\mathrm{acc}$ are shown for 0.1, 1 and 10 Myr (thin dotted lines).}
    \label{fig:mdisc_macc_varAlpha_lowFUV}
\end{figure*}

\begin{figure*}[h]
    \centering
    \includegraphics[width=0.9\hsize]{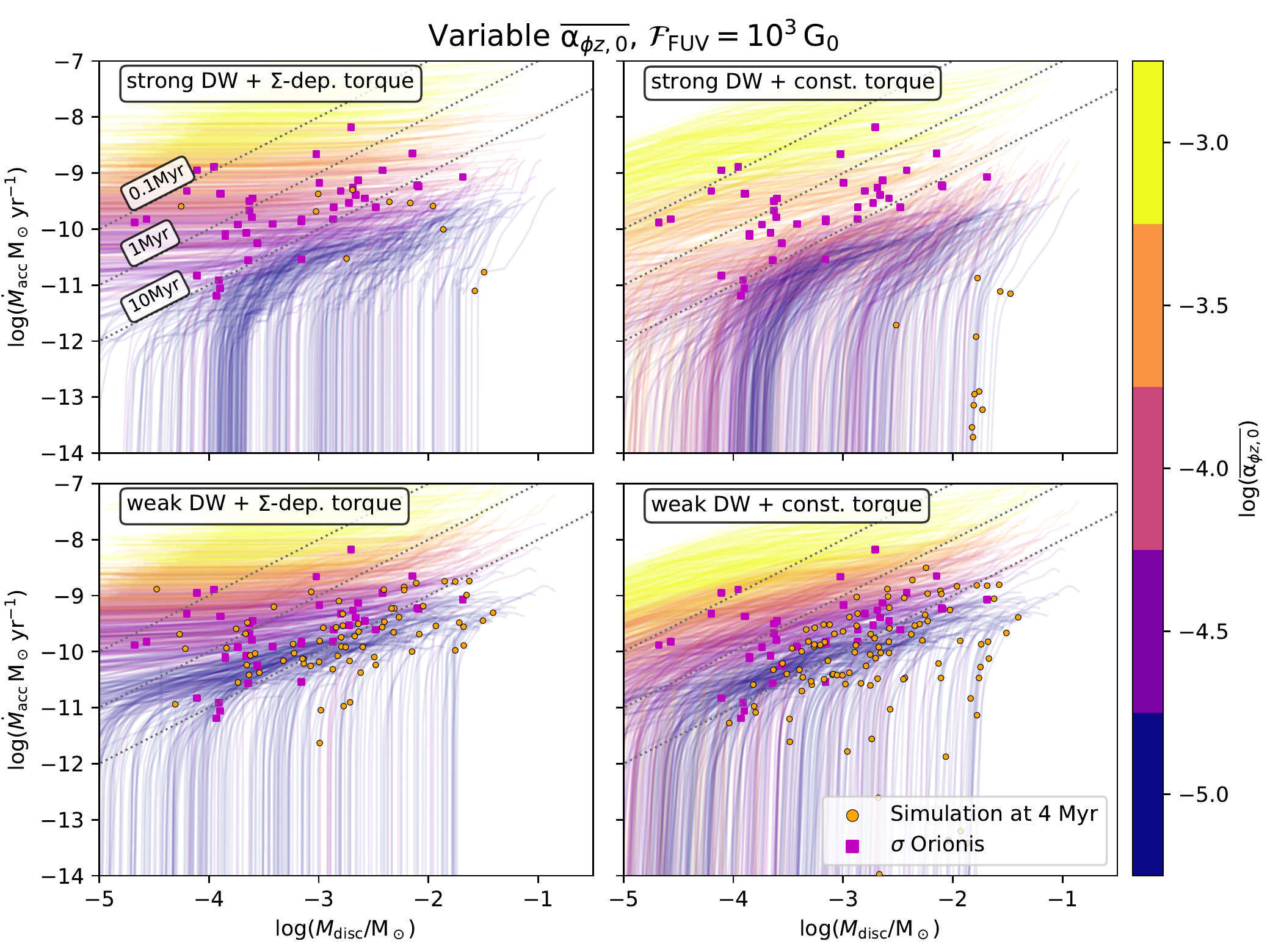}
   3 \caption{Stellar accretion rate $\dot{M}_\mathrm{acc}$ vs. gas disc mass $M_\mathrm{disc}$, analogous to Fig. \ref{fig:mass_acc_adams}, but for an external FUV field of $10^3\mathrm{G}_0$ and varying $\overline{\alpha_{\phi z,0}}$. Evolution tracks of individual systems are coloured by $\overline{\alpha_{\phi z, 0}}$. A snapshot for the individual systems is shown at $4\,\mathrm{Myr}$ (orange circles) in comparison with observational data from $\sigma$ Orionis. Observational dust disc masses are taken from \cite{Ansdell2017a} and converted to gas masses by assuming the standard dust-to-gas ratio of $0.01$. Stellar accretion rates are taken from \cite{Rigliaco2011}.
   }
    \label{fig:mdisc_macc_varAlpha_highFUV}
\end{figure*}

Here, we show additional cases with varying initial torque strength, $\overline{\alpha_{\phi z, 0}}$, to assess its influence on disc evolution on a population level. It is varied as an initial condition, with values distributed uniformly in log between $10^{-5.5}$ and $10^{-2.5}$. We further evolved populations in a weak and a strong FUV field environment with values of $10\mathrm{G}_0$ and $10^3\mathrm{G}_0$, respectively. We chose not to vary the FUV field alongside with $\overline{\alpha_{\phi z,0}}$ in order to get a better understanding on how the torque strength influences the evolution. We show disc lifetimes in Table \ref{fig:discfraction_varAlpha} and $M_\mathrm{disc}$-$\dot{M}_\mathrm{acc}$ evolutions for both FUV field strengths in Figs. \ref{fig:mdisc_macc_varAlpha_lowFUV} and \ref{fig:mdisc_macc_varAlpha_highFUV}. The characteristic numbers of the populations are shown in Tables \ref{tab:mdisc_macc_varAlpha_lowFUV} and \ref{tab:mdisc_macc_varAlpha_highFUV}.

The evolution tracks of the simulations show a clear separation by initial torque strength. Here, even high accretors are reached by simulation tracks with high initial torques. However, we still did not recover simulations at these high accretion rates after a few million years (i.e. see snapshots), especially beyond the \SI{0.1}{\mega\year} line of constant accretion. Discs beyond this point are rapidly dispersed. The simulation tracks show a much larger spread in accretion rates, more similar to the observations. This comes to no surprise, as the $\overline{\alpha_{\phi z,0}}$ has a direct influence on the accretion rate. Looking at the characteristic numbers of the $10\mathrm{G}_0$ populations in Table \ref{tab:mdisc_macc_varAlpha_lowFUV}, we also observe fewer discs opening an inner cavity for $\Sigma$-dependent torque cases (23\% to 35\%), compared to the constant torque cases (44\% to 57\%).
The same can be seen in the $10^{3}\mathrm{G}_0$ populations (Table \ref{tab:mdisc_macc_varAlpha_highFUV}). The torques start initially with the same strength, namely, $\overline{\alpha_{\phi z,0}}$. As the surface density declines, the torques get stronger in the $\Sigma$-dependent cases, leading to higher accretion rates. This makes it less likely for internal photoevaporation opening up a gap.

The evolutions of disc fractions show similar features as our nominal cases in Fig. \ref{fig:discfraction}. Strong disc winds have short lifetimes, regardless of the ambient FUV field strength. Discs exposed to weak winds have slightly longer lifetimes and show a dependency on the ambient FUV field. Lifetimes of weak disc wind scenarios for the \num{10}$\mathrm{G}_0$ cases are in line with observations. We also note that for strong and weak disc winds, the disc fractions evolve in pairs, regardless of the torque scenario.

\begin{table*}
      \begin{center}
         \caption{Characteristics for populations with $10\mathrm{G}_0$ FUV field environment and varying $\overline{\alpha_{\phi z, 0}}$.}
         \label{tab:mdisc_macc_varAlpha_lowFUV}
         \begin{tabular}{l c c c r r r r}
            \hline\hline
               DW scenario & $t_{1/2}$\textsuperscript{a)} & $f_\mathrm{disc,2Myr}$\textsuperscript{b)} & $f_\mathrm{cavity}$\textsuperscript{c)} & $f_{M_\mathrm{acc}}$\textsuperscript{d)} & $f_{M_\mathrm{PEW,int}}$\textsuperscript{d)} & $f_{M_\mathrm{PEW,ext}}$\textsuperscript{d)} & $f_{M_\mathrm{MDW}}$\textsuperscript{d)} \\
                            & [\SI{}{\mega\year}] & [\SI{}{\percent}] & [\SI{}{\percent}] & [\SI{}{\percent}] & [\SI{}{\percent}] & [\SI{}{\percent}] & [\SI{}{\percent}] \\
            \hline
               strong DW + $\Sigma$-dep. torque  & \SI{0.88}{}   &   \SI{22}{} &  \SI{35}{} & \SI{11.2}{}   & \SI{3.9}{} & \SI{63.7}{} & \SI{21.3}{} \\
               weak DW + $\Sigma$-dep. torque    & \SI{2.02}{}   &  \SI{50}{} &  \SI{23}{} &  \SI{17.5}{}   & \SI{7.0}{} & \SI{64.2}{} & \SI{11.4}{} \\
               strong DW + const. torque         & \SI{0.87}{}   &   \SI{18}{} &  \SI{57}{} &  \SI{5.9}{}   &  \SI{5.3}{} & \SI{64.2}{} & \SI{24.6}{} \\
               weak DW + const. torque           & \SI{2.69}{}   &  \SI{59}{} &  \SI{44}{} & \SI{14.1}{}   & \SI{8.8}{} & \SI{64.8}{} & \SI{12.3}{} \\
            \hline
         \end{tabular}
      \end{center}
      Notes: a) time at which half of the discs are dispersed, b) fraction of discs remaining at \SI{2}{\mega\year}, c) fraction of discs opening up an inner cavity, d) contribution of the different processes to disc dispersal in terms of mass loss percentage (mean value over all simulations)
   \end{table*}

\clearpage

\begin{table*}
      \begin{center}
         \caption{Characteristics for populations with a $10^3\mathrm{G}_0$ FUV field environment and varying $\overline{\alpha_{\phi z, 0}}$.}
         \label{tab:mdisc_macc_varAlpha_highFUV}
         \begin{tabular}{l c c c r r r r}
            \hline\hline
               DW scenario & $t_{1/2}$\textsuperscript{a)} & $f_\mathrm{disc,2Myr}$\textsuperscript{b)} & $f_\mathrm{cavity}$\textsuperscript{c)} & $f_{M_\mathrm{acc}}$\textsuperscript{d)} & $f_{M_\mathrm{PEW,int}}$\textsuperscript{d)} & $f_{M_\mathrm{PEW,ext}}$\textsuperscript{d)} & $f_{M_\mathrm{MDW}}$\textsuperscript{d)} \\
               & [\SI{}{\mega\year}] & [\SI{}{\percent}] & [\SI{}{\percent}] & [\SI{}{\percent}] & [\SI{}{\percent}] & [\SI{}{\percent}] & [\SI{}{\percent}] \\
            \hline
               strong DW + $\Sigma$-dep. torque  & 0.39\SI{}{}   &   5\SI{}{} &  24\SI{}{} & 3.9\SI{}{}   & 1.1\SI{}{} & 87.3\SI{}{} & 7.6\SI{}{} \\
               weak DW + $\Sigma$-dep. torque    & 0.66\SI{}{}   &  19\SI{}{} &  15\SI{}{} &  6.8\SI{}{}   & 1.8\SI{}{} & 87.7\SI{}{} & 3.7\SI{}{} \\
               strong DW + const. torque         & 0.47\SI{}{}   &   6\SI{}{} &  46\SI{}{} &  2.4\SI{}{}   &  1.5\SI{}{} & 87.5\SI{}{} & 8.6\SI{}{} \\
               weak DW + const. torque           & 0.86\SI{}{}   &  24\SI{}{} &  35\SI{}{} & 5.9\SI{}{}   & 2.2\SI{}{} & 87.8\SI{}{} & 4.0\SI{}{} \\
            \hline
         \end{tabular}
      \end{center}
      Notes: a) time at which half of the discs are dispersed, b) fraction of discs remaining at \SI{2}{\mega\year}, c) fraction of discs opening up an inner cavity, d) contribution of the different processes to disc dispersal in terms of mass loss percentage (mean value over all simulations)
   \end{table*}

\clearpage

   \section{EUV photoevaporation always active (i.e. no shielding)}
   \label{app:noshielding}
   
\begin{figure}
      \centering
      \includegraphics[width=0.9\hsize]{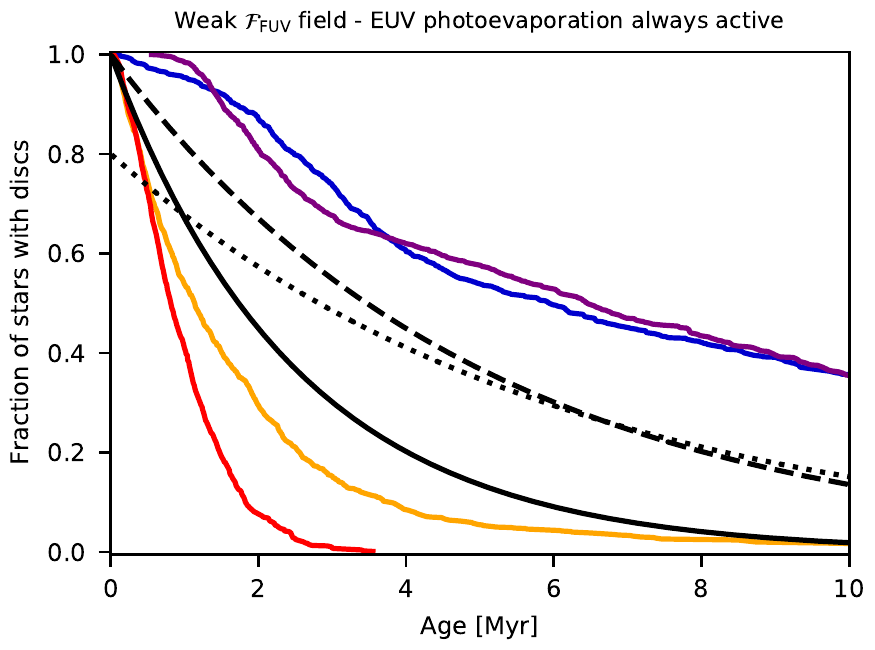} \\
      \includegraphics[width=0.9\hsize]{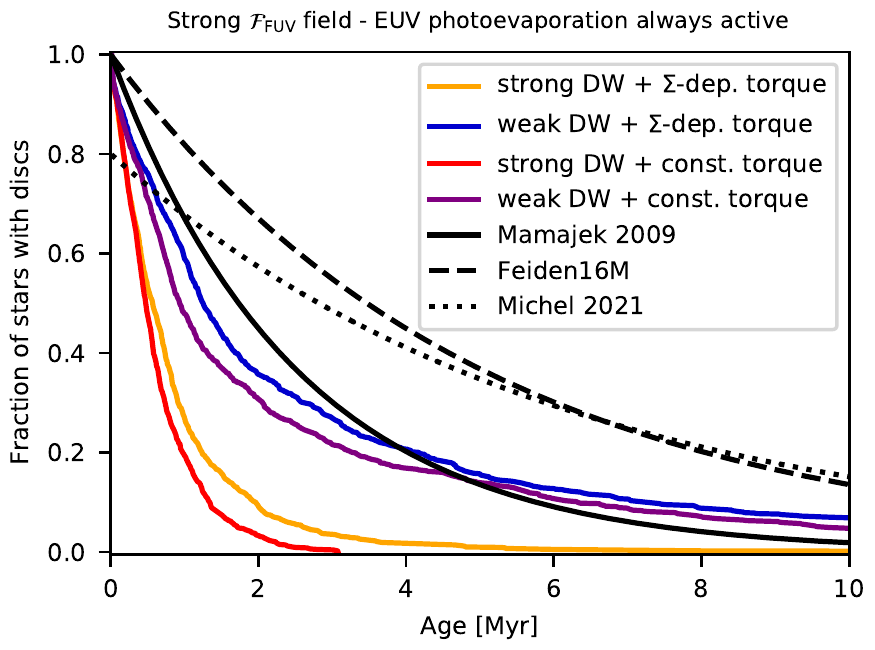}
      \caption{Fraction of stars possessing a circumstellar disc as a function of time, analogous to Fig. \ref{fig:discfraction}, but for populations with EUV photoevaporation always active (i.e. no shielding).}
      \label{fig:discfraction_noshielding}
   \end{figure}

\begin{figure*}
      \centering
      \includegraphics[width=0.9\hsize]{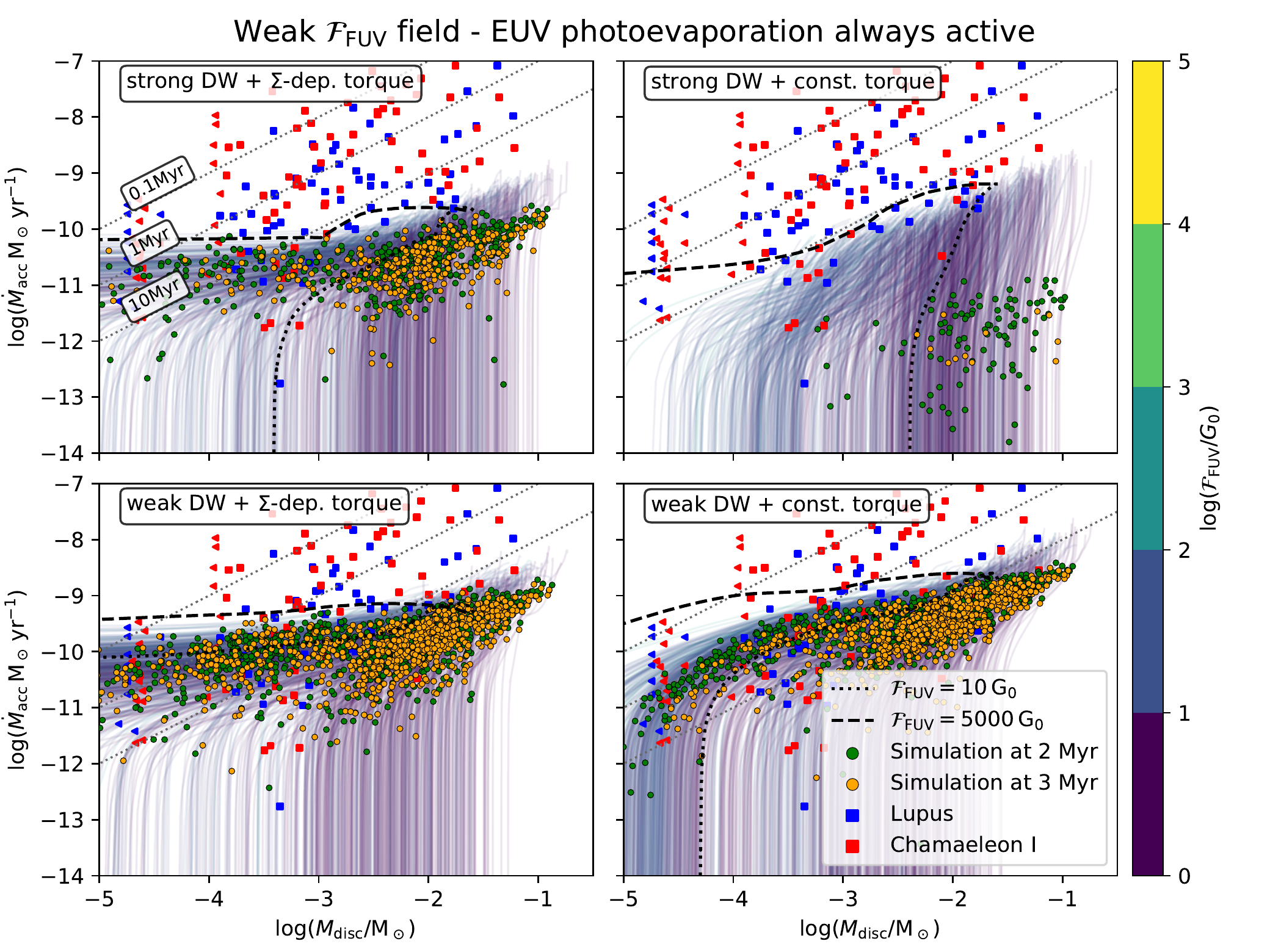}
      \caption{Stellar accretion rate $\dot{M}_\mathrm{acc}$ vs. gas disc mass $M_\mathrm{disc}$, analogous to Fig. \ref{fig:mass_acc_lowFUV}, but without EUV radiation shielding. Exemplary cases are the same as in Figs. \ref{fig:exemplary_case_10g0} and \ref{fig:exemplary_case_5e3g0}, but without EUV radiation shielding. Snapshots at $2\,\mathrm{Myr}$ (green circles) and $3\,\mathrm{Myr}$ (orange circles) are displayed. We compare our simulations with observed populations in Lupus and Chamaeleon I. Observational dust disc masses and stellar accretion rates are taken from \cite{Manara2019} and dust masses are converted to gas masses by assuming the standard dust-to-gas ratio of $0.01$. Triangles denote upper limits on disc mass. Lines of constant $M_\mathrm{disc}/\dot{M}_\mathrm{acc}$ are shown for 0.1, 1 and 10 Myr (thin dotted lines).}
      \label{fig:mass_acc_lowFUV_noshielding}
   \end{figure*}

\begin{figure*}
      \centering
      \includegraphics[width=0.9\hsize]{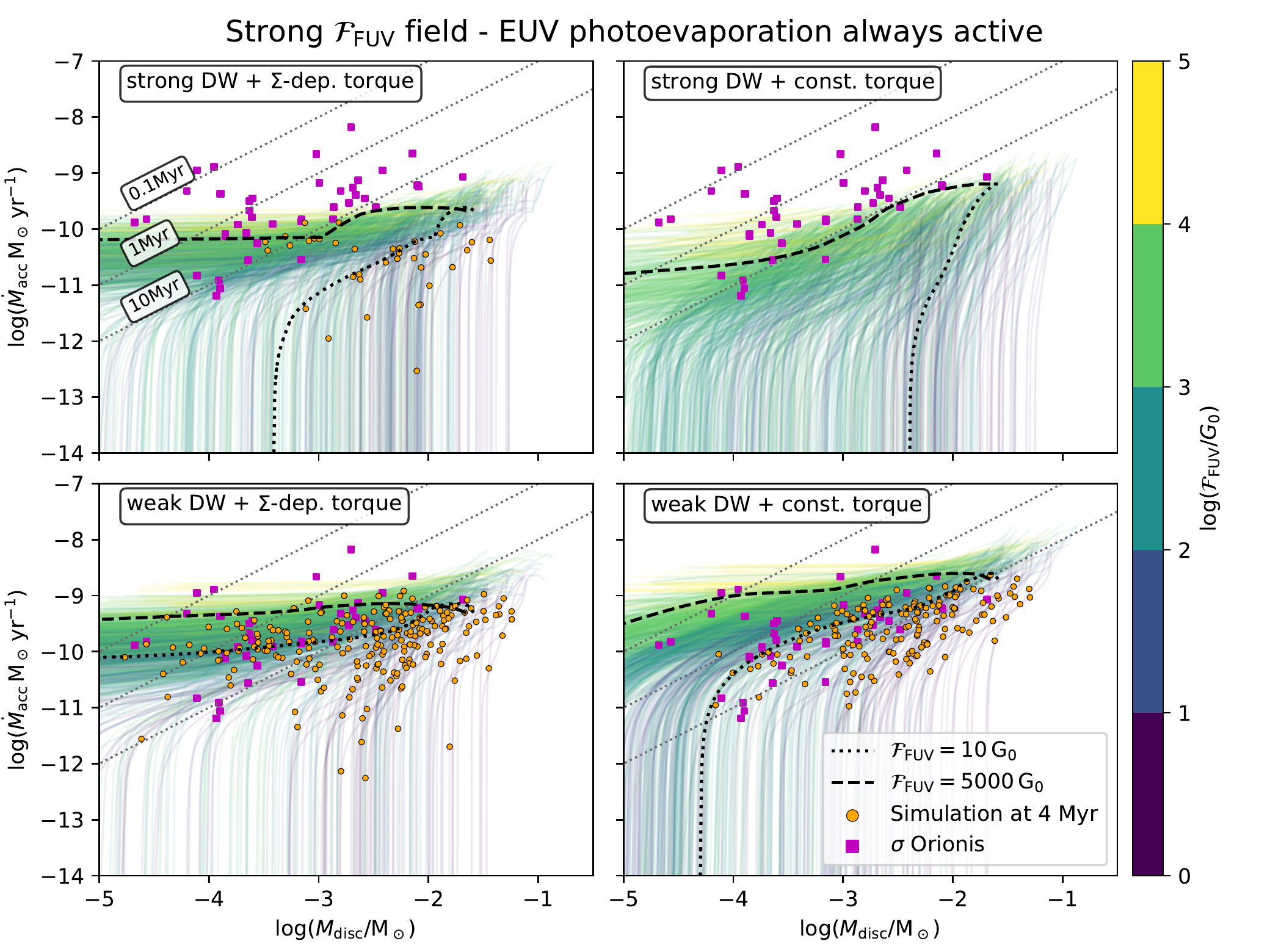}
      \caption{Stellar accretion rate $\dot{M}_\mathrm{acc}$ vs. gas disc mass $M_\mathrm{disc}$, analogous to Fig. \ref{fig:mass_acc_adams}, but without EUV radiation shielding. Exemplary cases are the same as in Figs. \ref{fig:exemplary_case_10g0} and \ref{fig:exemplary_case_5e3g0} but without EUV radiation shielding. A snapshot for the individual systems is shown at $4\,\mathrm{Myr}$ (orange circles) in comparison with observational data from $\sigma$ Orionis. Observational dust disc masses are taken from \cite{Ansdell2017a} and converted to gas masses by assuming the standard dust-to-gas ratio of $0.01$. Stellar accretion rates are taken from \cite{Rigliaco2011}.}
      \label{fig:mass_acc_adams_noshielding}
   \end{figure*}

   To investigate the effect of our prescription for the shielding of internal photoevaporation by MHD disc winds, we compare the nominal results with where internal photoevaporation is active throughout the whole evolution of the disc. We provide the disc lifetimes in Fig.~\ref{fig:discfraction_noshielding} and the mass and accretion rate evolution for the case with weak FUV field in Fig.~\ref{fig:mass_acc_lowFUV_noshielding}, with the quantitative results in Table~\ref{tab:numbers_lowFUV_noshielding}, while the same for the strong FUV environment are provided in Fig.~\ref{fig:mass_acc_adams_noshielding} and Table~\ref{tab:numbers_adams_noshielding}.
   
   The results are almost identical to those presented in the main text. The only noticeable differences are: slightly shorter disc lifetimes, a lower number of discs with non-zero stellar accretion rate, and a higher percentage of mass removal by internal photoevaporation. Nevertheless, the main results are unaffected by the inclusion of internal photoevaporation through the disc lifetimes. We still obtain the result that weak disc winds in combination with a strong constant torque are necessary to reproduce stellar accretion rates comparable with observations and observed disc lifetime distributions. This confirms that our prescription for internal photoevaporation is sufficiently weak to avoid affecting disc evolution before the dispersal stage.

   \begin{table*}
      \begin{center}
         \caption{Characteristics for different weak FUV field populations without EUV radiation shielding.}
         \label{tab:numbers_lowFUV_noshielding}
         \begin{tabular}{l c c c r r r r}
            \hline\hline
               DW scenario & $t_{1/2}$\textsuperscript{a)} & $f_\mathrm{disc,2Myr}$\textsuperscript{b)} & $f_\mathrm{cavity}$\textsuperscript{c)} & $f_{M_\mathrm{acc}}$\textsuperscript{d)} & $f_{M_\mathrm{PEW,int}}$\textsuperscript{d)} & $f_{M_\mathrm{PEW,ext}}$\textsuperscript{d)} & $f_{M_\mathrm{MDW}}$\textsuperscript{d)} \\
               & [\SI{}{\mega\year}] & [\SI{}{\percent}] & [\SI{}{\percent}] & [\SI{}{\percent}] & [\SI{}{\percent}] & [\SI{}{\percent}] & [\SI{}{\percent}] \\
            \hline
               strong DW + $\Sigma$-dep. torque  & 1.14\SI{}{}   &   30\SI{}{} &  58\SI{}{} & 0.6\SI{}{}   & 12.2\SI{}{} & 52.4\SI{}{} & 34.8\SI{}{} \\
               weak DW + $\Sigma$-dep. torque    & 5.95\SI{}{}   &  87\SI{}{} &  37\SI{}{} &  8.4\SI{}{}   & 22.9\SI{}{} & 53.8\SI{}{} & 15.0\SI{}{} \\
               strong DW + const. torque         & \SI{0.75}{}   &   \SI{8}{} &  \SI{63}{} &  \SI{0.1}{}   &  \SI{9.3}{} & \SI{49.3}{} & \SI{41.3}{} \\
               weak DW + const. torque           & \SI{6.29}{}   &  \SI{81}{} &  \SI{43}{} & \SI{13.3}{}   & \SI{18.8}{} & \SI{49.8}{} & \SI{18.0}{} \\
            \hline
         \end{tabular}
      \end{center}
      Notes: a) time at which half of the discs are dispersed, b) fraction of discs remaining at \SI{2}{\mega\year}, c) fraction of discs opening up an inner cavity, d) contribution of the different processes to disc dispersal in terms of mass loss percentage (mean value over all simulations)
   \end{table*}

   \clearpage

   \begin{table*}
      \begin{center}
         \caption{Characteristics for different strong FUV field populations without EUV radiation shielding.}
         \label{tab:numbers_adams_noshielding}
         \begin{tabular}{l c c c r r r r}
            \hline\hline
               DW scenario & $t_{1/2}$\textsuperscript{a)} & $f_\mathrm{disc,2Myr}$\textsuperscript{b)} & $f_\mathrm{cavity}$\textsuperscript{c)} & $f_{M_\mathrm{acc}}$\textsuperscript{d)} & $f_{M_\mathrm{PEW,int}}$\textsuperscript{d)} & $f_{M_\mathrm{PEW,ext}}$\textsuperscript{d)} & $f_{M_\mathrm{MDW}}$\textsuperscript{d)} \\
               & [\SI{}{\mega\year}] & [\SI{}{\percent}] & [\SI{}{\percent}] & [\SI{}{\percent}] & [\SI{}{\percent}] & [\SI{}{\percent}] & [\SI{}{\percent}] \\
            \hline
               strong DW + $\Sigma$-dep. torque  & 0.57\SI{}{}   &   9\SI{}{} &  32\SI{}{} & 0.3\SI{}{}   & 3.1\SI{}{} & 86.4\SI{}{} &  10.2\SI{}{} \\
               weak DW + $\Sigma$-dep. torque    & 1.25\SI{}{}   &  36\SI{}{} &  12\SI{}{} &  3.3\SI{}{}   & 5.5\SI{}{} & 87.4\SI{}{} &  3.9\SI{}{} \\
               strong DW + const. torque         & \SI{0.42}{}   &   \SI{3}{} &  \SI{34}{} &  \SI{0.1}{}   & \SI{2.5}{} & \SI{84.6}{} & \SI{12.9}{} \\
               weak DW + const. torque           & \SI{0.88}{}   &  \SI{29}{} &  \SI{22}{} &  \SI{5.3}{}   & \SI{4.4}{} & \SI{85.2}{} &  \SI{5.1}{} \\
            \hline
         \end{tabular}
      \end{center}
      Notes: a) time at which half of the discs are dispersed, b) fraction of discs remaining at \SI{2}{\mega\year}, c) fraction of discs opening up an inner cavity, d) contribution of the different processes to disc dispersal in terms of mass loss percentage (mean value over all simulations)
   \end{table*}

   \clearpage
   
   \section{No internal photoevaporation}
   \label{app:nointPEW}
   To asses how internal photoevaporation affect the dispersal of protoplanetary discs, we computed the other extreme case where internal photoevaporation has been removed altogether. We proceeded as detailed in the main text and Appendix~\ref{app:noshielding}, namely, by analysing disc lifetimes and the evolution of disc masses and stellar accretion rates.
   
\begin{figure}[h]
      \centering
      \includegraphics[width=0.9\hsize]{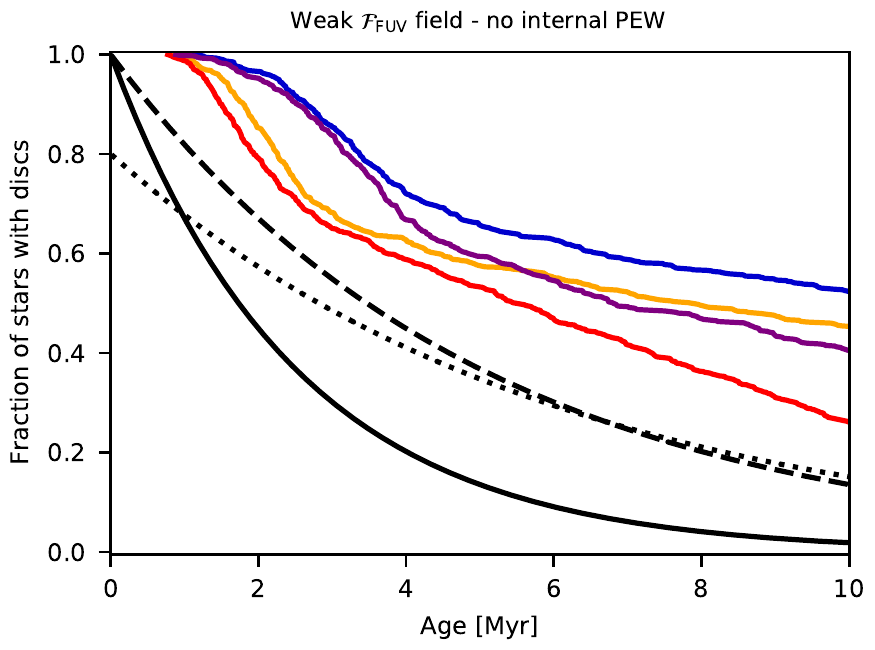} \\
      \includegraphics[width=0.9\hsize]{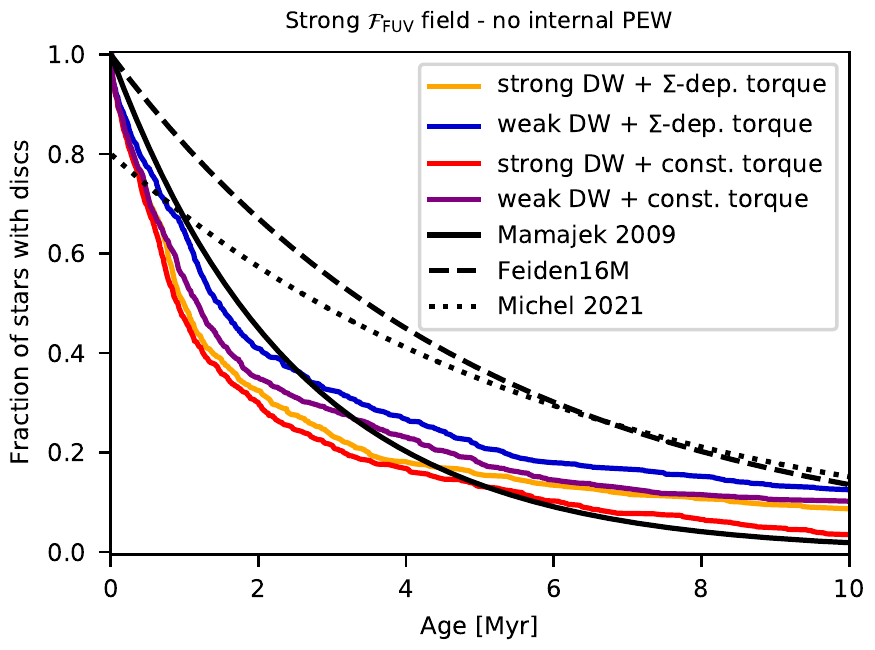}
      \caption{Fraction of stars possessing a circumstellar disc as a function of time, analogous to Fig. \ref{fig:discfraction}, but for populations without internal photoevaporation.}
      \label{fig:discfraction_nointPEW}
   \end{figure}

   In Fig. \ref{fig:discfraction_nointPEW}, we show the time evolution of the fraction of stars showing NIR signatures of a disc. When comparing to the simulations including internal photoevaporation it is evident that NIR lifetimes are generally longer without internal photoevaporation. This comes as no surprise as internal photoevaporation is removing mass from the disc which is then lacking to resupply the inner disc. In our simulations, discs are primarily dispersed by external photoevaporation, where as for the weak FUV field, MHD winds remove substantial fractions of gas. This combined with the longevity of the discs, we come to the conclusion that disc lifetimes cannot be reproduced by MHD winds only for the chosen initial conditions and disc wind scenarios. However, things may look different for stronger disc wind torques or radially dependent $\overline{\alpha_{r \phi}}$.
   
\begin{figure*}
      \centering
      \includegraphics[width=0.9\hsize]{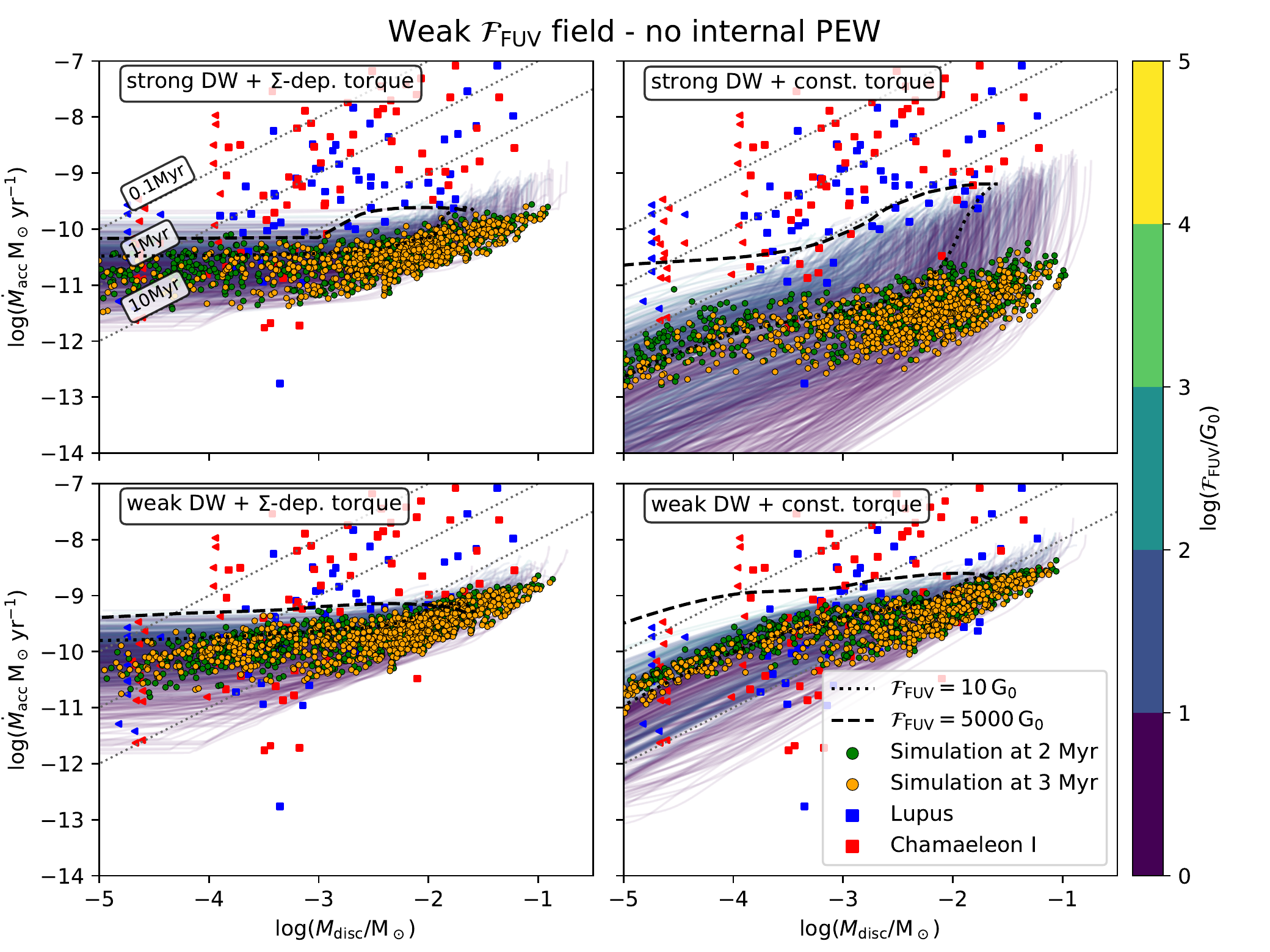}
      \caption{Stellar accretion rate $\dot{M}_\mathrm{acc}$ vs. gas disc mass $M_\mathrm{disc}$, analogous to Fig. \ref{fig:mass_acc_lowFUV}, but without internal photoevaporation. Exemplary cases are the same as in Figs.~\ref{fig:exemplary_case_10g0} and \ref{fig:exemplary_case_5e3g0} but without internal photoevaporation. Snapshots at $2\,\mathrm{Myr}$ (green circles) and $3\,\mathrm{Myr}$ (orange circles) are displayed. We compare our simulations with observed populations in Lupus and Chamaeleon I. Observational dust disc masses and stellar accretion rates are taken from \cite{Manara2019} and dust masses are converted to gas masses by assuming the standard dust-to-gas ratio of $0.01$. Triangles denote upper limits on disc mass. Lines of constant $M_\mathrm{disc}/\dot{M}_\mathrm{acc}$ are shown for 0.1, 1 and 10 Myr (thin dotted lines).}
      \label{fig:mass_acc_lowFUV_nointPEW}
   \end{figure*}

\begin{figure*}
      \centering
      \includegraphics[width=0.9\hsize]{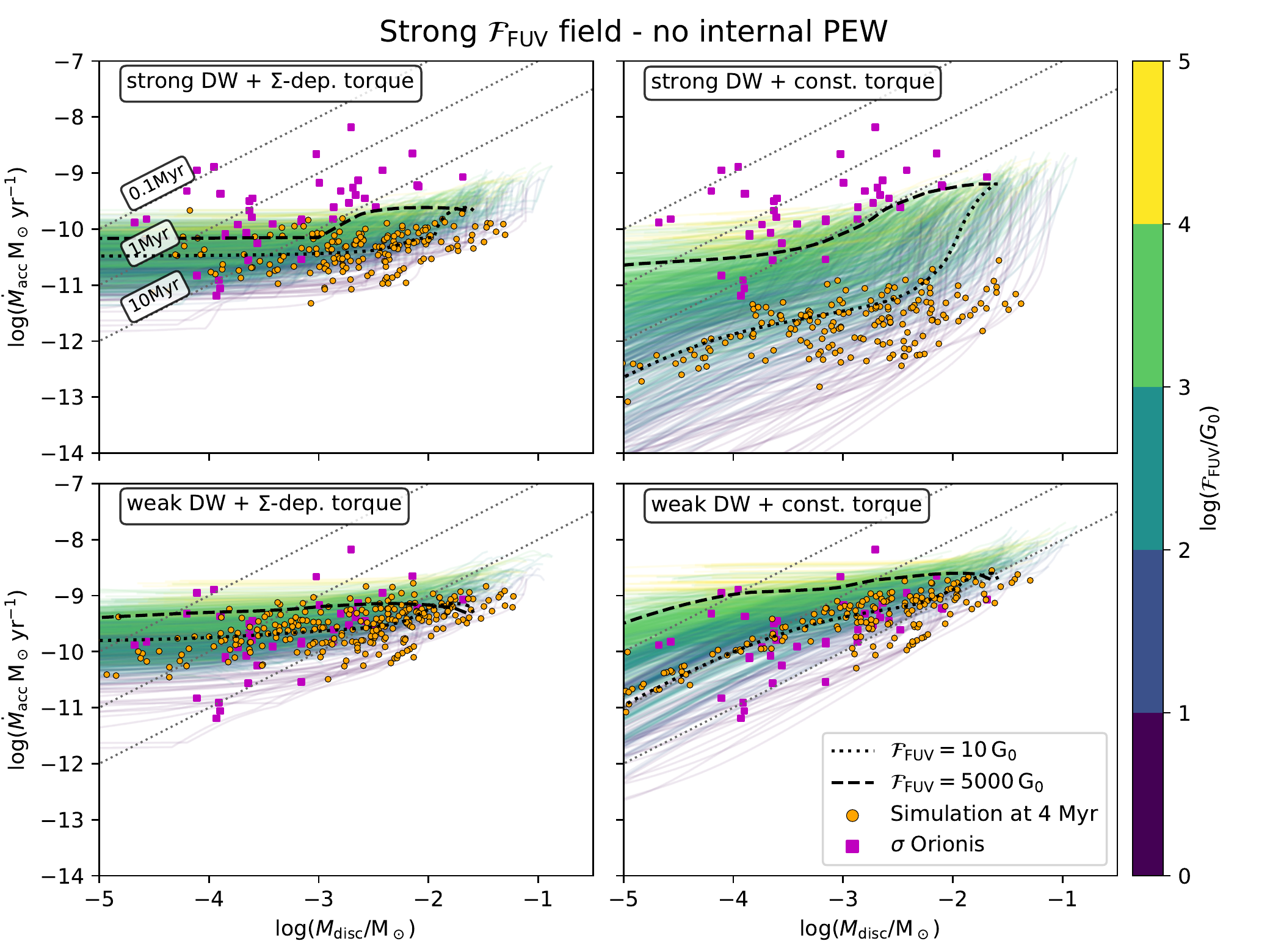}
      \caption{Stellar accretion rate $\dot{M}_\mathrm{acc}$ vs. gas disc mass $M_\mathrm{disc}$, analogous to Fig. \ref{fig:mass_acc_adams}, but without internal photoevaporation. Exemplary cases are the same as in Figs.~\ref{fig:exemplary_case_10g0} and \ref{fig:exemplary_case_5e3g0} but without internal photoevaporation. A snapshot for the individual systems is shown at $4\,\mathrm{Myr}$ (orange circles) in comparison with observational data from $\sigma$ Orionis. Observational dust disc masses are taken from \cite{Ansdell2017a} and converted to gas masses by assuming the standard dust-to-gas ratio of $0.01$. Stellar accretion rates are taken from \cite{Rigliaco2011}.}
      \label{fig:mass_acc_adams_nointPEW}
   \end{figure*}

   Analogously to Sect. \ref{subsec:macc_disc}, we show evolution tracks for disc populations in the $M_\mathrm{disc}-\dot{M}_\mathrm{acc}$ plane for different disc wind scenarios and ambient FUV field strength distributions in Figs. \ref{fig:mass_acc_lowFUV_nointPEW} and \ref{fig:mass_acc_adams_nointPEW}. The corresponding characteristic numbers are listed in Tables \ref{tab:numbers_lowFUV_nointPEW} and \ref{tab:numbers_adams_nointPEW}.
   
   The most striking difference is that the stellar accretion rates stay high during the evolution of the disc and do not drop while the disc is still massive. This is because the internal photoevaporation is largely responsible for the opening of the inner cavity -- and it is the opening of the inner cavity that leads the stellar accretion rates to drop sharply. We note that inner cavities are only opened in the case of a strong wind paired with a $\Sigma$-dependent torque. The main difference between these populations and the ones shown in the main text is the sudden drop in accretion rate when it reaches $\sim 10^{-11} \mathrm{M}_\odot /\mathrm{yr}$. In the absence of internal photoevaporation, no such drop is observed. However, the location of the synthetic discs in the $M_\mathrm{disc}-\dot{M}_\mathrm{acc}$ plane in the snapshots does not change significantly in the absence of internal photoevaporation. As discussed above, disc lifetimes are strongly affected by the internal photoevaporation. As a consequence, some of the discs in the populations without internal photoevaporation exhibit a more extended disc dispersal stage, with relatively low masses and stellar accretion rates.
   
   We conclude that in our model, internal photoevaporation (i.e. EUV photoevaporation only) only influences the late stage of protoplanetary disc evolution. It therefore has a big influence on the presence of inner cavities and disc lifetimes, but it does not have strong influence on the location where we find snapshots of our populations in the $M_\mathrm{disc}$-$\dot{M}_\mathrm{acc}$ plane. Thus, our findings are not affected by the inclusion of internal photoevaporation.

   \clearpage

   \begin{table*}
      \begin{center}
         \caption{Characteristics for different weak FUV field populations without internal photoevaporation.}
         \label{tab:numbers_lowFUV_nointPEW}
         \begin{tabular}{l c c c r r r r}
            \hline\hline
               DW scenario & $t_{1/2}$\textsuperscript{a)} & $f_\mathrm{disc,2Myr}$\textsuperscript{b)} & $f_\mathrm{cavity}$\textsuperscript{c)} & $f_{M_\mathrm{acc}}$\textsuperscript{d)} & $f_{M_\mathrm{PEW,int}}$\textsuperscript{d)} & $f_{M_\mathrm{PEW,ext}}$\textsuperscript{d)} & $f_{M_\mathrm{MDW}}$\textsuperscript{d)} \\
               & [\SI{}{\mega\year}] & [\SI{}{\percent}] & [\SI{}{\percent}] & [\SI{}{\percent}] & [\SI{}{\percent}] & [\SI{}{\percent}] & [\SI{}{\percent}] \\
            \hline
               strong DW + $\Sigma$-dep. torque  &  7.93\SI{}{}   &  85\SI{}{}   &  21\SI{}{} &  3.0\SI{}{}   & 0.0\SI{}{} & 59.8\SI{}{} & 37.3\SI{}{} \\
               weak DW + $\Sigma$-dep. torque    & 11.47\SI{}{}    & 97\SI{}{}   &   0\SI{}{} & 19.6\SI{}{}   & 0.0\SI{}{} & 61.6\SI{}{} & 18.8\SI{}{} \\
               strong DW + const. torque         &  \SI{5.35}{}   &  \SI{79}{} &   \SI{0}{} &  \SI{0.3}{}   & \SI{0.0}{} & \SI{52.4}{} & \SI{47.3}{} \\
               weak DW + const. torque           &  \SI{6.87}{}   &  \SI{95}{} &   \SI{0}{} & \SI{20.5}{}   & \SI{0.0}{} & \SI{60.4}{} & \SI{19.1}{} \\
            \hline
         \end{tabular}
      \end{center}
      Notes: a) time at which half of the discs are dispersed, b) fraction of discs remaining at \SI{2}{\mega\year}, c) fraction of discs opening up an inner cavity, d) contribution of the different processes to disc dispersal in terms of mass loss percentage (mean value over all simulations)
   \end{table*}

   \clearpage

   \begin{table*}
      \begin{center}
         \caption{Characteristics for different strong FUV field populations without internal photoevaporation.}
         \label{tab:numbers_adams_nointPEW}
         \begin{tabular}{l c c c r r r r}
            \hline\hline
               DW scenario & $t_{1/2}$\textsuperscript{a)} & $f_\mathrm{disc,2Myr}$\textsuperscript{b)} & $f_\mathrm{cavity}$\textsuperscript{c)} & $f_{M_\mathrm{acc}}$\textsuperscript{d)} & $f_{M_\mathrm{PEW,int}}$\textsuperscript{d)} & $f_{M_\mathrm{PEW,ext}}$\textsuperscript{d)} & $f_{M_\mathrm{MDW}}$\textsuperscript{d)} \\
               & [\SI{}{\mega\year}] & [\SI{}{\percent}] & [\SI{}{\percent}] & [\SI{}{\percent}] & [\SI{}{\percent}] & [\SI{}{\percent}] & [\SI{}{\percent}] \\
            \hline
               strong DW + $\Sigma$-dep. torque  & 0.99\SI{}{}   &  32\SI{}{} &  9\SI{}{} & 0.8\SI{}{}   & 0.0\SI{}{} & 87.6\SI{}{} &  11.7\SI{}{} \\
               weak DW + $\Sigma$-dep. torque    & 1.47\SI{}{}   &  41\SI{}{} &   0\SI{}{} & 6.2\SI{}{}   & 0.0\SI{}{} & 88.7\SI{}{} &  5.1\SI{}{} \\
               strong DW + const. torque         & \SI{0.86}{}   &  \SI{28}{} &   \SI{0}{} & \SI{0.2}{}   & \SI{0.0}{} & \SI{85.2}{} & \SI{14.6}{} \\
               weak DW + const. torque           & \SI{1.10}{}   &  \SI{35}{} &   \SI{0}{} & \SI{7.1}{}   & \SI{0.0}{} & \SI{87.2}{} &  \SI{5.6}{} \\
            \hline
         \end{tabular}
      \end{center}
      Notes: a) time at which half of the discs are dispersed, b) fraction of discs remaining at \SI{2}{\mega\year}, c) fraction of discs opening up an inner cavity, d) contribution of the different processes to disc dispersal in terms of mass loss percentage (mean value over all simulations)
   \end{table*}

\end{appendix}

\end{document}